\documentclass[iop,apj,numberedappendix,appendixfloats]{emulateapj}
\usepackage{color}
\usepackage{amsmath}

\newcommand{\solar}{$_{\odot}$}
\newcommand{\tco}{$^{12}$CO}
\newcommand{\ttco}{$^{13}$CO}
\newcommand{\ceto}{C$^{18}$O}
\newcommand{\hcop}{HCO$^+$}

\newcommand{\joz}{$J$=1$\rightarrow$0}
\newcommand{\jto}{$J$=2$\rightarrow$1}

\newcommand{\kms}{\,km\,s$^{-1}$}
\newcommand{\degree}{$^{\circ}$}
\newcommand{\fdeg}{$^{\circ}$\hspace{-1mm}.}

\newcommand{\tex}{$T_{\rm ex}$}
\newcommand{\tmb}{$T_{\rm mb}$}
\newcommand{\tbg}{$T_{\rm bg}$}
\newcommand{\vlsr}{$V_{\rm LSR}$}

\newcommand{\htwo}{H$_2$}

\newcommand{\nhtwo}{$N_{\rm H_2}$}
\newcommand{\nco}{$N_{\rm ^{12}CO}$}
\newcommand{\nttco}{$N_{\rm ^{13}CO}$}
\newcommand{\ico}{$I_{\rm ^{12}CO}$}
\newcommand{\ittco}{$I_{\rm ^{13}CO}$}

\def\lapp{\ifmmode\stackrel{<}{_{\sim}}\else$\stackrel{<}{_{\sim}}$\fi}
\def\gapp{\ifmmode\stackrel{>}{_{\sim}}\else$\stackrel{>}{_{\sim}}$\fi}

\shorttitle{CHaMP IV. Molecular Clump Dynamical Evolution}
\shortauthors{Barnes et al.}

\begin{document}

\title{The Galactic Census of High- and Medium-mass Protostars. \\
    IV.  Molecular Clump Radiative Transfer, Mass Distributions, \\
    Kinematics, and Dynamical Evolution}

\author{Peter J. Barnes\altaffilmark{1,2}, Audra K. Hernandez\altaffilmark{3}, Erik Muller\altaffilmark{4}, and Rebecca L. Pitts\altaffilmark{1} \\
}
\email{pjb@ufl.edu}

\altaffiltext{1}{Astronomy Department, University of Florida, P.O. Box 112055, Gainesville, FL 32611, USA}
\altaffiltext{2}{School of Science and Technology, University of New England, Armidale NSW 2351, Australia}
\altaffiltext{3}{Astronomy Department, University of Wisconsin, 475 North Charter St., Madison, WI 53706, USA}
\altaffiltext{4}{National Astronomical Observatory of Japan, Chile Observatory, 2-21-1 Osawa, Mitaka, Tokyo 181-8588, Japan}

\begin{abstract}
We present \tco, \ttco, \& \ceto\ data as the next major release for the CHaMP project, an unbiased sample of Galactic molecular clouds in $l$ = 280\degree--300\degree.  From a radiative transfer analysis, 
we self-consistently compute 3D cubes of optical depth, excitation temperature, and column density for 
$\sim$300 massive clumps, and update the \ico-dependent CO$\rightarrow$\htwo\ conversion law of \citet{bm15}.  For $N$ $\propto$ $I^p$, we find $p$ = 1.92$\pm$0.05 for the velocity-resolved conversion law aggregated over all clumps. 
A practical, integrated conversion law is \nco\ = (4.0$\pm$0.3)$\times$10$^{19}$m$^{-2}$\,\ico$^{1.27\pm0.02}$, confirming an overall 2$\times$ higher total molecular mass for Milky Way clouds, compared to the standard $X$ factor.

We use these laws to compare the kinematics of clump interiors with their foreground \tco\ envelopes, and find evidence that most clumps are not dynamically uniform: irregular portions seem to be either slowly accreting onto the interiors, or dispersing from them.  We compute the spatially-resolved mass accretion/dispersal rate across all clumps, and map the local flow timescale.  While these flows are not clearly correlated with clump structures, the inferred accretion rate is a statistically strong function of the local mass surface density $\Sigma$, suggesting near-exponential growth or loss of mass over effective timescales $\sim$30--50\,Myr.  At high enough $\Sigma$, accretion dominates, suggesting gravity plays an important role in both processes.  If confirmed by numerical simulations, this sedimentation picture would support arguments for long clump lifetimes mediated by pressure confinement, 
with a terminal crescendo of star formation, suggesting a resolution to the 40-yr-old puzzle of the dynamical state of molecular clouds and their low star formation efficiency.
\end{abstract} 


\section{Introduction}

Understanding the origin, lifetimes, and star-forming efficiencies of molecular clouds in gas-rich galaxies is of crucial importance for understanding our cosmic origins.  Despite more than four decades of work on this subject, the physical processes that drive molecular cloud evolution, and the initial conditions leading to star formation, are still widely debated.  Outstanding issues include the timescale for conversion of atomic to molecular hydrogen \citep[e.g.,][]{gm07a,gm07b,bsl15,kk18}, and whether the kinematics of cloud formation are dominated by bulk gas flow (usually suggesting shorter evolutionary timescales) or quasi-static turbulence (typically accompanied by longer timescales) \citep{mo07,pf14}.  Also, although CO has long been thought a reliable tracer of molecular clouds and their dynamics, we now recognise the role of CO-dark gas in the intermediate stages of molecular cloud formation \citep{p13,s14,hvb15}.  Furthermore, recent revisions to mass conversion laws also change derived star formation efficiencies, leading to longer global timescales for cloud evolution \citep{bm15}.

Central to many of the above issues is having accurate molecular cloud masses and kinematics available for study.  Large, systematic, and sensitive observational samples of star-forming clouds are needed, in order to provide the statistical basis for comparing their physical properties with analytical theory or numerical simulations.  But until recently, few such surveys existed, in particular when adding the requirement of high spatial dynamic range, i.e., wide-field ($>$10\,pc) maps with sub-pc resolution, so that one could examine individual cluster-forming molecular clumps, while surveying a large population in their complete environmental setting.  Although the CfA survey of \tco\ across the whole Milky Way \citep{dht01} provided fundamental data on all Giant Molecular Clouds (GMCs) in our Galaxy, at 8$'$ resolution (7\,pc at typical distances of 3\,kpc) it could not resolve individual pc-scale clumps.

Studies like the Galactic Ring Survey in \ttco\ \citep[GRS][]{j06}, however, at 0.8$'$ resolution, could do so, but dilemmas remained.  Even the CfA and GRS data are in single spectral lines, so analysis of these data depends on making certain assumptions about the virial state of clouds, or the excitation and/or opacity of emission lines, the molecular abundances, and so on.  In other words, the central dilemma is the conversion of emission measures (whether from molecular spectroscopy or dust continuum) to masses, since in general emissivity $\neq$ mass.  As a consequence of this conversion problem, single tracers are also susceptible to biases in deriving the kinematics, dynamics, and evolution of molecular clouds, since each may emphasise different components of cloud structure and kinematics, yielding potentially confusing or conflicting information about cloud physics.  These difficulties propagate into efforts to accurately determine the connection between the natal gas in star-forming clouds, and the collective properties of their stellar end-product.  In contrast, the MALT90 survey \citep{j13} solves the single-tracer problem, but does so only for small maps around the highest column density submm clumps, even though it covers $\sim$2,000 of them throughout the fourth quadrant of the Milky Way.  Thus, MALT90's deliberate selection bias limits its exposure to the full range of molecular clump properties.

For these and other reasons, we designed the Galactic {\em Census of High- and Medium-mass Protostars} (CHaMP), to survey all massive, pc-scale, cluster-forming clumps within a 20\degree$\times$6\degree\ area of the southern Milky Way, and to do so in multiple ($>$30) molecular species at the same resolution (0.6$'$) and line sensitivity ($\sim$0.5\,K).  CHaMP results reported previously include the first complete dense gas data in \hcop\ on the molecular clump population of $\sim$300 clouds in the Nanten Master Catalogue \citep[][hereafter Paper I]{b11}; systematic near-IR broad- and narrowband imaging of the clumps' embedded stellar content plus nebular diagnostics \citep{b13}; preliminary SED analysis of the clumps' dust emission \citep{mtb13}; and the first analysis of \tco\ emission from the clumps' embedding envelopes \citep[][hereafter Paper III]{bh16}.  Paper I found evidence for long (up to 100\,Myr) timescales for clump evolution, based on the population demographics, supporting similar conclusions from an extragalactic setting \citep{kss09}.  Paper III provided a physical basis to support the long lifetime scenario, based on the measured pressure confinement of clumps' denser interiors by the less dense but more massive envelopes.

In this paper, we report results and analysis of the complete iso-CO mapping (i.e., \tco, \ttco, and \ceto) of the CHaMP clumps (\S\S\ref{obsred}--\ref{radxfer}).  Using these data, we present a more sensitive analysis of the implied mass conversion laws first described by \citet{bm15} as part of the ThrUMMS project (\S\S\ref{mass}--\ref{conv}).  With these new laws we examine the dynamics and implied evolution of the massive dense clump population (\S\ref{kinem}), and for the first time, may directly measure the timescales associated with mass accretion and dispersal in the clump population.  For accretion we find that these match well with the demographically-implied values from Paper I.  We conclude with a discussion comparing these results with current models (\S\ref{disc}).

We also provide complete sets of data products for all clumps in 4 Appendices, but to illuminate the discussion in the body of the paper, we show either sample maps/plots of single Regions, or aggregated results.  The organisation of the Appendices by Region is similar to that in Papers I and III. 
Appendix \ref{rgbimages} contains the basic map and line ratio data: 3-colour overlays of both ($l$,$b$) and position-velocity integrated intensities, line ratio diagrams, $X$ vs $I$ plots, and derived $X$ maps (see below for definitions).  Appendix \ref{physmaps} contains integrated or averaged maps of the four basic radiative transfer quantities, $\tau_{12}$, \tex, \nco, and $R_{18}$.  Appendix \ref{zmaps} shows mass surface density weighted moment maps of each field, i.e., integrated $\Sigma$, \vlsr, and $\sigma_V$.  Appendix \ref{dynmaps} gives the maps of momentum, differential envelope motion, mass flux, and timescales from the dynamical analysis in \S\ref{kinem}.

\section{Observations and Data Reduction}\label{obsred}

The molecular line data were obtained with the Mopra radio telescope\footnote{The Mopra telescope is part of the Australia Telescope, funded by the Commonwealth of Australia for operation as a National Facility managed by CSIRO. The University of New South Wales Digital Filter Bank used for observations with the Mopra telescope was provided with support from the Australian Research Council.} 
during 2004--12, using the standard on-the-fly (OTF) mapping control software, and the MOPS digital filterbank.  The design and execution of the observational strategy, data reduction and analysis procedures, and later updates to the processing pipeline and analysis techniques, were extensively described in Papers I and III.  Briefly, the 2004--07 seasons (Phase I) were used to make maps of 16 molecular emission lines in the range 85--93\,GHz across $\sim$30 Regions in the 20\degree$\times$6\degree\ CHaMP survey window.  This frequency range captured simultaneous maps of many standard dense gas tracers of these Regions, including the \hcop\ results reported in Paper I (physical data on 303 massive pc-scale dense clumps, based on the prior Nanten Master Catalogue of 209 larger clouds, designated by their ``BYF'' number).  Each map is of size $\sim$0.1\degree--0.9\degree, and covers all molecular clouds in Vela, Carina, and Centaurus (300\degree$>$$l$$>$280\degree) with detectable \hcop\ or \ceto\ emission, i.e., all those of sufficient mass or density to be capable of forming stars.

\notetoeditor{}
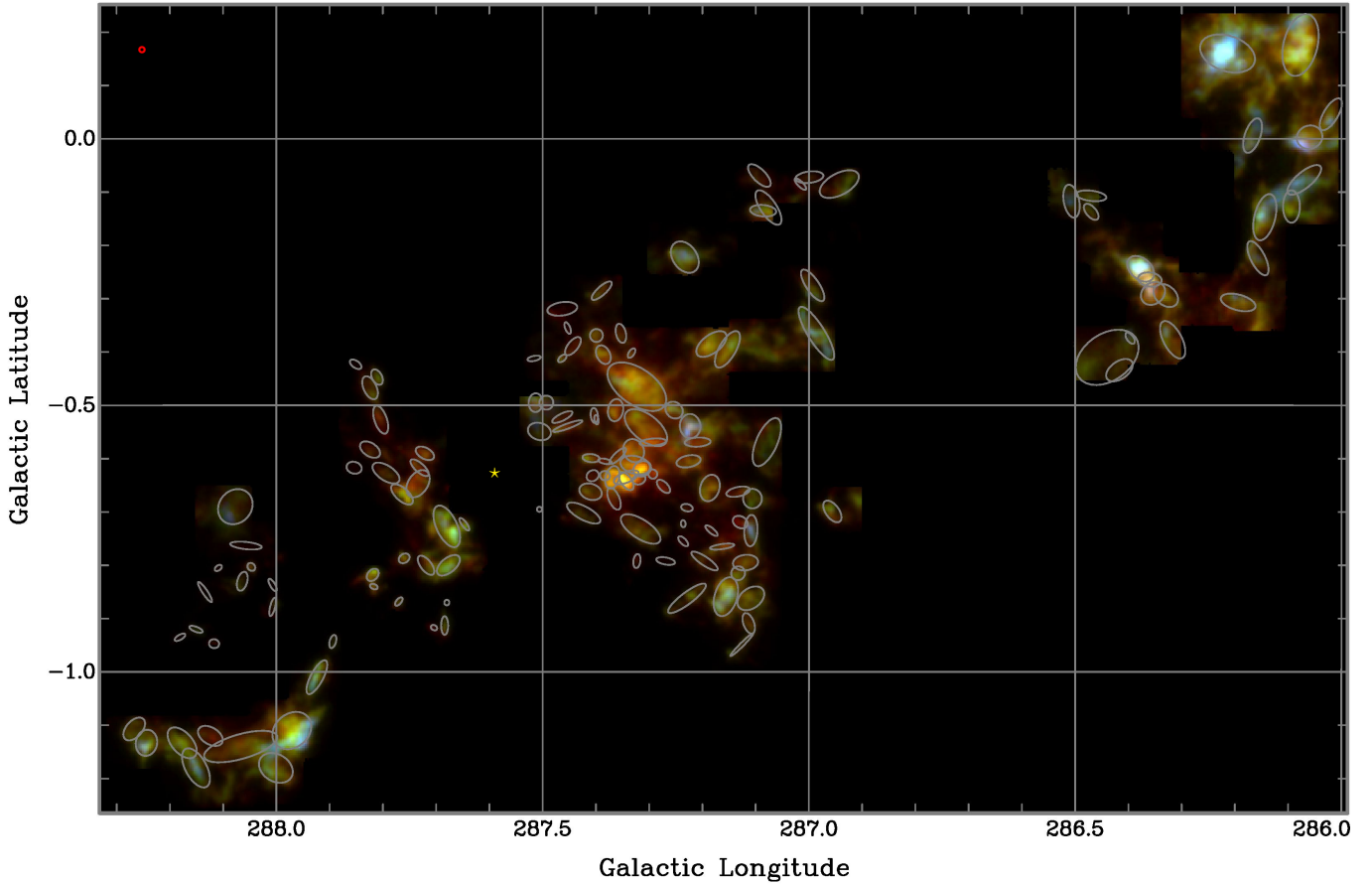
\begin{figure*}[t]
\includegraphics[angle=-90,scale=0.65]{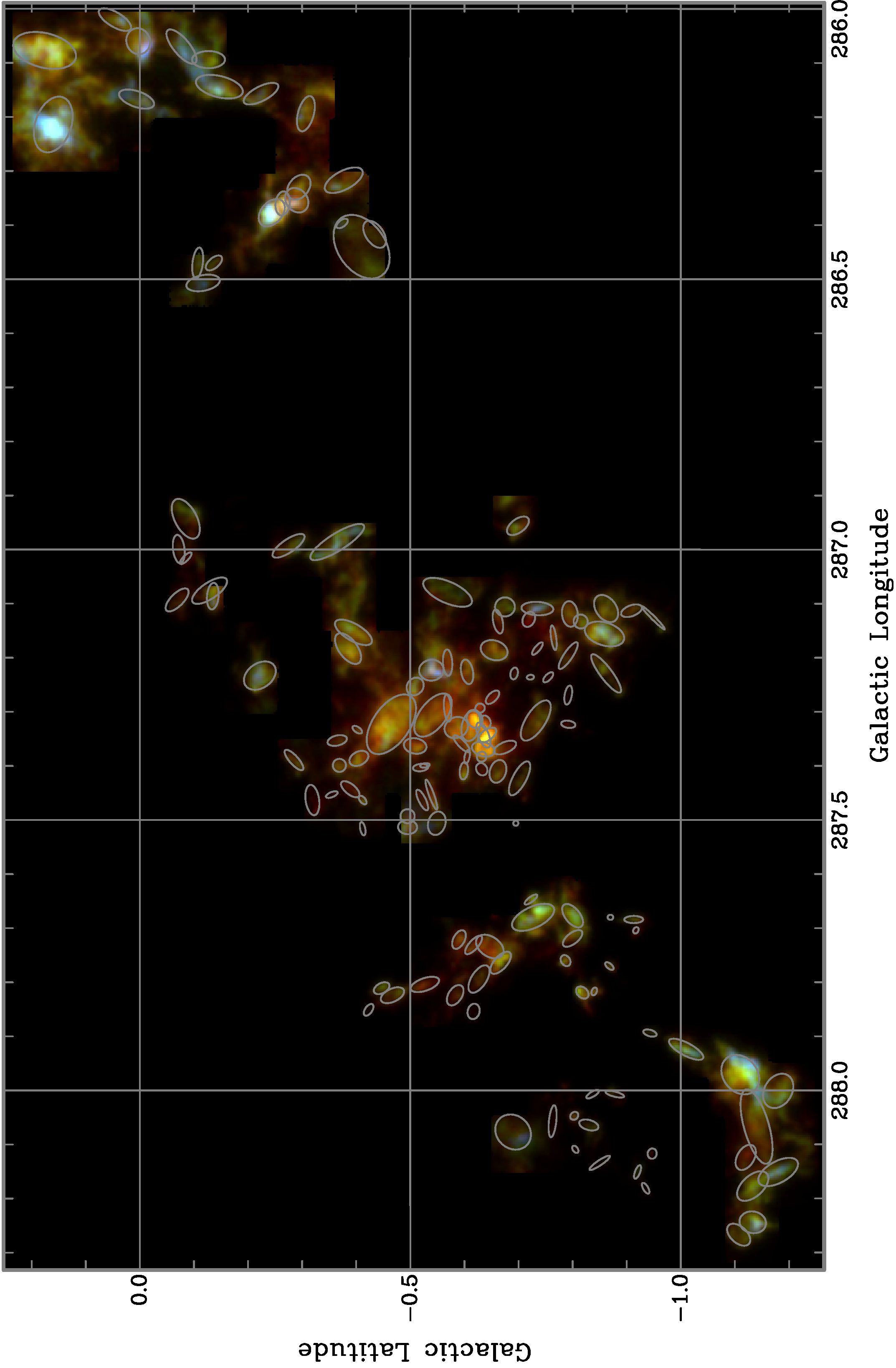}
\caption{
Pseudocolour image of \tco\ (red), \ttco\ (green), and \ceto\ (blue) \joz\ integrated intensities for the $\eta$ Carinae GMC, a mosaic of Regions 11, 10, \& 9 from left to right.  The location of $\eta$ Car itself is indicated by the yellow star at $l$=287.60, $b$=--0.63.  The ellipses are the same as in the bottom panel.  The Mopra HPBW at these frequencies is 37$''$ $\approx$ 0\fdeg01, shown as a red circle in the TL corner.  
}
\label{etacar}
\begin{picture}(1,1)
\thicklines
{\color{yellow}\put(185,195){$\star$}}
{\color{red}\put(50,358){\circle{2}}}
\end{picture}
\vspace*{-3mm}
\end{figure*}

The 2009--12 seasons (Phase II) saw mapping of these same Regions at 107--115\,GHz, simultaneously in the 3 main CO isotopologues plus 13 more exotic emission lines.  Paper III described the improvements and updates implemented for Phase II via the \tco\ results presented therein, including improved calibration, new error-correction and mitigation routines for the data processing, and improved imaging techniques such as smooth-and-mask (SAM).  Readers are referred to Papers I and III for more details.

Here we present the third major molecular line data release for CHaMP, the \ttco\ and \ceto\ maps.  The data processing procedures for these lines are essentially the same here as for \tco\ in Paper III.  Due to the simultaneity of the Mopra mapping through the MOPS spectrometer, the data quality in the two new species presented here is a close match to that in Paper III for \tco.  The only meaningful difference is that, because of the atmospheric O$_2$ emission, the rms noise levels in the $\sim$110\,GHz data (i.e., for \ttco\ and \ceto) are about 40\% lower than in the \tco\ data at 115\,GHz, with typical values of 0.4\,K/channel at 110\,GHz rather than 0.7\,K/ch at 115\,GHz.

The combination of the 3 species means that we can now perform more advanced analyses than were possible with the \tco\ data alone.  We detail these calculations in the following sections.

\section{Line Ratios and Radiative Transfer}\label{radxfer}

With emission lines of 3 isotopologues that should have a close chemical relation to each other, this very rich data set for a large, unbiased cloud sample allows two new types of analysis: a full radiative transfer study to derive the total CO mass distribution, and a dynamical analysis of the clump envelopes relative to their interiors.  We begin, however, with composite colour images of the integrated intensities in the 3 lines, for each Region and separate velocity component,  
plus 3-colour overlays for position-velocity (PV) maps as well.  A sample ($l$,$b$) mosaic for Regions 9--11 is shown in Figure \ref{etacar}.

This colour presentation gives us an intuitive feel for the line ratio analysis to follow.  By rendering the \tco\ emission as the red image, \ttco\ as green, and \ceto\ as blue, clumps which have a relatively low opacity and/or high excitation will appear reddish, since in either condition the \tco\ line will be significantly brighter than either the \ttco\ or \ceto\ lines.  In contrast, clumps that appear bluer or greener than average signify gas where the \ttco\ and/or \ceto\ emission is more comparable in brightness to the \tco, indicating higher opacity and/or lower excitation.  This is true for both the ($l$,$b$) and PV images.  The renderings in these figures are each adjusted to produce whitish clouds when the line ratios are more average, $I_{18}$/$I_{13}$ $\sim$ 0.2, $I_{13}$/$I_{12}$ $\sim$ 0.5, although this varies somewhat from Region to Region (see further below).

\notetoeditor{}
\begin{figure*}[t]
\vspace{-1mm}
\centerline{\includegraphics[angle=0,scale=0.19]{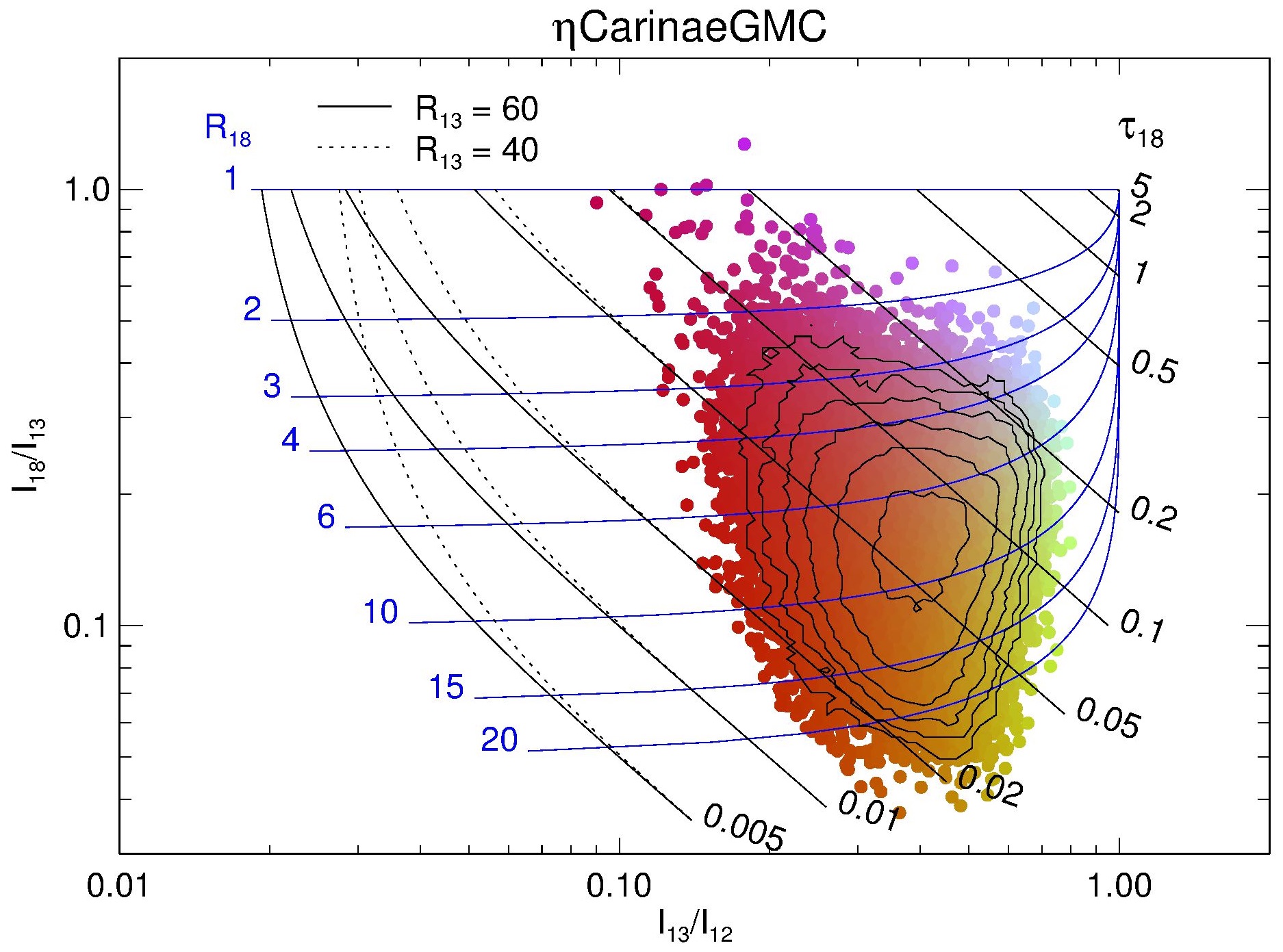}\hspace{4mm}}
\vspace{-1mm}
\caption{The iso-CO ratio-ratio diagram (RRD), combining line ratios from the three data cubes for all of Regions 9--11 (covering the $\eta$ Car GMC, the same area and 3-colour rendering as appears in Fig.\,\ref{etacar}).  The coloured dots are contoured by voxel incidence at 2,4,8...64, and overlaid by 2 grids (solid \& dotted, labelled by $R_{13}$) of $\tau$(\ceto) and $R_{18}$ = [\ttco]/[\ceto] from the radiative transfer analysis, Eqs.\,1--6.
}
\label{rrd}\vspace{-1mm}
\end{figure*}

From Figure \ref{etacar}, we can clearly see that the central portion of the GMC most closely surrounding $\eta$ Car and the various Trumpler clusters (i.e., most of Region 10 and the northern parts of Region 11) has higher excitation (redder colour) that the more distal clumps.  We can also see that clumps BYF\,73, 77, and 111 (the bluer clumps to the E and W) have particularly high opacity, even compared to their neighbours.

This description is quantified with the same radiative transfer analysis as performed by \citet{bm15}.  The approach is to assume, at each velocity channel and pixel (i.e., we do this calculation in 3D), that all lines are formed in Local Thermodynamic Equilibrium (LTE) at a single excitation temperature \tex.  Because of the high opacity in the \tco\ line, \tex\ is well-traced by the \tco\ brightness temperature:
\begin{equation} 
	T_{\rm mb} = [S_{\nu}(T_{\rm ex})-S_{\nu}(T_{\rm bg})](1-e^{-\tau}) \\
\end{equation}
\begin{displaymath}
	\hspace{18mm}\approx (T_{\rm ex}-T_{\rm bg})~,~~~\tau \gg 1~{\rm and}~h\nu \ll kT,
\end{displaymath}
where $\tau$ is the optical depth in the line, $T_{\rm mb}$ is the observed main beam brightness temperature corrected for the relevant antenna efficiencies, $T_{\rm bg}$ = 2.726\,K is the cosmic background temperature, and the source function $S_{\nu}$($T$) $\equiv$ $h\nu/k$(${\rm e}^{h\nu/kT}-1)$ at frequency $\nu$, with $h$ and $k$ as Planck's and Boltzmann's constants, respectively.  We show the Rayleigh-Jeans approximation to $S$ in the second version of Eq.\,(1) for illustrative purposes only: in all calculations herein, we use the full definition of $S$.  We also assume that all CHaMP clouds have a single abundance ratio $R_{13}$ $\equiv$ [\tco]/[\ttco] = 60, a number typical of molecular clouds near the solar circle \citep{gwb14}, including those in the Carina Arm of the Milky Way.  As explained by \citet{bm15}, however, the results of the calculations that follow are relatively insensitive to this assumption; this insensitivity is also illustrated in Figure \ref{rrd} (see next) via grids for both $R_{13}$ = 60 and 40.

The result is that we can evaluate not only the optical depth in all lines at every ($l$,$b$,$V$) position in the data cubes, but also the inherent $R_{18}$ $\equiv$ [\ttco]/[\ceto] abundance ratio at every coordinate, via
\begin{eqnarray} 
	\frac{T_{\rm 13}}{T_{\rm 12}} & = & \left[ \frac{S_{13}(T_{\rm ex})-S_{13}(T_{\rm bg})}{S_{12}(T_{\rm ex})-S_{12}(T_{\rm bg})} \right] \frac{1-e^{-\tau_{\rm 13}}}{1-e^{-R_{\rm 13}\tau_{\rm 13}}}~~, \\
	\frac{T_{\rm 18}}{T_{\rm 13}} & = & \left[ \frac{S_{18}(T_{\rm ex})-S_{18}(T_{\rm bg})}{S_{13}(T_{\rm ex})-S_{13}(T_{\rm bg})} \right] \frac{1-e^{-\tau_{\rm 18}}}{1-e^{-\tau_{\rm 13}}}~~, \\
	\tau_{\rm 12} & = & R_{\rm 13}\tau_{\rm 13}~~,~~{\rm and} \\
	\tau_{\rm 13} & = & R_{\rm 18}\tau_{\rm 18}~~,
\end{eqnarray}
where the subscripts 12, 13, and 18 refer to the $T_{\rm mb}$, $\tau$, and $S$ of the relevant isotopologue.  With $T_{\rm ex}$ from Eq.\,1, $R_{13}$ assumed, and $T_{\rm bg}$ known, these equations solve for $\tau_{13}$, $\tau_{18}$, $\tau_{12}$, and $R_{18}$ in turn.  The basic physics is illustrated in what we call the {\em iso-CO ratio-ratio diagram} (RRD): a composite example is given in Figure \ref{rrd} for all voxels in Regions 9--11. 
This diagram can be thought of analogously to optical and near-IR colour-colour diagrams (CCDs) widely used for various diagnostics and studies of stellar photometry and evolution.

Having solved for the $\tau$ and \tex\ at each voxel in a given Region, we can then simply evaluate the column density for any species over the same volume via\footnote{This formula corrects a typographical error from the version that appears in Paper I.}
\begin{equation} 
	N = \frac{3h}{8\pi^3\mu^2}~\frac{Q(T_{\rm ex})e^{E_u/kT_{\rm ex}}}{J_u(e^{h\nu/kT_{\rm ex}}-1)}~\int\tau{\rm d}V~~,
\end{equation}
where $\mu$ is the molecule's dipole moment, $Q$ is the rotational partition function, $E_u$ and $J_u$ are the energy and quantum number of the upper level of the transition at frequency $\nu$, and the integral is over each cloud's emission profile as a function of velocity $V$.  This expression can also be evaluated at each channel (or voxel) rather than as an integral, giving a cube of $N$ in each species as a function of ($l$,$b$,$V$).  The $\tau$, \tex, $R_{18}$, and $N$ cubes can then be processed with the usual moment calculations, or reprojected into position-velocity diagrams, as for regular data cubes.  Figure \ref{phsample} gives examples of such maps, 
while the same moments for all Regions appear in Appendix \ref{physmaps}.  

\notetoeditor{}
\begin{figure}[t]
\includegraphics[angle=-90,scale=0.317]{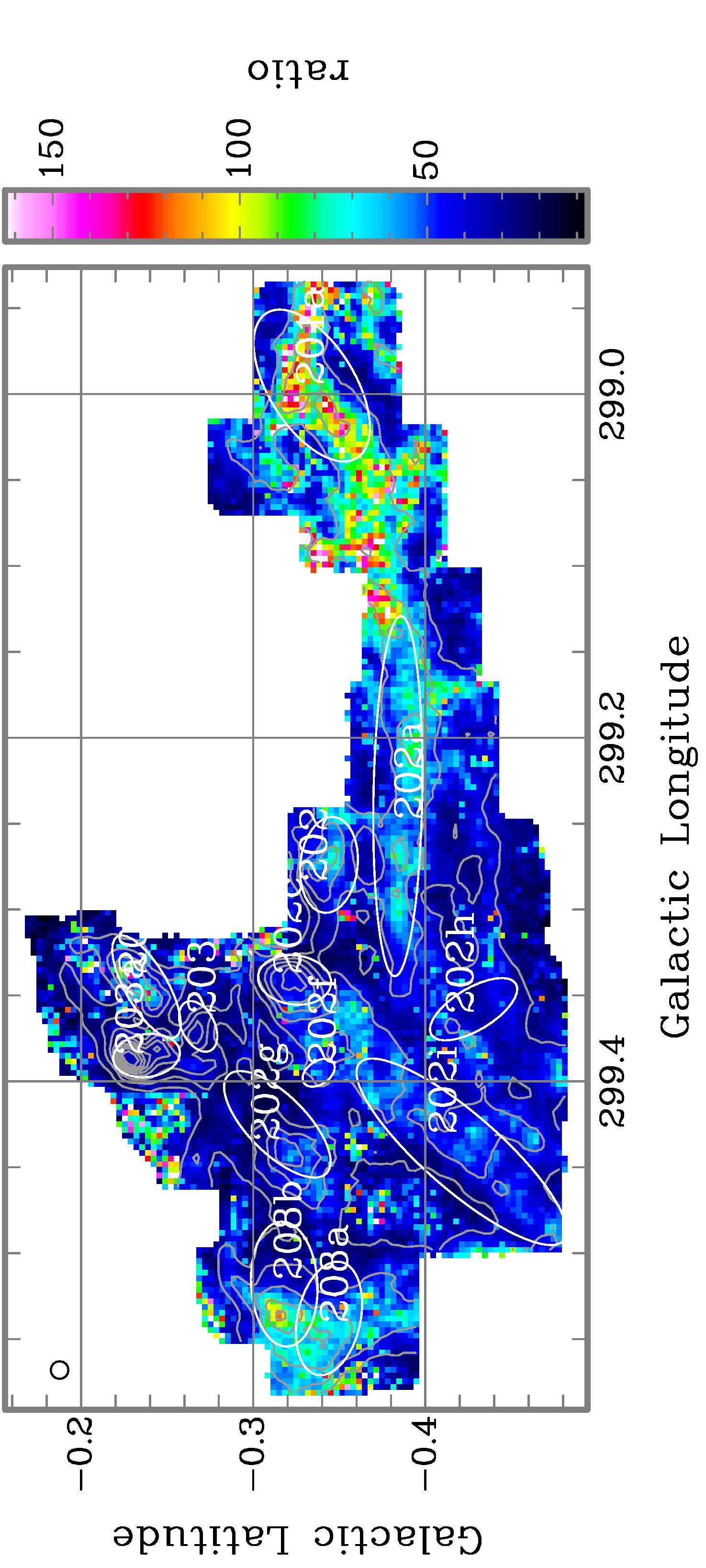} \vspace{-7.5mm}\\
\includegraphics[angle=-90,scale=0.317]{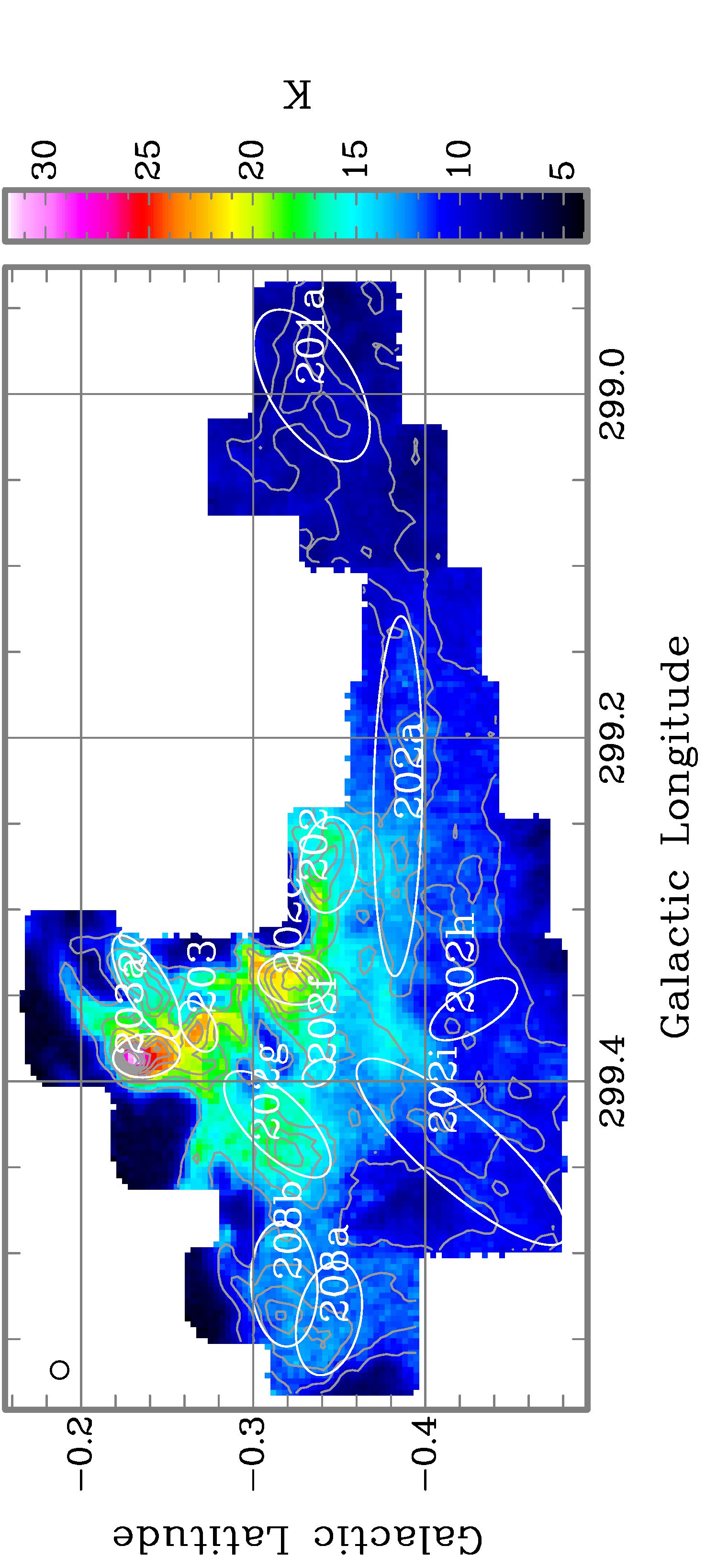} \vspace{-7.5mm}\\
\includegraphics[angle=-90,scale=0.317]{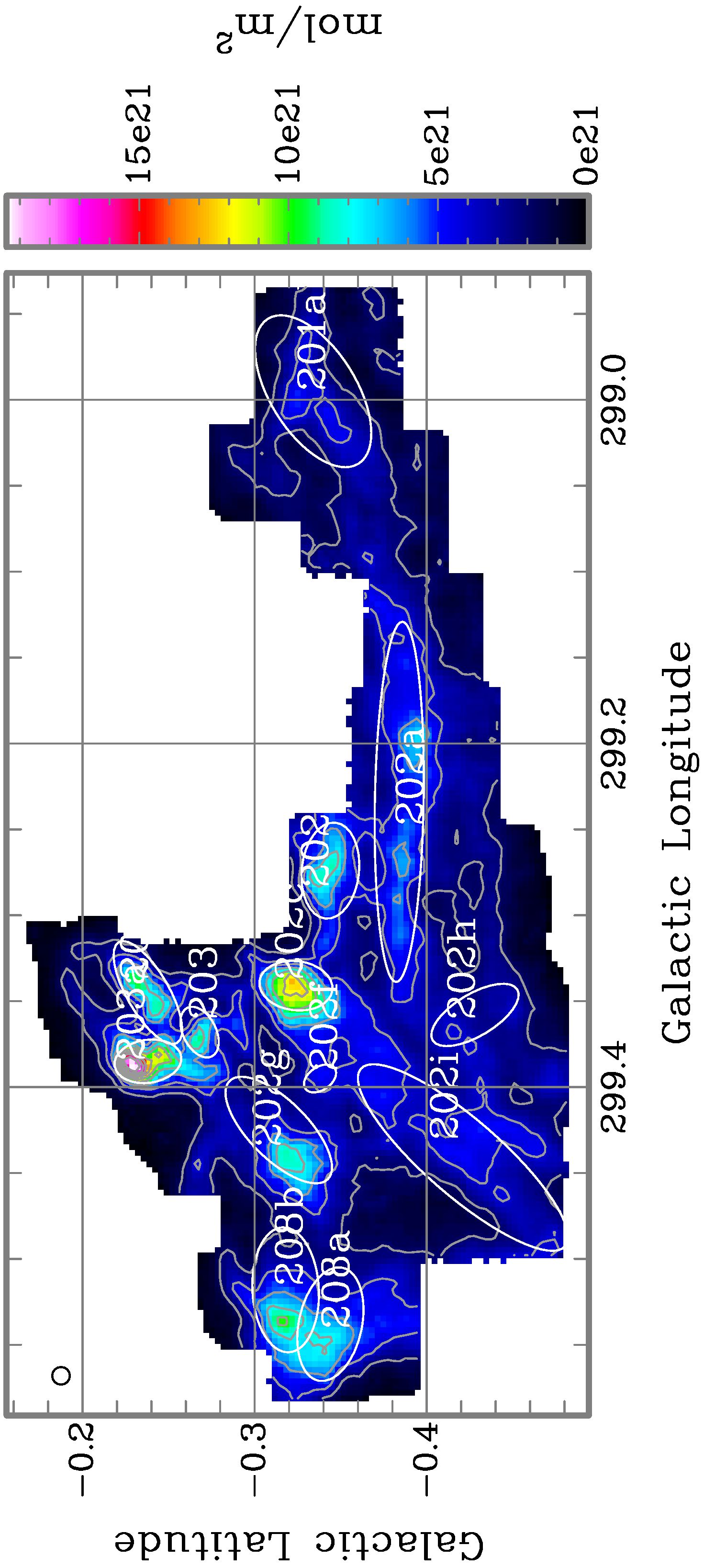} \vspace{-7.5mm}\\
\includegraphics[angle=-90,scale=0.317]{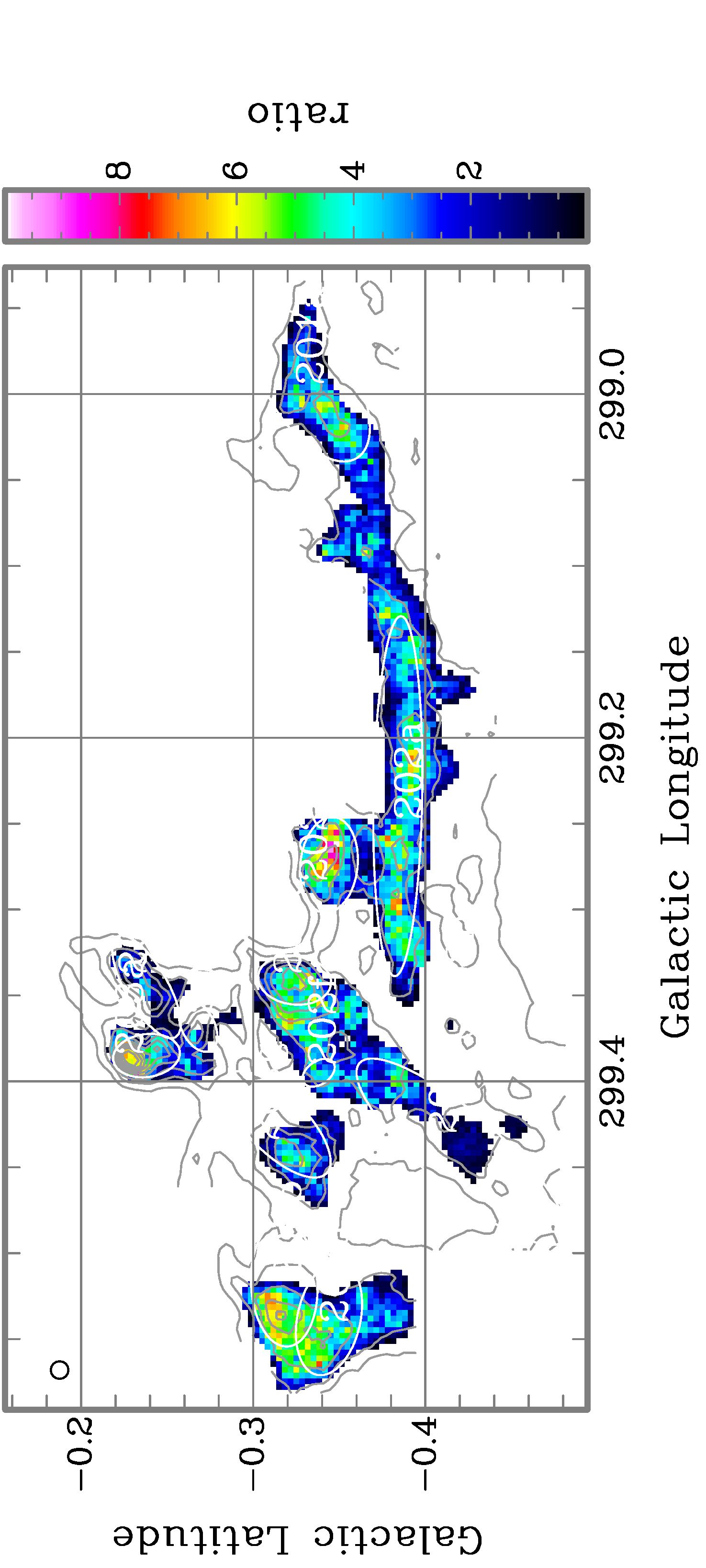} \vspace{-7.5mm}\\
\includegraphics[angle=-90,scale=0.264]{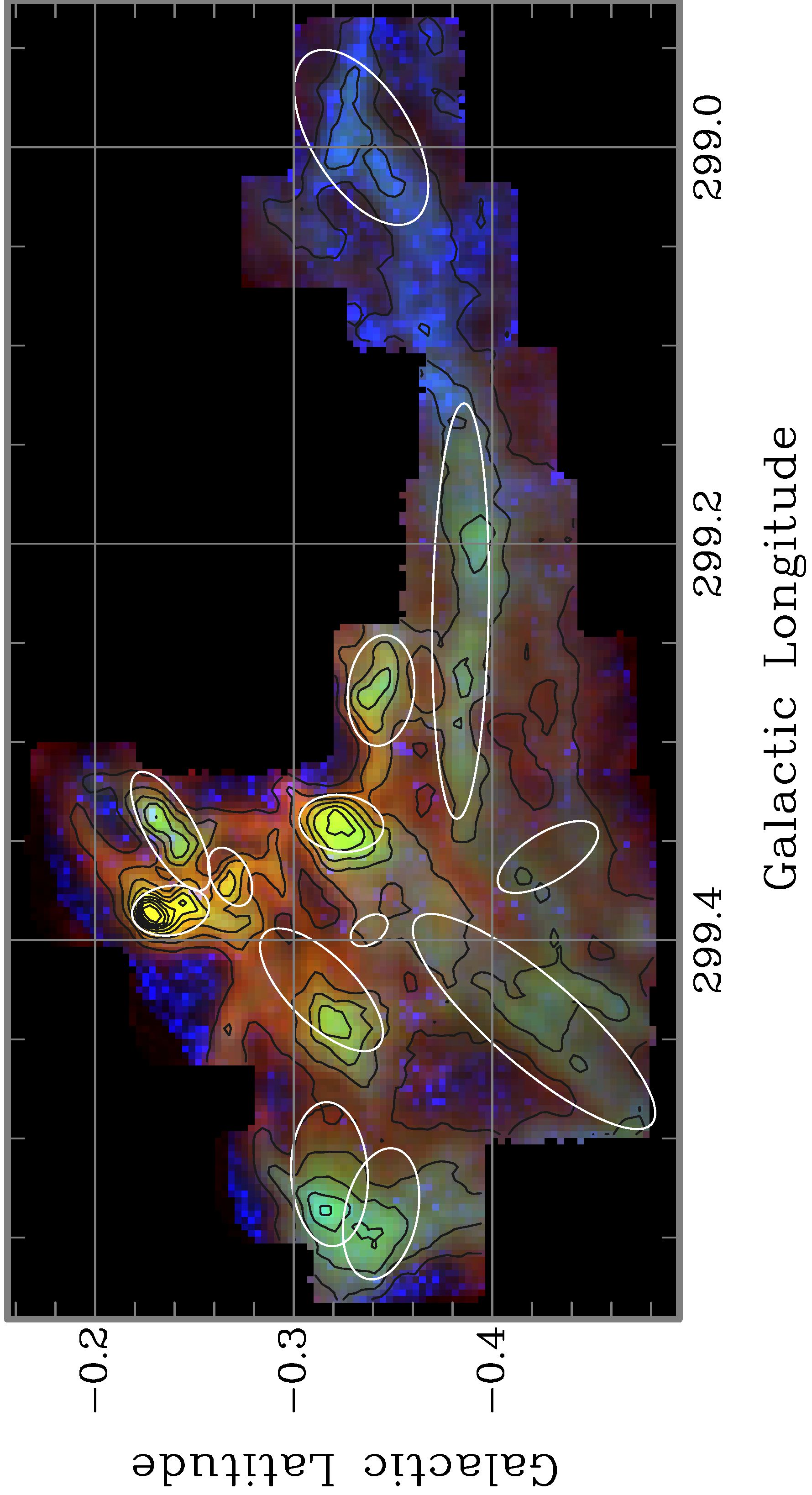} \vspace{-4.5mm}\\
\includegraphics[angle=-90,scale=0.264]{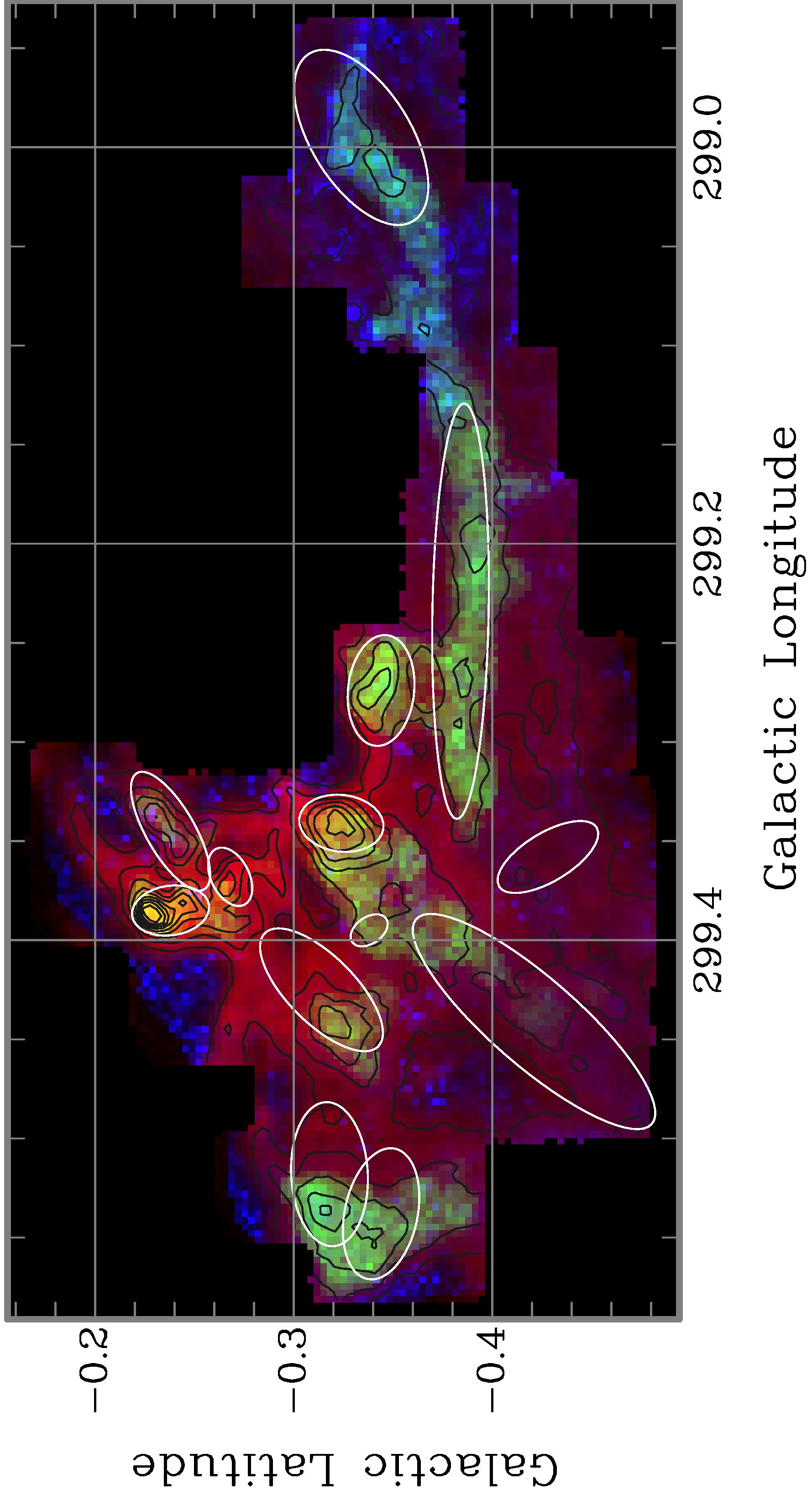} \vspace{-211mm}\\
\centerline{\hspace{25mm}{\large Peak $\tau_{12}$}}\vspace{31.5mm} \\
\centerline{\hspace{25mm}{\large Peak \tex}}		\vspace{29mm} \\
\centerline{\hspace{25mm}{\large $\int$\nco\,d$V$}}	\vspace{29mm} \\
\centerline{\hspace{25mm}{\large Avg.~$R_{18}$}}	\vspace{98mm}
\caption{Sample maps of physical quantities from radiative transfer analysis (Eqs.\,1--6) in Region 26b, as labelled. 
The 5$^{\rm th}$ and 6$^{\rm th}$ panels are RGB colour overlays of \tex,\nco,$\tau_{12}$ and \tex,$R_{18}$,$\tau_{12}$ (resp.), highlighting the differences in the spatial distributions of the 4 quantities.}
\label{phsample}
\end{figure}

The approach here derives \nco\ and \nttco\ based on the \tco\ and \ttco\ data, and doesn't depend on $R_{18}$ directly.  This means that we can solve for $R_{18}$ independently via Eq.\,5.  Thus, $R_{18}$ is found to vary widely, by a factor of \gapp20 across the different clumps, even while $R_{13}$ is assumed not to.  The derived $R_{18}$ variations would hardly change if we chose different values for $R_{13}$, but the latter are unlikely to vary by more than a factor of 2 since almost all clouds (except the farthest) share the Sun's Galactocentric distance, $R_0$.  Therefore, even considering only $R_{18}$, CHaMP delivers an interesting new result which has consequences for astrochemical models.

\notetoeditor{}
\begin{figure*}[t]
			\includegraphics[angle=-90,scale=0.31]{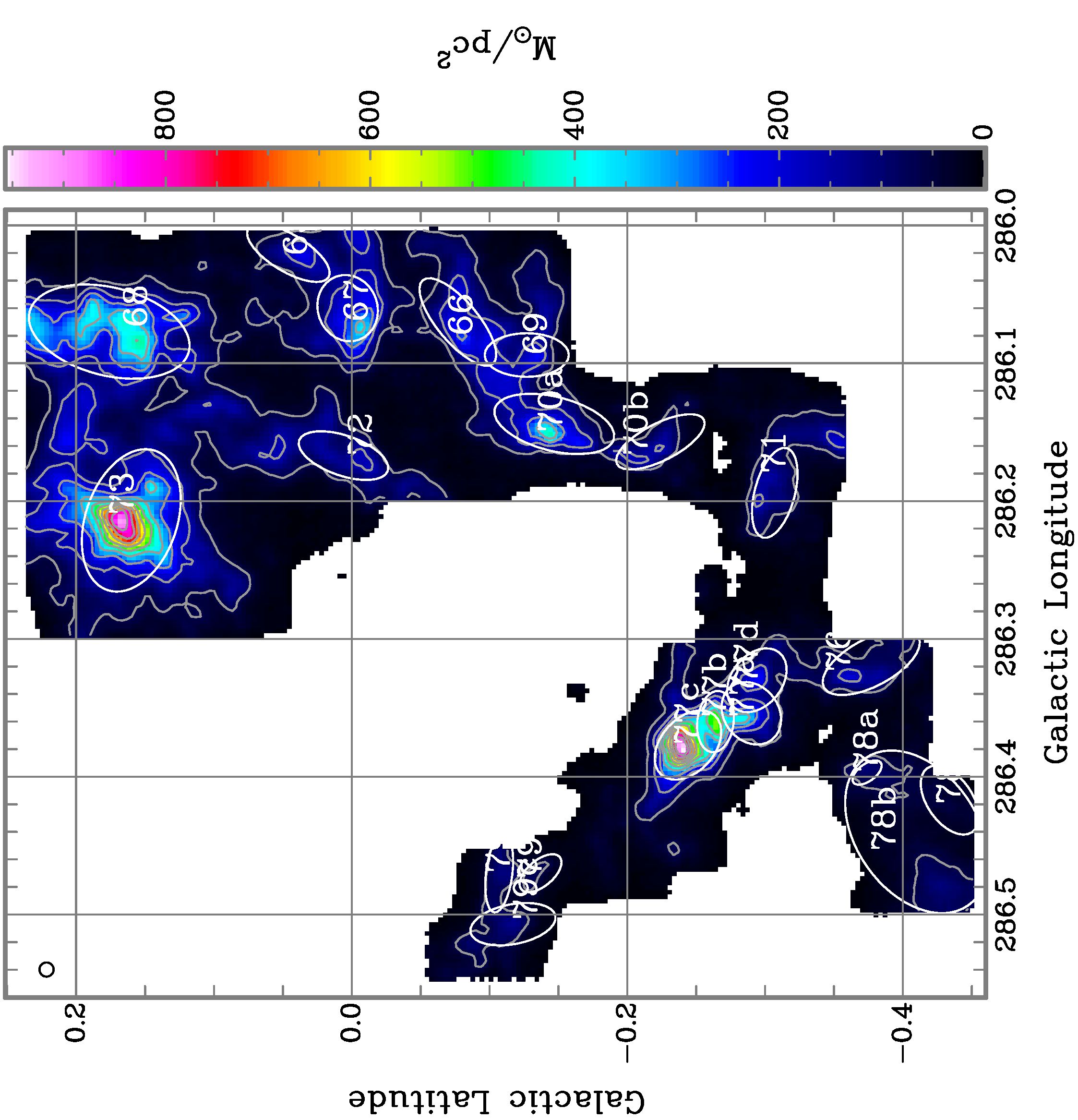}
\hspace{-1mm}  \includegraphics[angle=-90,scale=0.31]{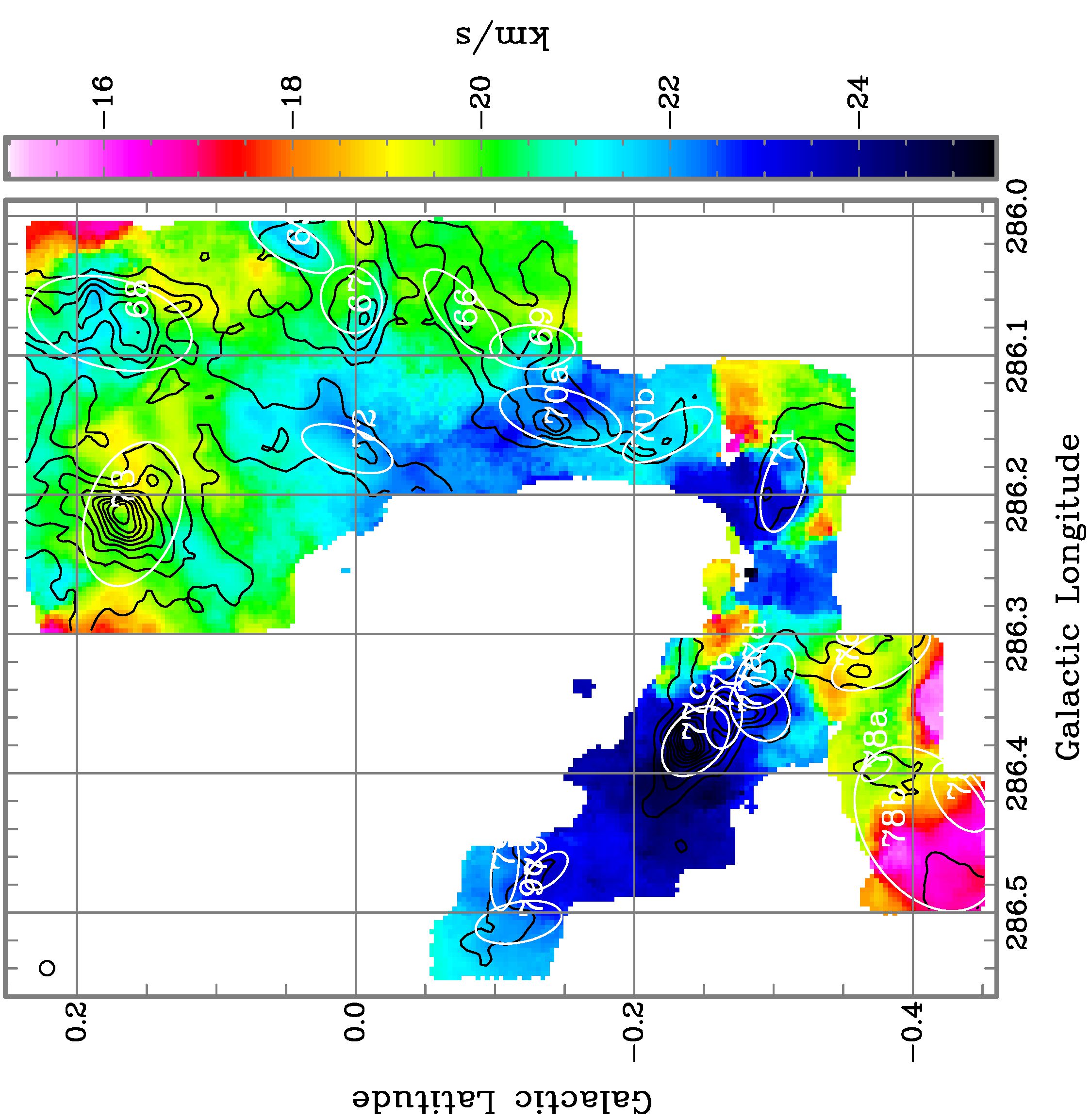}
\hspace{-1mm}  \includegraphics[angle=-90,scale=0.31]{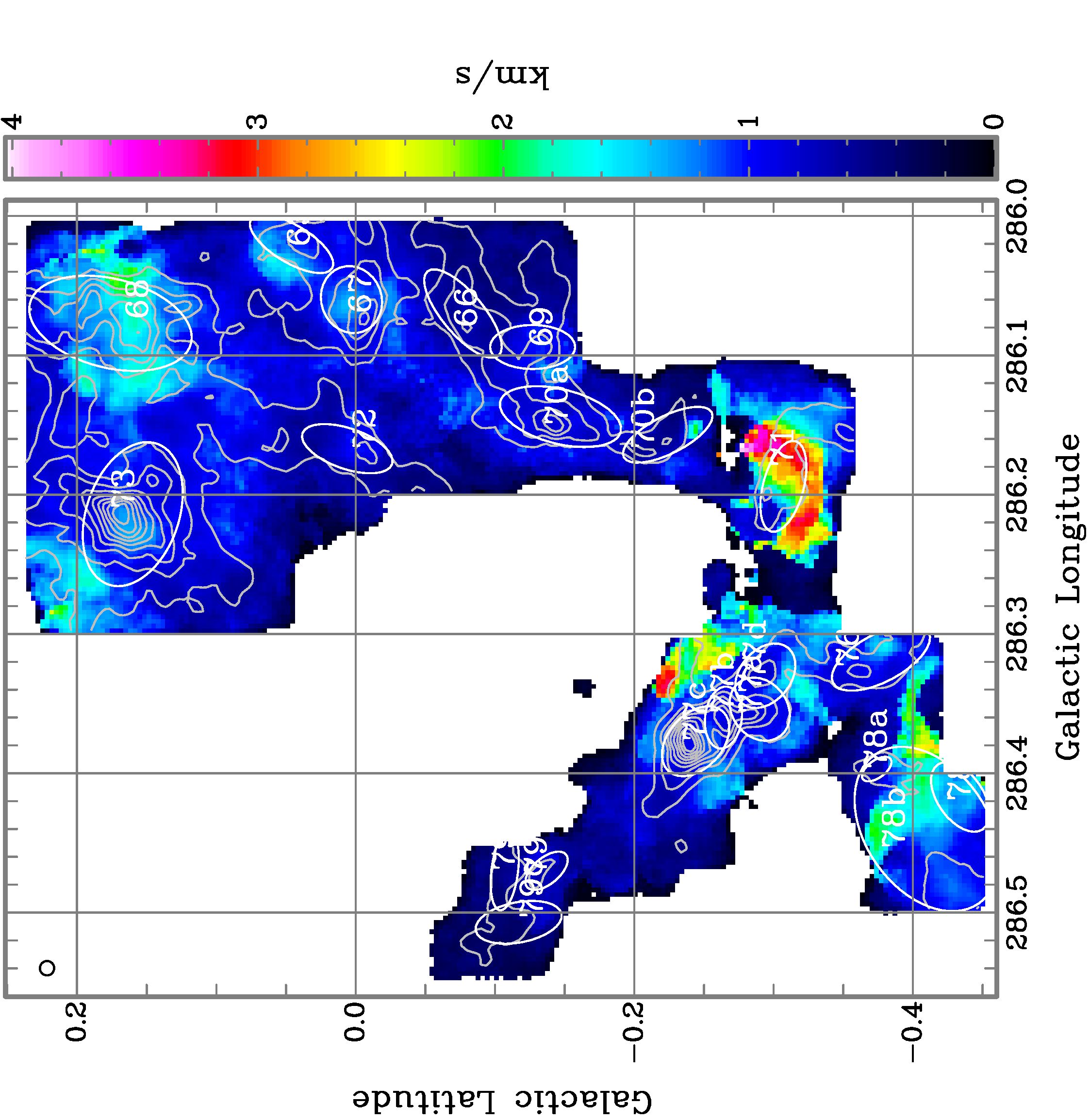}
\caption{Sample moment maps of mass surface density (Eq.\,7) in Region 9, from left to right: the velocity-integrated total surface density $\Sigma_{\rm mol}$, velocity field \vlsr, and velocity dispersion $\sigma_V$.}
\label{zgsample}
\end{figure*}

The \nco\ cubes in particular, being derived directly from the radiative transfer solutions, are equivalent to total mass surface density $\Sigma_{\rm mol}$ cubes\footnote{We make a distinction between $\Sigma_{\rm mol}$ defined from \nco, and $\Sigma_{12}$ defined from \ico: see \S\ref{formalkin}.} via
\begin{eqnarray}   
	\Sigma_{\rm mol} & = & N_{\rm H_2}~\mu_{\rm mol}~m_{\rm H} \\ 
	& = & 1.88\,{\rm M}_{\odot}\,{\rm pc}^{-2}~N_{\rm^{12}CO}~R_{12}/(10^{24}{\rm molec\,m}^{-2})~,~~ \nonumber 
\end{eqnarray}
where $\mu_{\rm mol}$ = 2.35 for 9\% He by number, $m_{\rm H}$ is the mass of the H atom, and  $R_{12}$ $\equiv$ [\htwo]/[\tco] (see \S\ref{mass}--\S\ref{conv}).  Because they derived from an intrinsic property of the gas in each cloud, these cubes' moments will be a truer representation of the cloud's physical state than equivalent moments of any individual spectral line's \tmb\ cube.  As can be seen by comparing Eqs.\,1 and 6, the spectral line cubes give instead moments of the emissivity of each species, a complex convolution of that line's optical depth, excitation temperature, and chemistry (via the abundance) at different velocities, compared to the {\em different} combination of $\tau$ and \tex\ contributing to the cloud's actual column density.  Our LTE plane-parallel radiative transfer calculation at least recovers, to first order and within the limits of the assumptions above, the actual mass traced by each line.

\section{Clump Mass Distributions}\label{mass}

With the physical ($l$,$b$,$V$) cubes from \S\ref{radxfer}, we have an opportunity to analyse these 
in preference to the equivalent observed data 
for each of the iso-CO lines.  As Figures \ref{etacar}--\ref{rrd} make clear, the isotopologues have line ratios that vary markedly with position and velocity.  If one is interested in true mass or velocity distributions, moments of these emission line cubes will have large inherent biases, compared to the intrinsic column density distribution, due to the variability in the lines' optical depth, excitation, and abundance.  With the radiative transfer analysis, we at least remove these effects to first order, and so the resulting \nco\ (or equivalent $\Sigma_{\rm mol}$) cubes should give a much more reliable measure of the true gas column density distribution and kinematics.

Another important point about this procedure is that the resulting $N$ or $\Sigma$ cubes are of very high dynamic range, with S/N ratios of several hundreds.  This comes about from peeling away layers of optical depth in molecular clouds via the isotopologues' different opacities.  The \tco\ maps provide a sensitive measure of column density even when this is quite low, due to the \tco's very high abundance.  Where the \tco\ opacity is high (i.e., in most voxels) and would normally be a heavily saturated measure of the true mass column, the \ttco\ and \ceto\ lines give reasonable measures of the column density.

This is not to suggest that all uncertainties have been removed: at the very least, the \tco\ abundance with respect to \htwo\ is also likely to vary, possibly by as much as a factor of 10.  For example, astrochemical studies like that of \citet{gbs14} actually show that the typical range of $R_{12}$ runs from about 10$^4$ in hot cores and near HII regions, to about 10$^5$ in dark molecular clouds, with a median around 3$\times$10$^4$ across all environments.  Unless otherwise stated, in this work we assume a standard value for $R_{12}$ 
= 10$^{4}$, which means that \tco\ $\rightarrow$ \htwo\ mass conversions will still give a conservative lower limit to molecular masses, despite our conversion laws driving these masses up compared to a standard single $X$ factor (see \S\ref{conv}).  Using the median value for $R_{12}$ instead would give total masses in our maps about 3$\times$ larger than shown in Figures \ref{phsample} or \ref{zgsample}, 
but the situation is more complex than that, since the actual $R_{12}$ likely varies with location (see \S\ref{genlaw}). 
Similarly, our assumptions of LTE and a common \tex\ between the three iso-CO species are likely to be violated to some degree as well, but our contention is that such effects will be of second order, compared to the more dramatic absolute variations in $\tau$ and \tex.

We therefore treat the $N/\Sigma$ cubes as regular data cubes, insofar as they can be analysed with the same SAM-based moment calculation techniques as in \citet{bm15,bh16}.  
Samples of higher $N/\Sigma$ moment maps for Region 9 are shown in Figure \ref{zgsample}. 

\section{Conversion Laws}\label{conv}
\subsection{Overall Pattern and Physical Basis}

\begin{figure*}[t]
\centerline{\includegraphics[angle=0,scale=0.21]{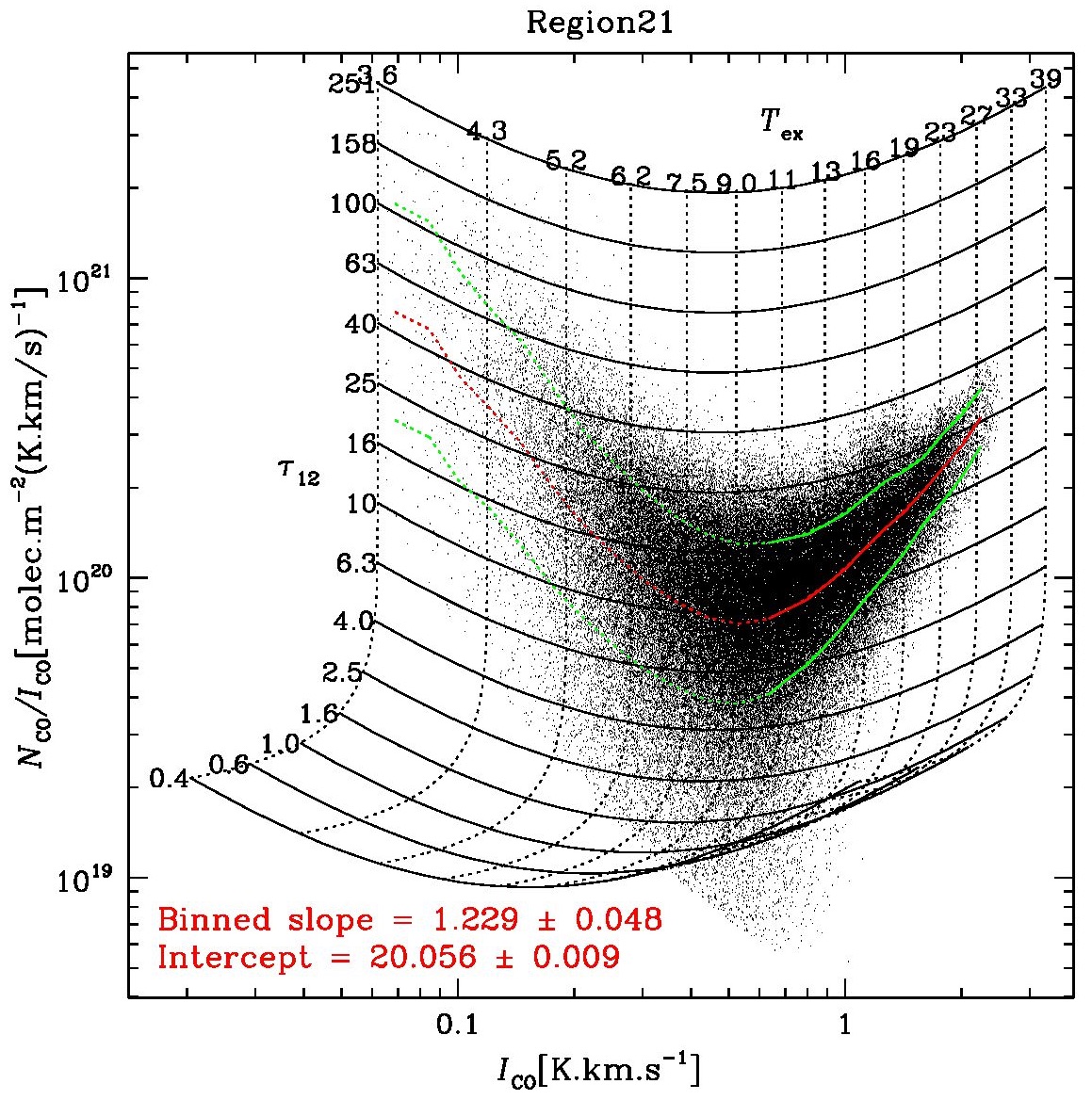}\hspace{4mm}
		\includegraphics[angle=0,scale=0.21]{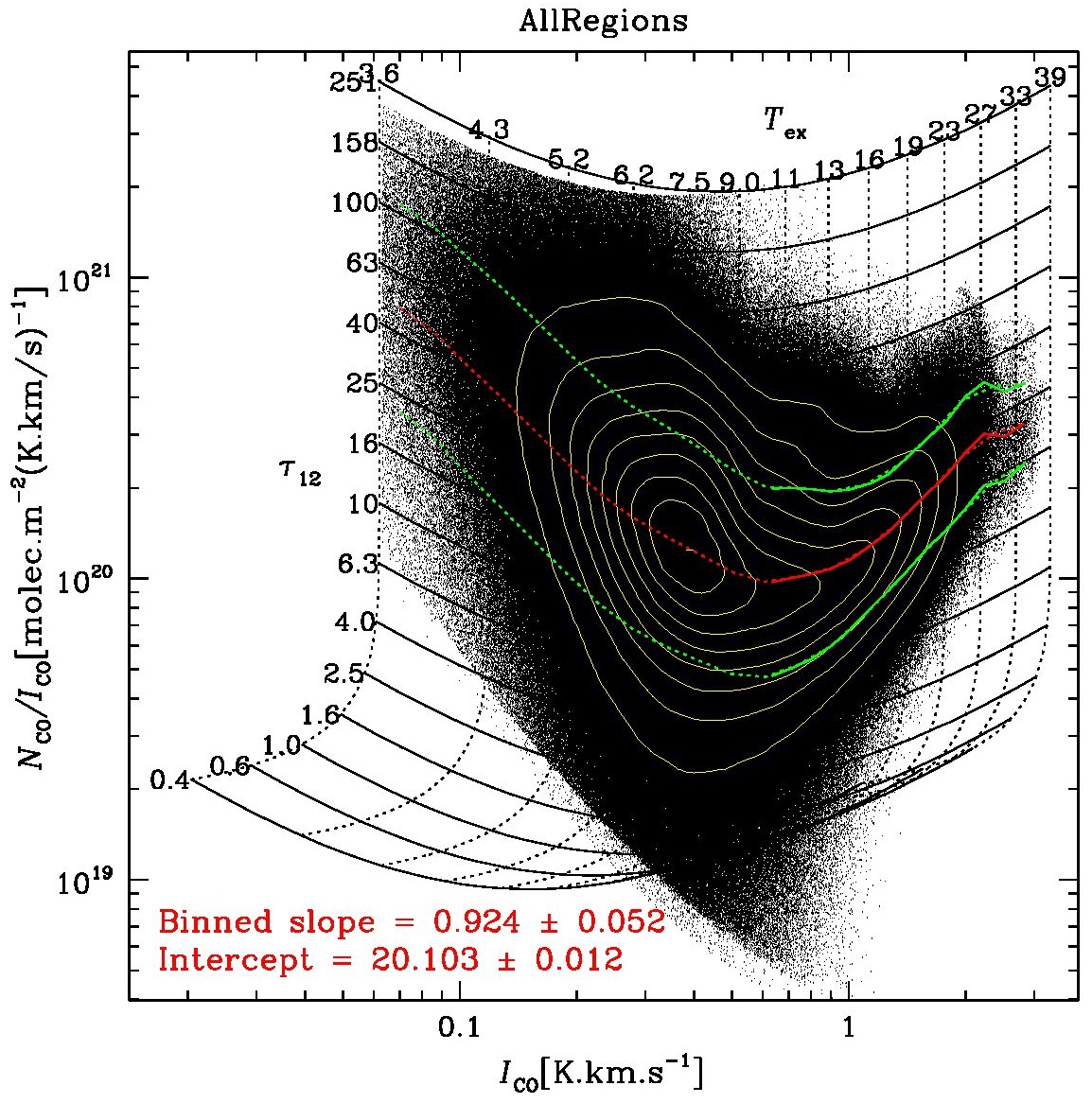}}
\vspace{-5mm}\hspace{5mm}(a)\hspace{88mm}(b)
\caption{(a) Sample plot of computed \nco/\ico\ ratio vs \ico, equivalent to the $X$-factor divided by the abundance ratio $R_{12}$ = [\htwo]/[\tco], shown as black dots for all voxels in Region 21 with S/N$>$1 in \ico\ ($\sigma_{\rm rms}$ = 0.063\,K\kms).  The \ttco\ data also impose a S/N cutoff, apparent in the curved edge to the points' distribution to the lower left.  To derive an appropriate power-law relation (the solid red and green curves), we limit analysis to \ico\ $>$ 0.6\,K\kms\ $\sim$ 10$\sigma$.  Although the S/N below this limit is still useful, the data are less complete, partly because we focus on clouds with detectable \hcop\ or \ceto\ emission, and partly because the SAM technique blanks low-S/N voxels, especially in the \ttco\ cubes (the curved cutoff).  To derive the bright-\ico\ law, we also need to avoid the negative trend for low \ico\ (see text for discussion).  The underlying grid is again from the radiative transfer analysis, Eqs.\,1--6, but now labelled by the \tco\ optical depth and common \tex\ in K.  See text for further discussion of this grid.  \\
\hspace*{6mm}The points are binned in log \ico, the mean (red) and $\pm\sigma$ (green) values of log($N$/$I$) in each bin are computed, and these are then connected by straight line segments, solid above the \ico\ $>$ 0.6\,K\kms\ limit and dotted below.  Then above this limit, a least-squares slope (= power-law index) and intercept (at log \ico\ = 0) are fitted to the binned means, as labelled in red.  Because we plot \nco/\ico\ vs \ico, the actual exponent in a conversion law of the type \nco\ $\propto$ \ico$^p$ is $p$ = slope+1, or 2.23$\pm$0.05 in the case of Region 21.  \\
\hspace*{6mm}Plots like this for each Region are given in Appendix \ref{rgbimages}.  In {\em this} Figure, we also indicate by a thin blue horizontal line the equivalent ``standard'' $X$-factor = 1.8$\times$10$^{24}$m$^{-2}$/(K\kms) with $R_{12}$ = 10$^4$ \citep{dht01}, while larger values of $R_{12}$ = 3$\times$10$^4$ and 10$^5$ \citep{gbs14}, more representative of dark cloud conditions, place the same $X$-factor at the thick cyan and magenta lines, resp.  \\
\hspace*{6mm}(b) A similar plot to panel $a$ (i.e., at the full velocity and angular resolution of the data), but here aggregated for all CHaMP data in all Regions ($\sim$10$^7$ voxels); the blue lines, grid, and fit have the same meaning as in panel $a$, but here we also add yellow contours of voxel incidence at 10(10)90\% of the maximum, due to the high density of points.  
Another cutoff can be seen at the top of the points' distribution, at \gapp2$\times$10$^{21}$molec\,m$^{-2}$/(K\kms): this is due to self-absorption in the \tco\ emission, precluding a solution for \nco\ with the simple radiative transfer calculation we use, where \ittco\ $>$ \ico.  As suggested by the points' distribution, the fraction of voxels so affected by self-absorption is low, $<$1\%.  Thus, aggregated across all CHaMP data of sufficient completeness and sensitivity, $p$ = 1.92$\pm$0.05 for \ico\ $>$ 0.6\,K\kms, or equivalently for \tmb(\tco) \gapp\ 7\,K\,chan$^{-1}$.}
\label{convlaw}
\begin{picture}(1,1)
\thinlines
{\color{blue}
\put(35,343){\line(2,0){211}}
\put(293,343){\line(2,0){211}}}
\thicklines
{\color{cyan}
\put(32,310){\line(2,0){211}}
\put(290,310){\line(2,0){211}}}
{\color{magenta}
\put(29,276){\line(2,0){211}}
\put(287,276){\line(2,0){211}}}
\end{picture}
\vspace*{-2mm}
\end{figure*}

In the ThrUMMS project, \citet{bm15} first used those data with the above radiative transfer analysis to examine the relation between the \tco\ integrated intensity \ico, and the column density \nco\ derived from the line ratios, as a proxy for the $X$-factor relating \nhtwo\ to \ico.  (One would need to additionally determine $R_{12}$ as described above, to get the actual $X$-factor.)  \citet{bm15} found that this pseudo-$X$-factor is not constant, but seems instead to define a non-linear conversion law.  Having here obtained column density cubes from the much higher-sensitivity CHaMP data, we can revisit this analysis and determine if the ThrUMMS result can be reproduced.  We therefore make similar plots of \nco/\ico\ vs \ico\ as in \citet{bm15} but for each CHaMP Region separately (see Appendix \ref{rgbimages}); a sample is provided in Figure \ref{convlaw}a.

\notetoeditor{}

While we broadly see a similar pattern of rising $X$ = \nco/\ico\ values with \ico\ compared to that manifest in the ThrUMMS data, we note four key differences with CHaMP.  First, CHaMP's higher sensitivity allows us to perform the analysis for each channel and pixel, as opposed to the ThrUMMS analysis, which binned the data in velocity to 1\kms\ in order to provide sufficient S/N for that case.  Since the native Mopra/MOPS channel width at these frequencies is $<$0.1\kms, this means that the horizontal scale of the plots in Figure \ref{convlaw} is shifted to lower \ico\ by a factor of 11 compared to Figure 15 in \citet{bm15}, but this is just a normalisation on the abscissa.  Since the \nco\ and \ico\ are both calculated here per channel, their ratio is properly normalised to the usual $X$-factor units.

Second, a more interesting result is that the fitted slopes to the derived conversion laws are typically steeper than in \citet{bm15}, at brighter \ico.  These correspond to power-law indices in $N$ $\propto$ $I^{p}$ ranging from $p$ = --0.3 to 2.8 in the different CHaMP 
Regions,\footnote{
While the power-law fit for voxels in each Region applies across a relatively small range of \ico\ ($\sim$half an order of magnitude), this is still significant since those voxels are at high S/N, so their impact on any $X$-factor based mass calculation is large.  Moreover, when integrated over discrete structures in angle and/or velocity (see \S\S\ref{vres}{\em ff}, Fig.\,\ref{fullint}b, Eq.\,8), the fitted power laws extend over a much broader range of \ico\ ($\sim$1--2 orders of magnitude).  Finally, these power laws at brighter \ico\ are important to distinguish from the different $X$ vs $I$ behaviour at fainter \ico\ (see next).} 
although the lower values for $p$ occur only in those maps with fainter overall CO emission (see next).  In most clouds $p$ clusters around 1.5 to 2.2, compared to $p$ = 1.4 from ThrUMMS.  An aggregate value over all CHaMP data is $p$ = 1.9 (shown in Fig.\,\ref{convlaw}b).  (Uncertainties for these indices are provided in each Figure, formally \lapp0.1, with dispersions in the normalisation $\sim$0.2\,dex.)

A third feature of these plots is that there seems to be a consistent inflection in the $N$/$I$ = $X$ trend at lower $I$.  That is, the true $X$ vs.\ $I$ relationship consists of two parts: a decreasing or flat $X$ as $I$ rises for faint $I$ (i.e., $p$ is near zero or slightly positive), but then an increasing $X$ vs.\ $I$ at brighter $I$ ($p$ positive definite).  Region 21 shows this effect most cleanly (Fig.\,\ref{convlaw}a), but it is apparent in all the $X$ vs.\ $I$ plots: the inflection in $p$ occurs near $I$ = 0.5\,K\kms (corresponding to \tmb(\tco) \gapp\ 5.5\,K in each 0.1\kms\ channel).  

We believe this low-\ico\ trend of a negative slope is real, although one could argue that it is partly an artifact of the completeness and S/N limits inherent in our data (described in the caption to Fig.\,\ref{convlaw}).  For example, the negative slopes in both panels of Figure \ref{convlaw} seem to parallel somewhat the curved \ttco\ S/N cutoff, so the actual value of $p$ below this inflection is uncertain.  Because of this limitation, we do not fit a formal power law to these fainter points.  However, the S/N of the \ttco\ data is still reasonably high: for the mean trend (the dotted red curve), \ttco\ begins around 15$\sigma$ at the \ico\ = 0.6\,K\kms\ fitting limit, is $\sim$5$\sigma$ at \ico\ = 0.2\,K\kms, and is still near 3$\sigma$ at \ico\ = 0.1\,K\kms.  Therefore we contend that the deficit of points below the mean trend is not purely a noise or selection effect.  This means that there is probably a large amount of high-opacity, low-excitation, and high-column-density CO emission prevalent throughout the Milky Way's molecular ISM, and not just in our maps, as discussed next.

\begin{figure}[t]
\includegraphics[angle=0,scale=0.49]{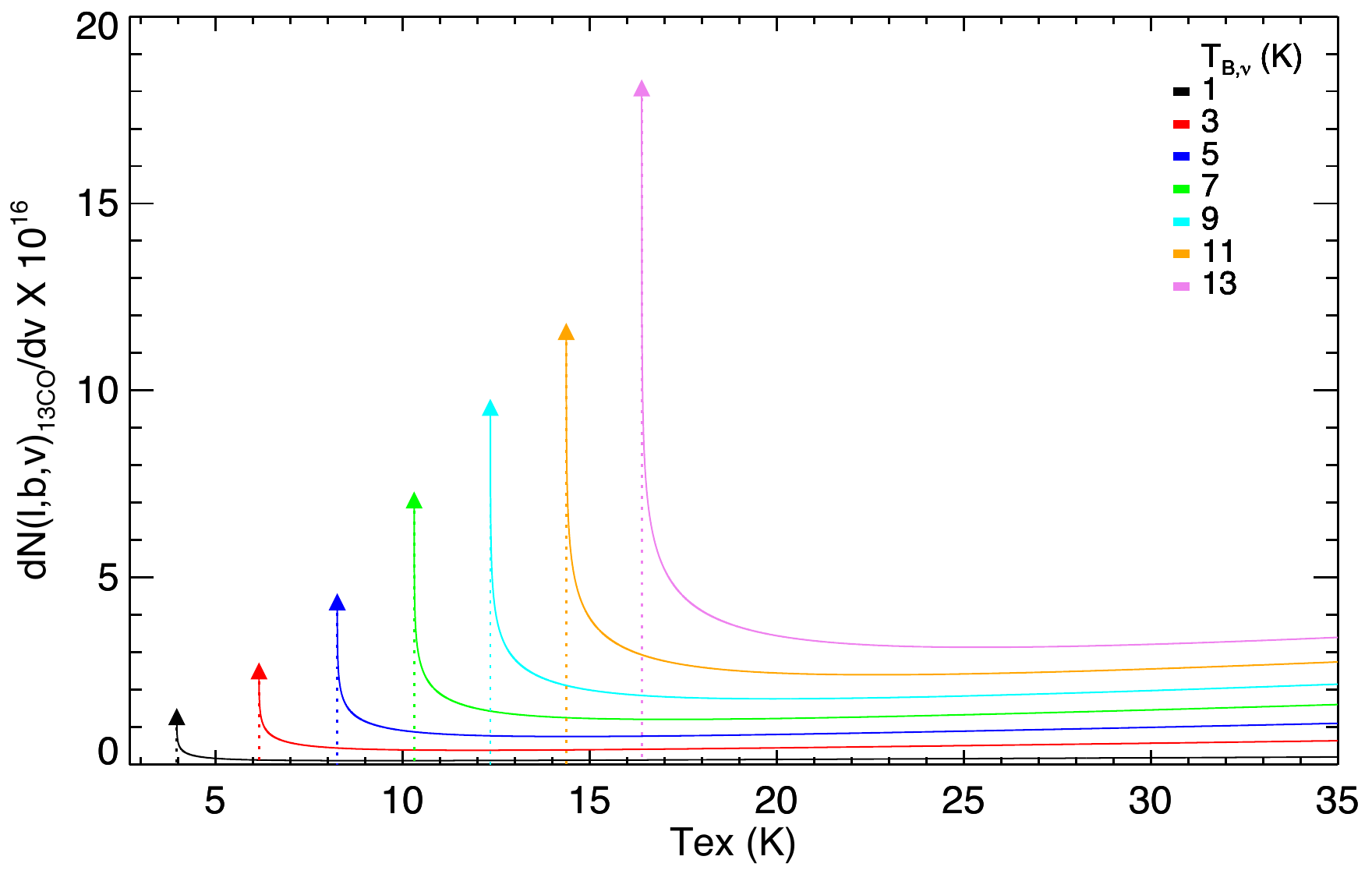}
\caption{Sample radiative transfer computation of \nttco\ as a function of \tex, at each of several different observed \tmb(\ttco) indicated by the coloured lines.  
The rise in $\tau$ and \nco\ at low \tex\ is analogous for both \tco\ and \ttco, and is a generic feature of molecular spectroscopic data; \citet{sb16} describe a similar case with HCN.  See also Fig.\,\ref{convlaw} and the text for further discussion.
}
\label{texlim}
\end{figure}

Simple modelling confirms that this behaviour can be expected at low \tex.  To understand this, note that for an observed emission line which is in the optically thin and Rayleigh-Jeans limits, Eq.\,1 implies that \tex\ is approximately $\propto$ $\tau^{-1}$, as long as \tex\ is also reasonably large compared to \tbg, which is usually true when $\tau$ $\ll$ 1.  In such cases, the column density $N$ remains relatively insensitive to these compensating changes in \tex\ and $\tau$, at fixed \tmb\ (see Paper I for an example and discussion of this insensitivity).  But eventually, as \tex\ is reduced sufficiently to approach (\tmb--\tbg), $\tau$ rises dramatically, since the optically thin limit no longer applies and the full non-linearity of Eq.\,1 comes into play.  Figure \ref{texlim} shows an example of this behaviour.

The two panels of Figure \ref{convlaw} also illustrate this property in the $N$/$I$ vs.\ \ico\ plane, via the labelled grid of ($\tau_{12}$, \tex) values.  Note the falling trends in this grid of $N$/$I$ with $I$ at low $I$ (asymptotic slope = --1), bottoming near \ico\ = 0.5\,K\kms, and then rising at high $I$ (asymptotic slope = +1), similar to the data trends seen in Fig.\,\ref{convlaw}a. 
But in the low-$\tau_{12}$ and bright-\ico\ limits (lower right of the plots), the grid curves become degenerate, i.e., \nco\ becomes relatively insensitive to the exact values of $\tau_{12}$ or \tex, as described above.  In contrast, for any given \ico\ or equivalent \tmb, where the fitted \tex\ drops sufficiently, it forces $\tau_{12}$ to very high values, indicated by the vertical dotted lines, which is consistent with the pattern in Figure \ref{texlim}.

\notetoeditor{}

The combination of the high-S/N CHaMP data with the radiative transfer grid displayed in Figure \ref{convlaw} 
reveals a fourth feature of this diagram which was not immediately apparent in the ThrUMMS data.  For bright \ico, we can see that the power-law conversion law is steeper than the curves of constant $\tau_{12}$.  This means that, in general, brighter \tco\ emission converts to an \nco\ above the standard $X$ factor due to {\em both} an opacity and an excitation term, or approximately as \nco\ $\propto$ $\tau_{12}$\tex.

Given these features of the data and analysis, we confirm that the inflection in $X$ vs $I$ seen in Figure \ref{convlaw} must be real, despite some S/N limitations at low \ittco.  The column density at low \ico\ can be unexpectedly high because there are large amounts of gas that are very subthermally excited (i.e., \tex\ $\ll$ $T_{\rm kin}$), even for \tco.  This will be the case where, for example, \tco\ is relatively faint, but \ttco\ is almost as bright as the \tco, requiring a low \tex\ and high $\tau$.  At the same time, some locations where \tco\ is faint will have very faint or undetectable \ttco, giving a low $\tau$ and larger \tex.  The point is that at small \ico, a wide range of opacities and column densities are possible, whereas at large \ico, the derived \nco\ is a very strong function of this brightness.  This means that the \ico\ to \nhtwo\ conversion law consists of two parts, one (for \ico\ \lapp\ 0.5\,K\kms) where $X$ is typically falling with \ico\ or at most approximately constant, and another (for \ico\ \gapp\ 0.5\,K\kms) where \nhtwo\ $\propto$ \ico$^2$, approximately.


Recent theoretical work confirms the importance of excitation and opacity variations in interpreting emission line data on molecular clouds.  For the \tco\ \jto/\joz\ line ratio, \citet{pc17,pc18} showed that clouds can contain a bimodal distribution of gas in (density, temperature) space, including a low-opacity, subthermally-excited component which can occupy a large fraction of the volume of the cloud.  This gas can also contribute significantly to the line emission for high opacity lines of sight, complicating the interpretation of line emission maps, unless some effort is made to understand the radiative transfer.  Also, when considering a large number of $J$ transitions for the CO energy ladder in an extragalactic context, \citet{w18} have derived similar CO$\rightarrow$\htwo\ conversion laws to ours, with even steeper powers $p$.  Their models also highlight the important contributions to the CO emission of both $\tau$$\ll$1 and $\gg$1 gas.  These studies follow earlier numerical work without radiative transfer post-processing \citep{ck12}, which found a bifurcation in the gas distribution between a lower-density, more turbulent state and a higher-density, gravity-dominated state.  Such works intriguingly parallel our results: in Figure \ref{convlaw} we also see a bimodality in the $X$ vs $I$ relationship, based on the radiative transfer in the \tco/\ttco\ line ratio.  At low \tex, we have contributions from both high- and low-opacity gas, and this complicates how \ico\ is used and interpreted, in ways that have not otherwise been discussed in the literature to date.

\subsection{Dependence on Velocity Resolution}\label{vres}

\notetoeditor{}
\begin{figure*}[t]
\centerline{
\includegraphics[angle=0,scale=0.147]{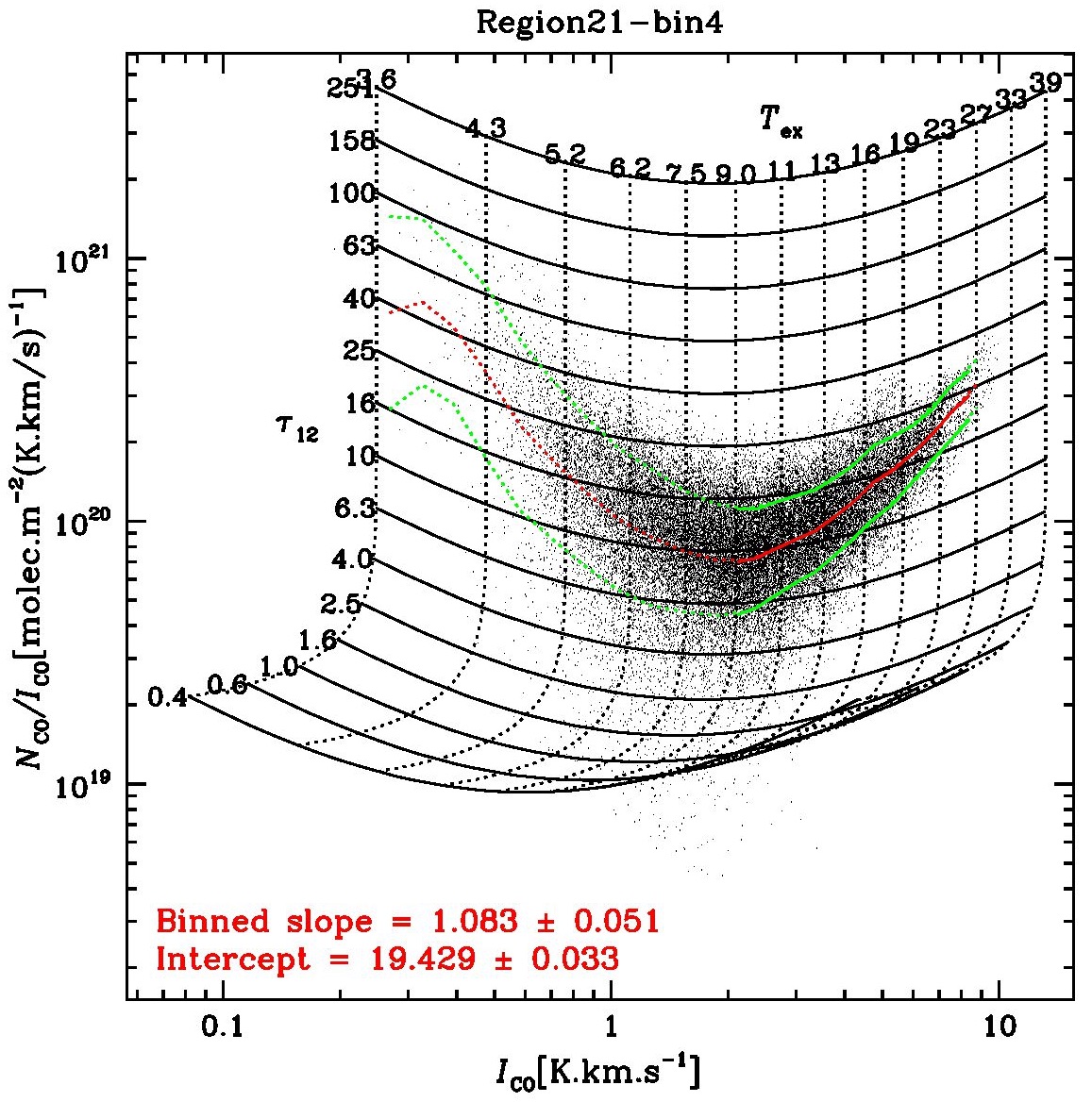}
\includegraphics[angle=0,scale=0.147]{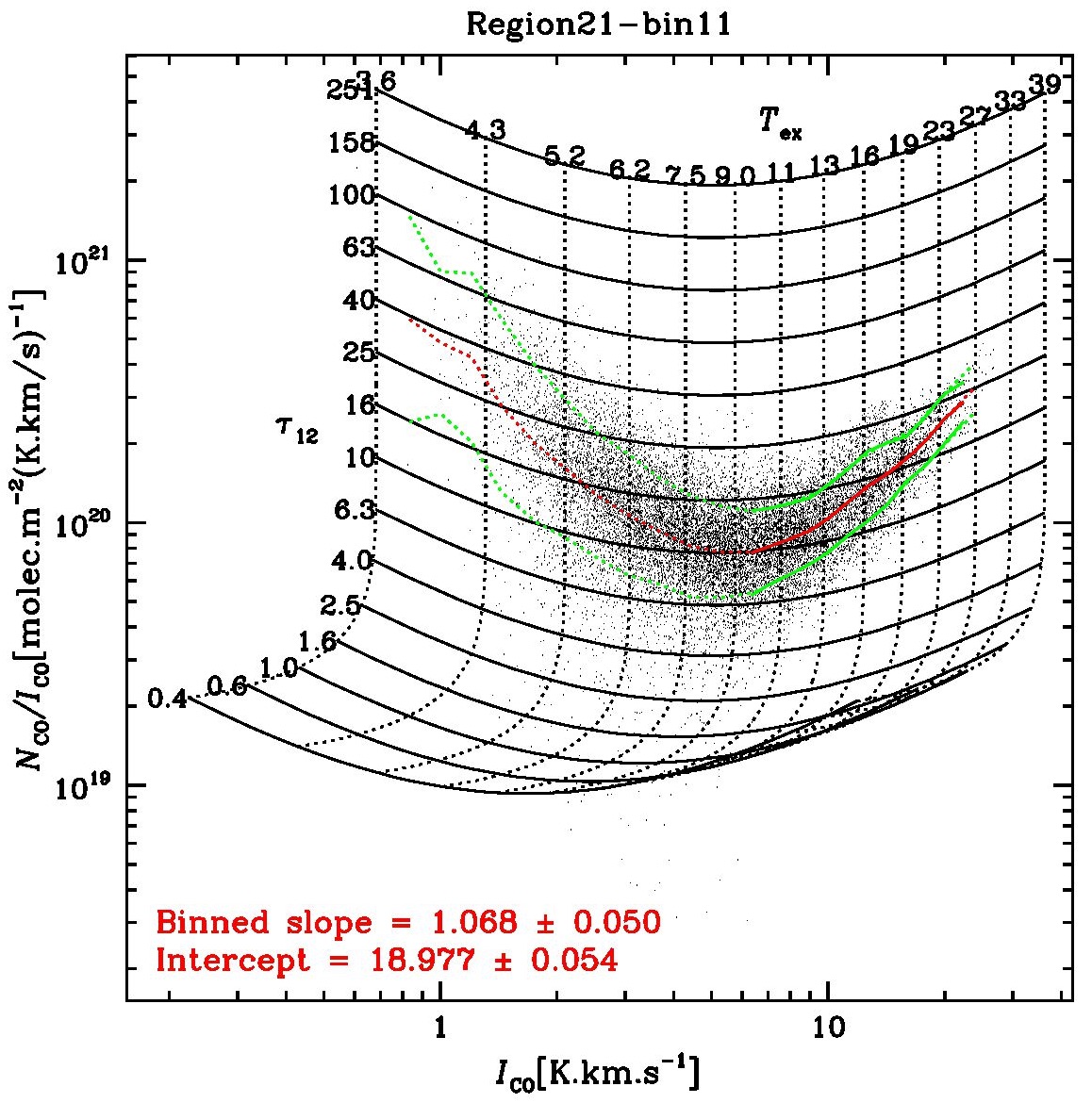}
\includegraphics[angle=0,scale=0.147]{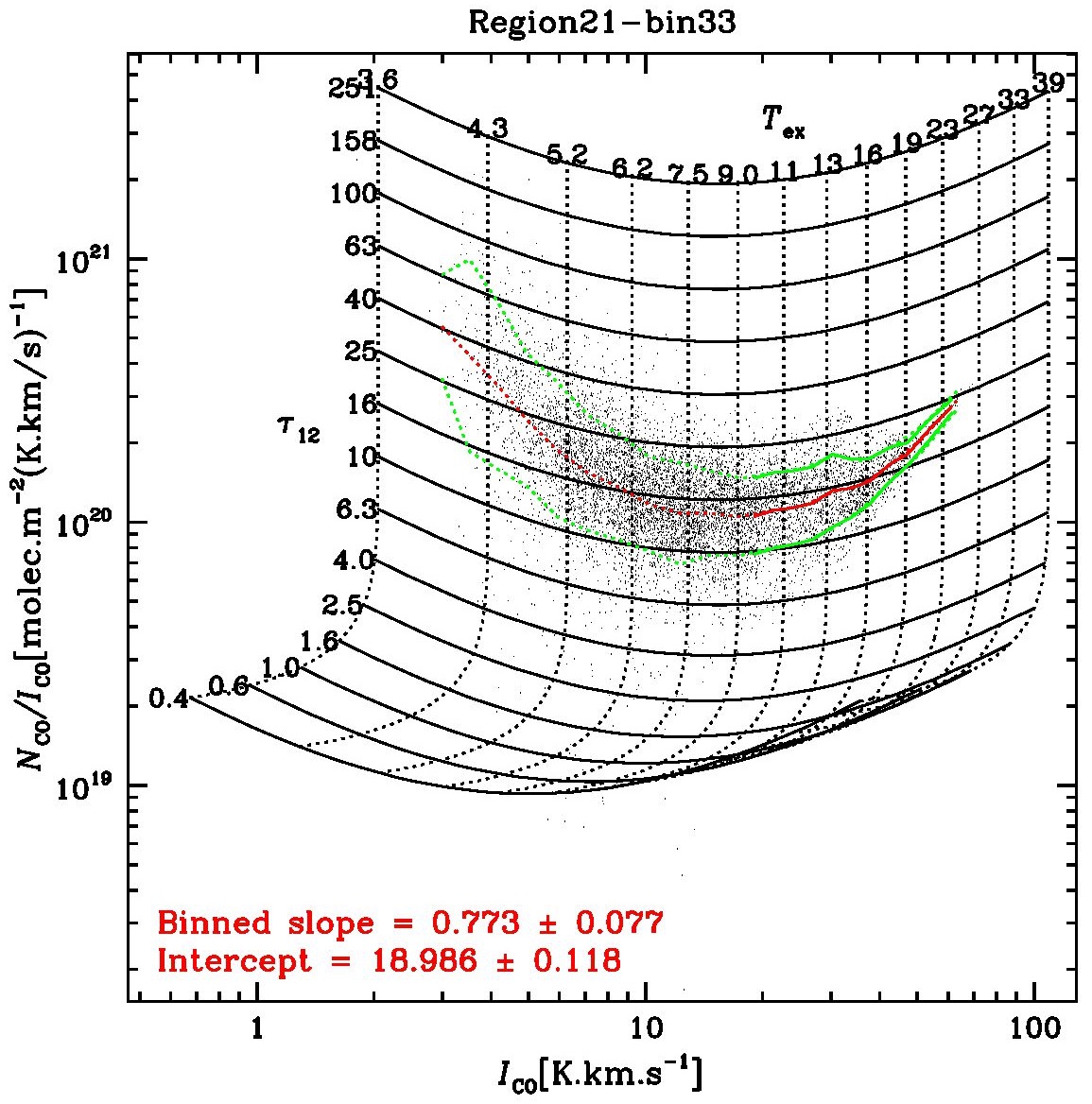}}
\centerline{
\includegraphics[angle=0,scale=0.147]{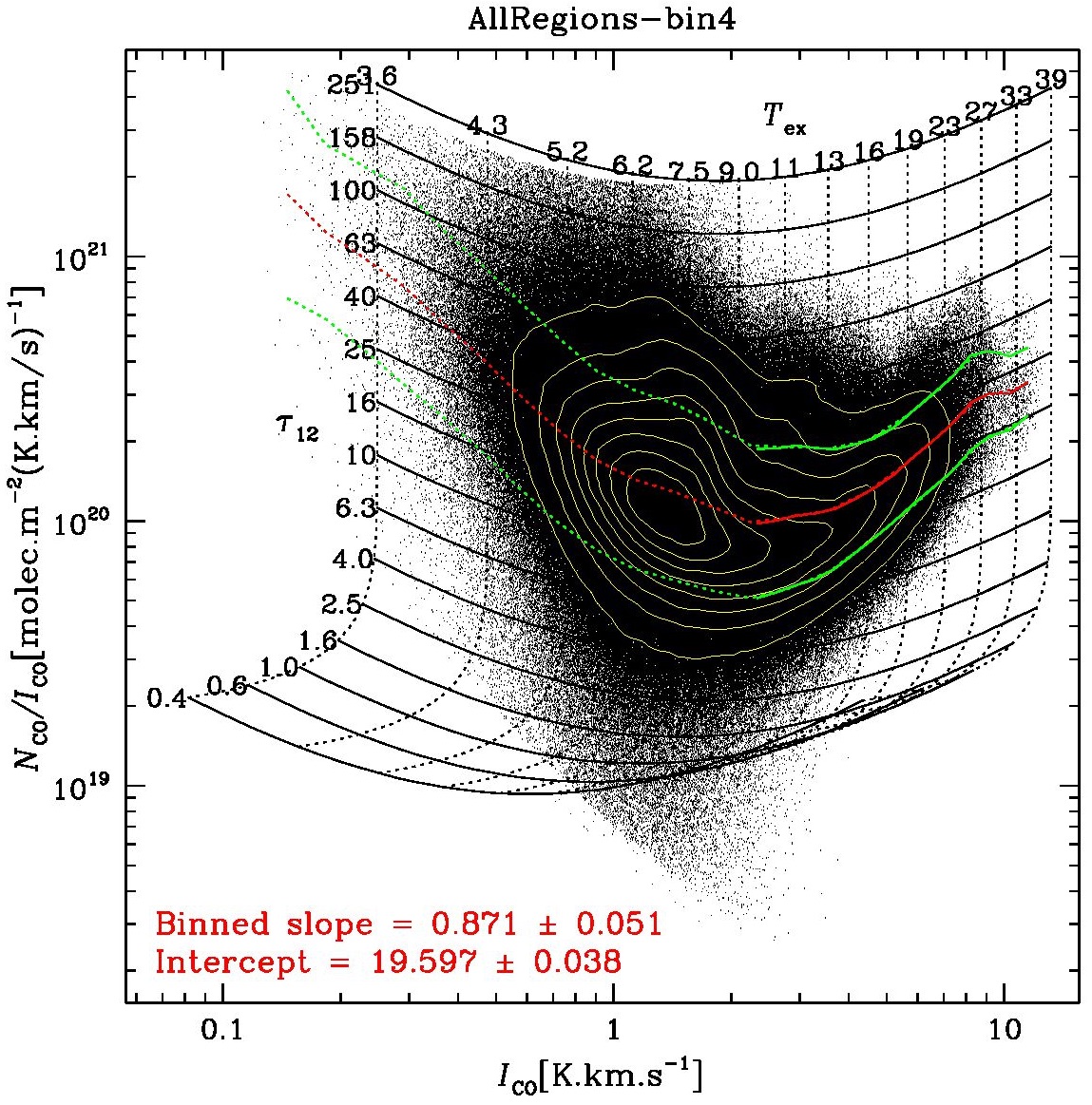}
\includegraphics[angle=0,scale=0.147]{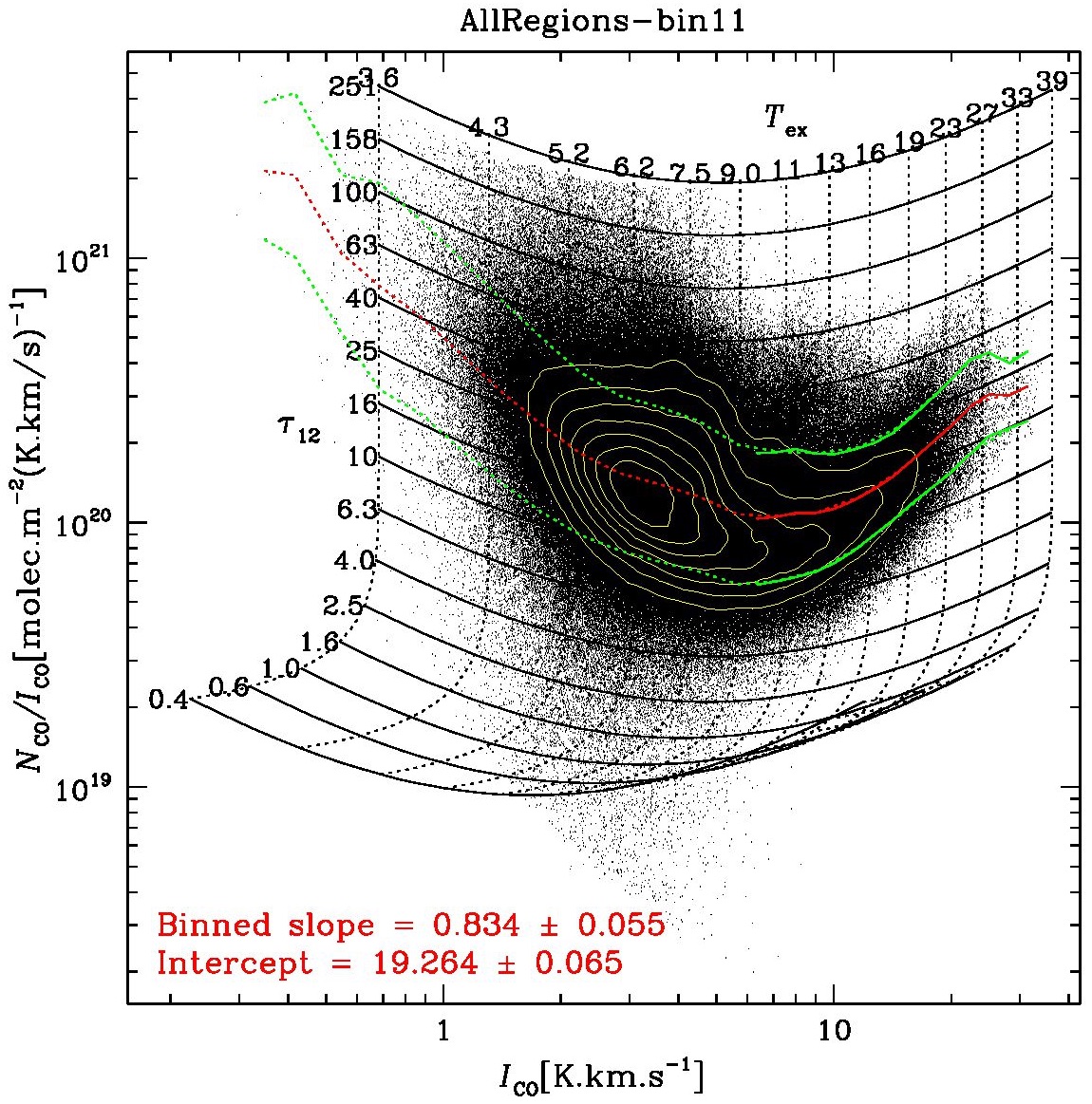}
\includegraphics[angle=0,scale=0.147]{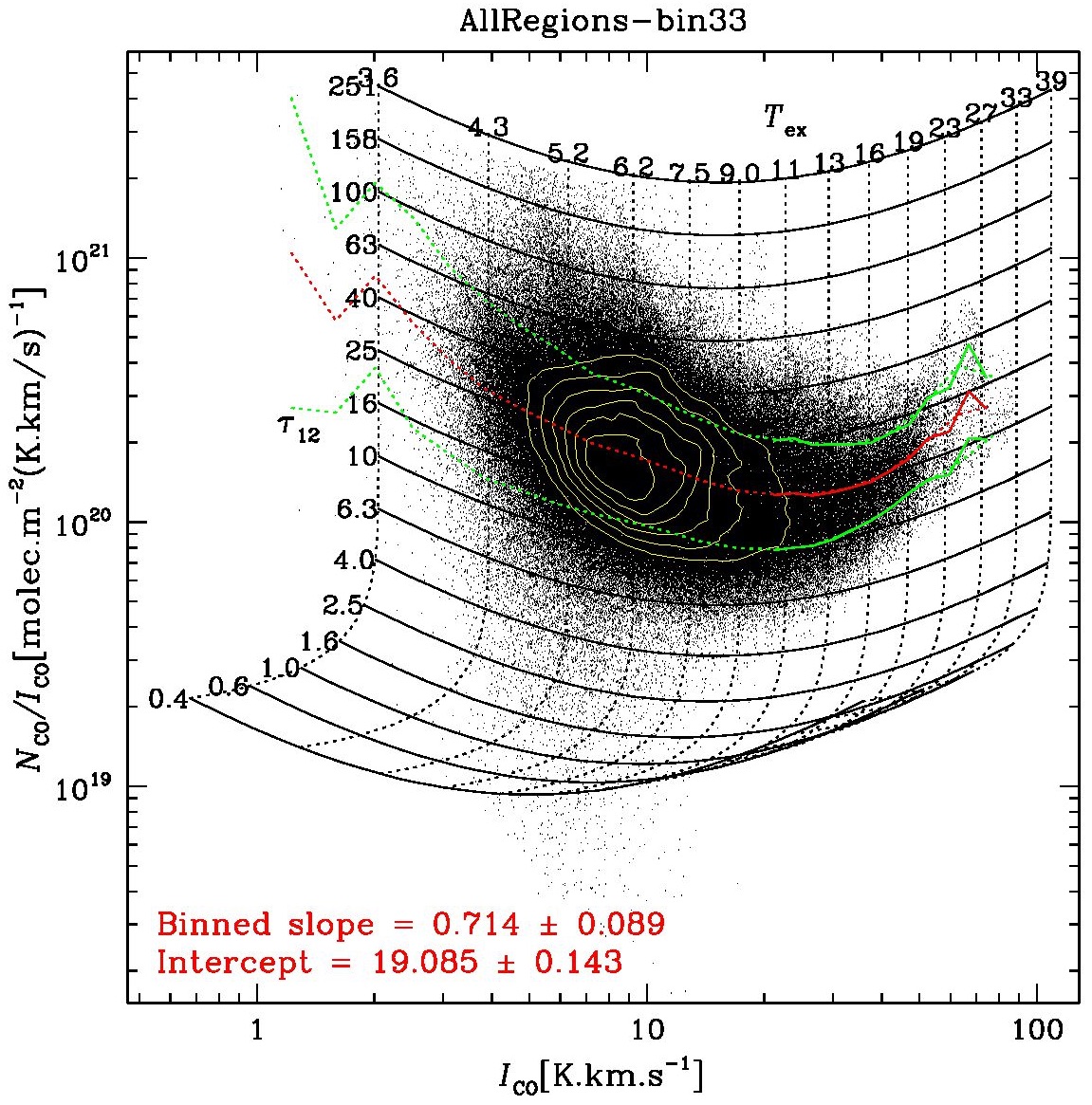}}
\vspace*{-4mm}\hspace{5mm}(a)\hspace{57mm}(b)\hspace{57mm}(c)
\caption{({\em Top row}) (a) \nco/\ico\ ratio vs \ico\ plot as in Fig.\,\ref{convlaw}a, but now for Region 21 data binned by 4 channels (i.e., to 0.37\kms) {\em before} the \nco\ calculation.  (b) Another such plot, but with the data binned by 11 channels (to 1.0\kms), and (c) another binned by 33 channels (to 3.0\kms). ({\em Bottom row}) (a--c) Same as top row, but now aggregated for all CHaMP data in all Regions.  Contours of voxel incidence are also shown where the points are too dense to indicate the distribution.}
\label{convagg}
\end{figure*}

For the brighter portion of this conversion law, we not only confirm its power-law form, but also find the result to be stronger (smaller dispersion) and steeper (larger $p$) than first found by \citet{bm15}.  In working through this analysis with the various data cubes, it is apparent that the \nco\ cubes are much more ``peaky'' than the line emission cubes, i.e., the conversion is highly non-linear, as already explained in Paper III, but also this is apparent as a function of velocity in the emission line profiles.  Therefore, a possible cause for the steeper conversion law found here is the velocity binning in the ThrUMMS analysis.  If we average or bin by velocity in the $I$ cubes, and then convert these to $N$, we are averaging together brighter and less bright voxels, short-circuiting the nonlinearity in the conversion law (arising in Eqs.\,1 \& 6).  We should rather first calculate $N$ cubes directly from the $I$ cubes, and then integrate $N$ in velocity.  In other words, when computing \nco\ from \ico, velocity-binning or integrating in \ico\ is not commutative with the $I$$\rightarrow$$N$ conversion.

If this is true, the shallower ThrUMMS law or standard flat ($X$ = constant, $N$ $\propto$ $I$) conversion, being averages over wider velocity ranges, may be smoothing together different column density to intensity relationships.  To test this, we also formed binned versions of each CHaMP iso-CO emission line cube, and redid the entire radiative transfer/column density calculation from \S\ref{radxfer}, for each of 0.37, 1.0, and 3.0\kms\ bins in each Region.

The aggregate results are shown in Figure \ref{convagg}, and our suspicion is confirmed: the broader the velocity binning/integration, the flatter is the derived conversion law. 
This means that prior determinations of $X$ in the literature, which usually measure some proxy for \nhtwo\ and compare this with \ico\ integrated over all emission from whatever cloud is under study, are inherently missing this physics.

Indeed, the plots in Figure \ref{convagg} illustrate this Central Limit Theorem aspect of our results.  While we see a wide variation in $X$ in all voxels, especially at low $I$, due primarily to the real differences in physical conditions (mainly opacity) between different parcels of gas in the clumps, when we average together many voxels, the mean result approaches the standard pseudo-$X$ value of $\sim$1.8$\times$10$^{20}$ CO molecules\,m$^{-2}$/(K\kms), assuming $R_{12}$ as above.  (However, if the median $R_{12}$ is triple the standard value, our $X$ factor normalisation would be correspondingly higher.)  This is best seen in Figure \ref{convagg}c, where the overall variance from the standard $X$ is only $\sim$0.2\,dex and the average $X$ is very close to the standard value, even though the true voxel variance in Figure \ref{convlaw}b is closer to 0.5\,dex, and the mean $X$ is similar to the standard value but varies systematically with \ico.

\notetoeditor{}
\begin{figure*}[ht]
\centerline{\includegraphics[angle=0,scale=0.21]{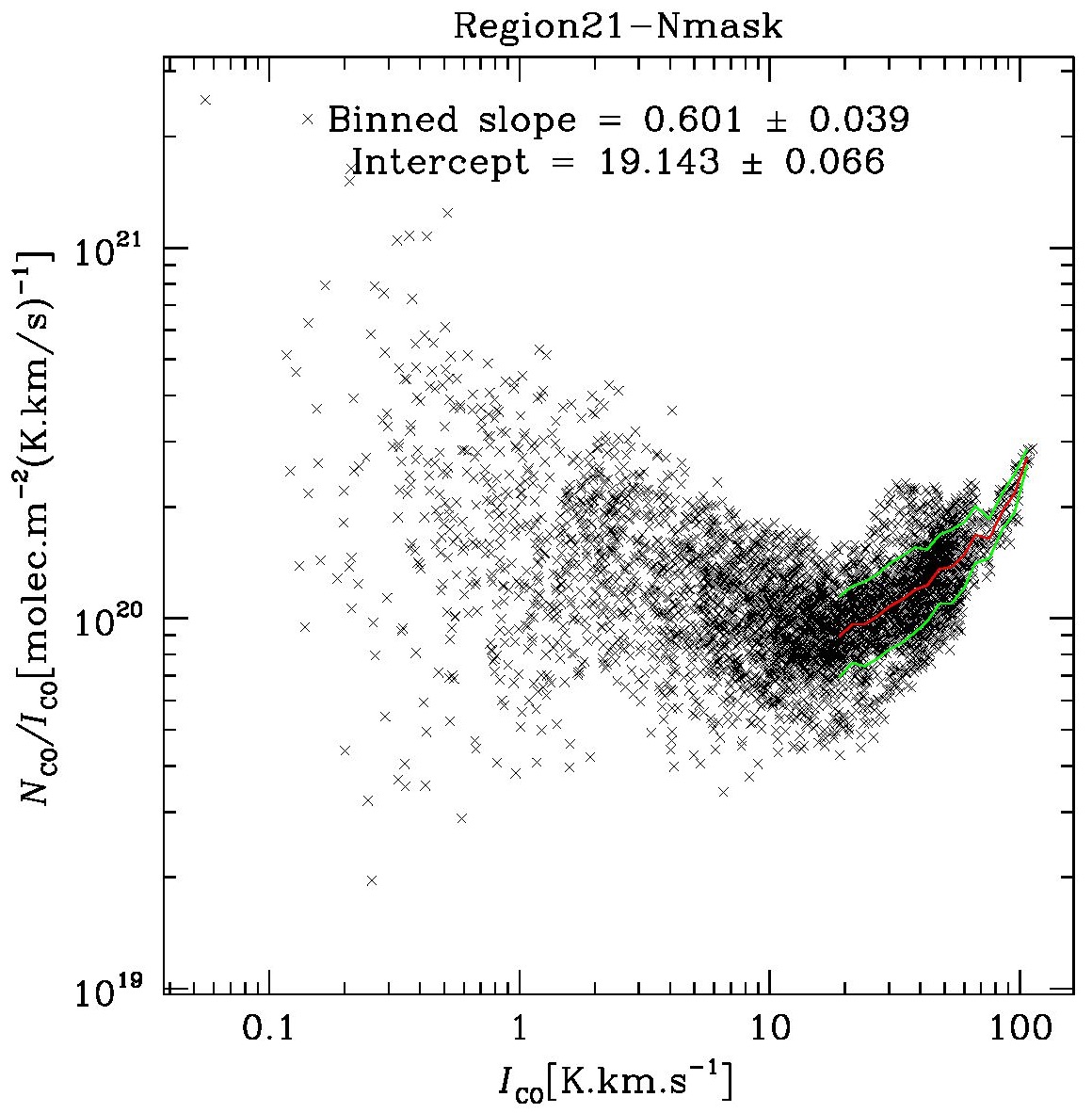}\hspace{4.2mm}
		\includegraphics[angle=0,scale=0.21]{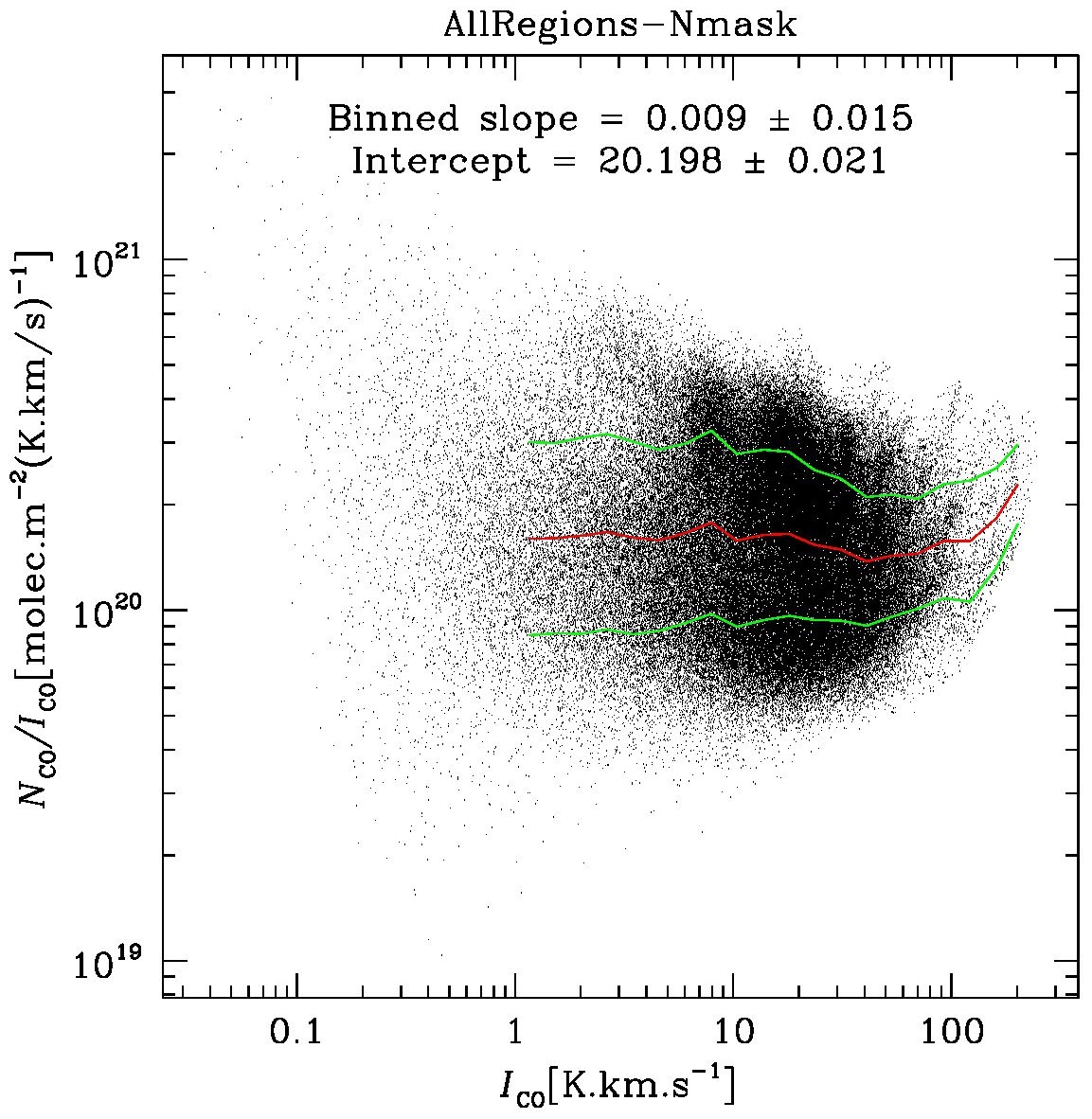}}
\vspace{-5mm}\hspace{5mm}(a)\hspace{88mm}(b)
\caption{(a) \nco/\ico\ ratio vs \ico\ as in Fig.\,\ref{convlaw}a, but where each quantity is integrated over all velocity channels, except that \ico\ is counted only in those channels where \nco\ is defined.  We also show the same coloured horizontal lines as in Fig.\,\ref{convlaw} to indicate the locus of the standard $X$-factor for three different values of $R_{12}$, but do not show the ($\tau_{12}$, \tex) grid since the integration over different velocity ranges for each pixel smears this grid along the $x$ axis.  (b) Same as panel $a$, but an aggregated plot for all Regions.
}
\label{maskint}
\begin{picture}(1,1)
\thinlines
{\color{blue}
\put(42,178){\line(2,0){204}}
\put(300,179.5){\line(2,0){204}}}
\thicklines
{\color{cyan}
\put(39,137){\line(2,0){204}}
\put(298,141){\line(2,0){204}}}
{\color{magenta}
\put(36,95){\line(2,0){204}}
\put(295,101){\line(2,0){204}}}
\end{picture}
\vspace*{-5mm}
\end{figure*}

\subsection{A General Law}\label{genlaw}

Figures \ref{convlaw} and \ref{convagg} also show that the normalisation on the abscissa, and hence the fitted intercept on the ordinate, depends on the velocity resolution/binning.  These results, plus those of \citet{bm15}, demonstrate the need to replace the standard $X$-factor with some more general conversion, which takes into account its dependency on velocity resolution.

However, in many studies that explore or use the \tco\ $\rightarrow$ \htwo\ conversion, the velocity resolution is not always a matter of choice: the usual approach is to use the total \tco\ line integral.  Given this, what matters is the relationship between the total column density, $\int$\nco\,d$V$, and the \tco\ line integral $\int$\ico\,d$V$.  The CO column density is related to \nhtwo\ via astrochemistry, modulated by the weaker self-shielding of \tco\ compared to \htwo\ that gives rise to a layer of CO-dark molecular gas \citep{p13}.  The CO line integral has also been extensively studied as a link to \nhtwo\ via the $X$ factor, which is calibrated from dust emission or extinction measurements, or other non-spectroscopic methods \citep[see][and references therein]{bwl13}.  Dust-related \htwo\ column density measurements, however, have no velocity resolution and therefore can't be included in this comparison.

In order to define a better conversion, we start by dividing \nco\ moment-0 maps like those in Figure \ref{phsample} by \ico\ moment-0 maps like those in Paper III (the same as the red channel in images like Figure \ref{etacar} and in Appendix \ref{rgbimages}).  However, in order to preserve the intrinsic physics afforded by our sensitivity and velocity resolution, we should mask the \ico\ moment calculation to only those channels where \nco\ is defined.  These channel ranges are different because \nco\ is only defined where we have signal in both \tco\ and \ttco.  The \tco\ in general extends across more channels, and is still reasonably bright over many of these, even where \ttco\ becomes undetectable.  Without this masking, the quotient $\int$\nco\,d$V$ / $\int$\ico\,d$V$ would be artificially lowered to inherently unphysical values by potentially large factors in the denominator.  We performed this calculation on all our Regions, and show a sample result (Region 21) in Figure \ref{maskint}a, with the CHaMP aggregate in Figure \ref{maskint}b.

From this comparison, we find a very interesting result.  Figure \ref{maskint}a is similar to most Regions: we see the same low-$I$ and high-$I$ dichotomy in the column density conversion as in Figure \ref{convlaw}, but at higher S/N due to the masked integration.  In other words, the low-$I$ conversion to column density can be as large or larger than the high-$I$ conversion, with a minimum conversion factor around \ico\ = 20\,K\kms, give or take a factor of 2.  {\em However}, when we aggregate all these masked conversion laws, Figure \ref{maskint}b shows that the low-$I$/high-$I$ distinction has been almost completely erased, due to variations in both the absolute \nco/\ico\ normalisation and the \ico\ value at which the conversion reaches a minimum.  This aggregate therefore yields essentially the same single $X$ factor (apart from an absolute $R_{12}$ normalisation) that has been discussed in the literature for so many years \citep{bwl13}, despite the fact that nearly all Regions show a distinct pattern of $N$/$I$ variations with $I$, similar to Figure \ref{maskint}a.  Hints of these underlying trends can still be seen in the Figure \ref{maskint}b aggregate, especially at large $I$.

\notetoeditor{}
\begin{figure*}[ht]
\centerline{\includegraphics[angle=0,scale=0.21]{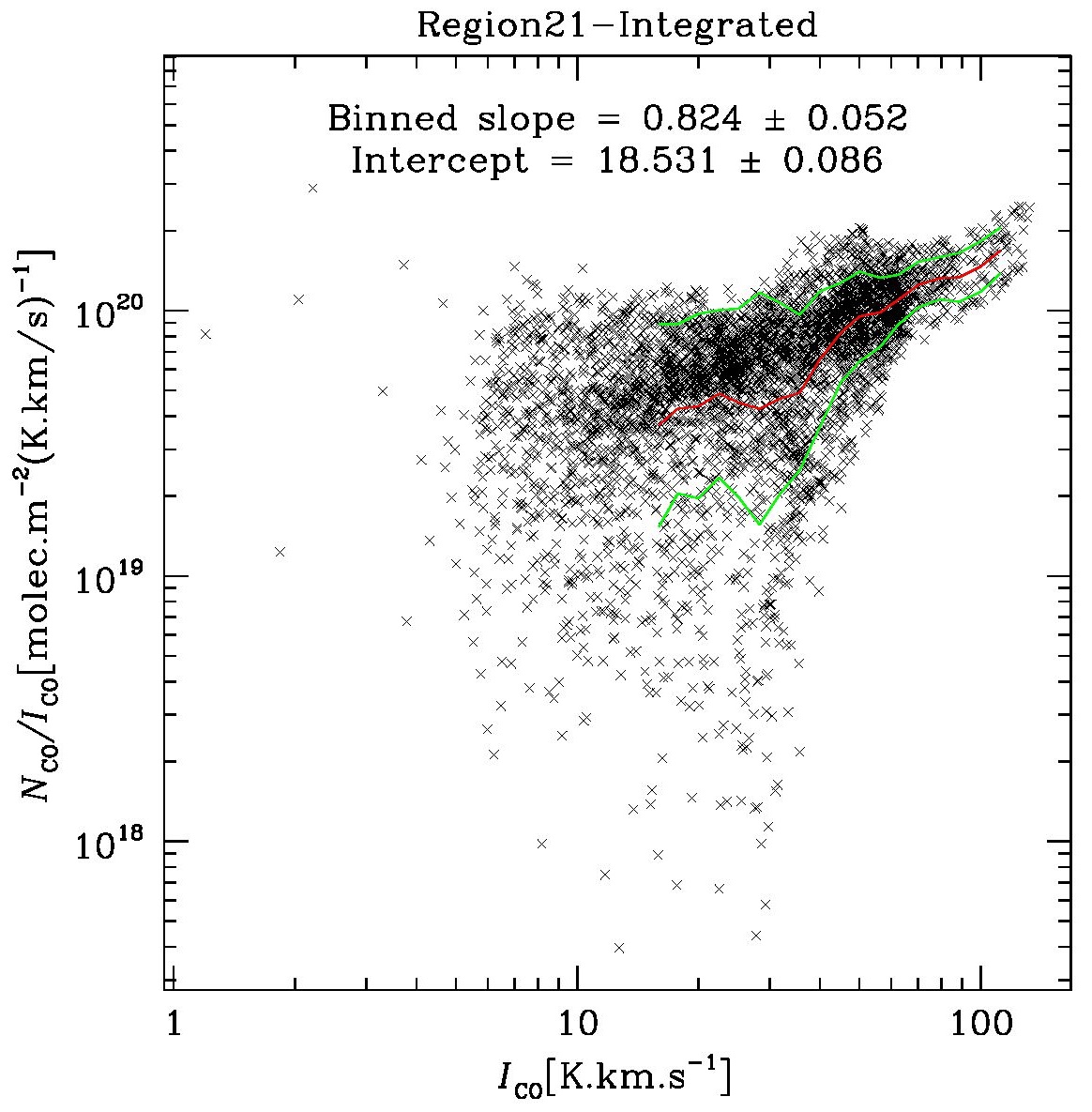}\hspace{4mm}
		\includegraphics[angle=0,scale=0.21]{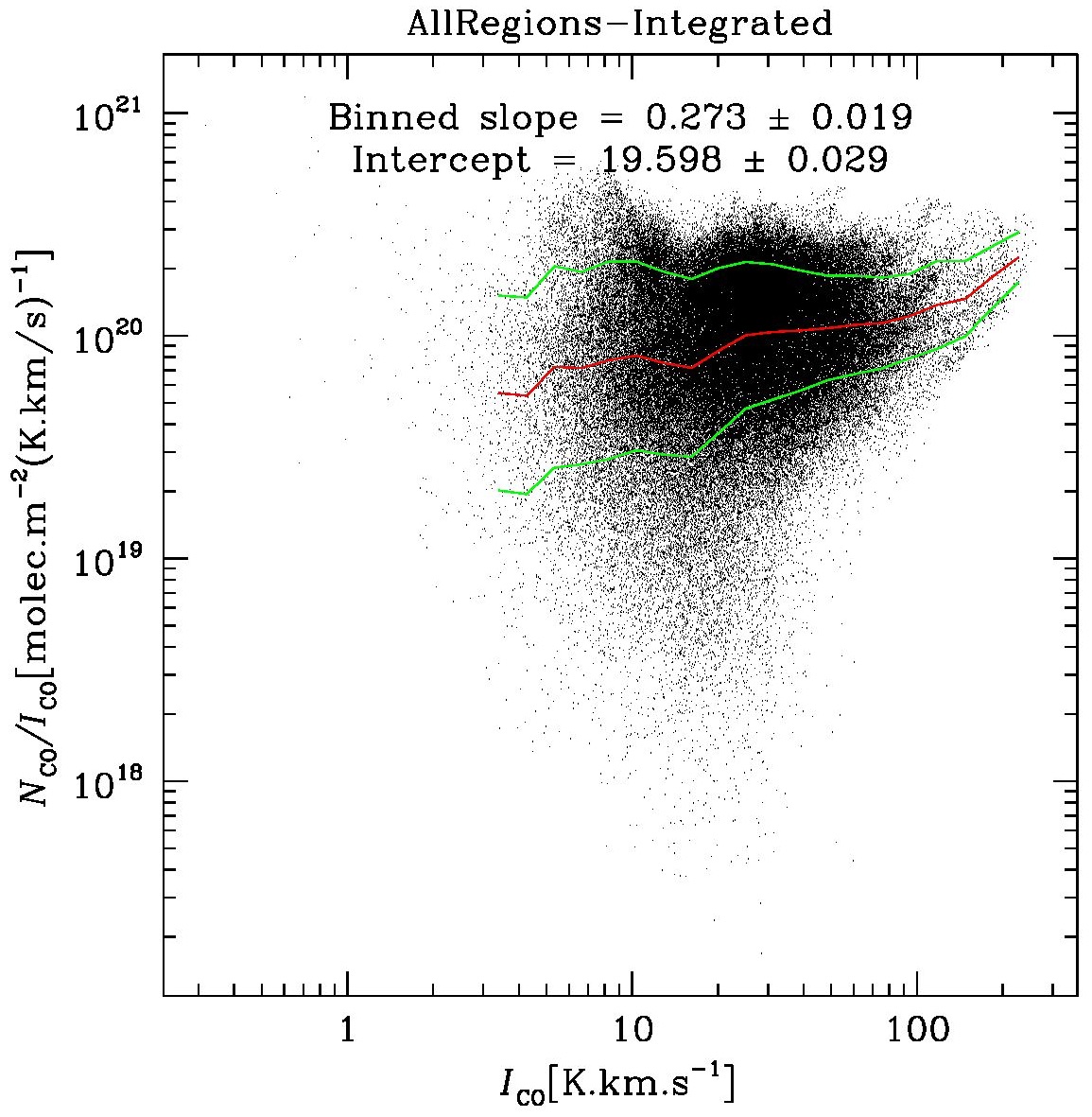}
}
\vspace{-5mm}\hspace{5mm}(a)\hspace{88mm}(b)
\caption{(a) \nco/\ico\ ratio vs \ico\ as in Figs.\,\ref{convlaw}a and \ref{maskint}a, but where each quantity is integrated over {\bf all} velocity channels, regardless of whether \nco\ is defined in all the channels where \ico\ is detected.  We show again the same horizontal lines as in Figs.\,\ref{convlaw} and \ref{maskint} to indicate the locus of the standard $X$-factor for different values of $R_{12}$, but without the ($\tau_{12}$, \tex) grid since this integration blurs the grid.  (b) Same as panel $a$, but an aggregated plot for all Regions.
}
\label{fullint}
\begin{picture}(1,1)
\thinlines
{\color{blue}
\put(41,240){\line(2,0){204}}
\put(300,231){\line(2,0){204}}}
\thicklines
{\color{cyan}
\put(38,211){\line(2,0){204}}
\put(297,207.1){\line(2,0){204}}}
{\color{magenta}
\put(35,180.4){\line(2,0){204}}
\put(294,181.6){\line(2,0){204}}}
\end{picture}
\vspace*{-5mm}
\end{figure*}
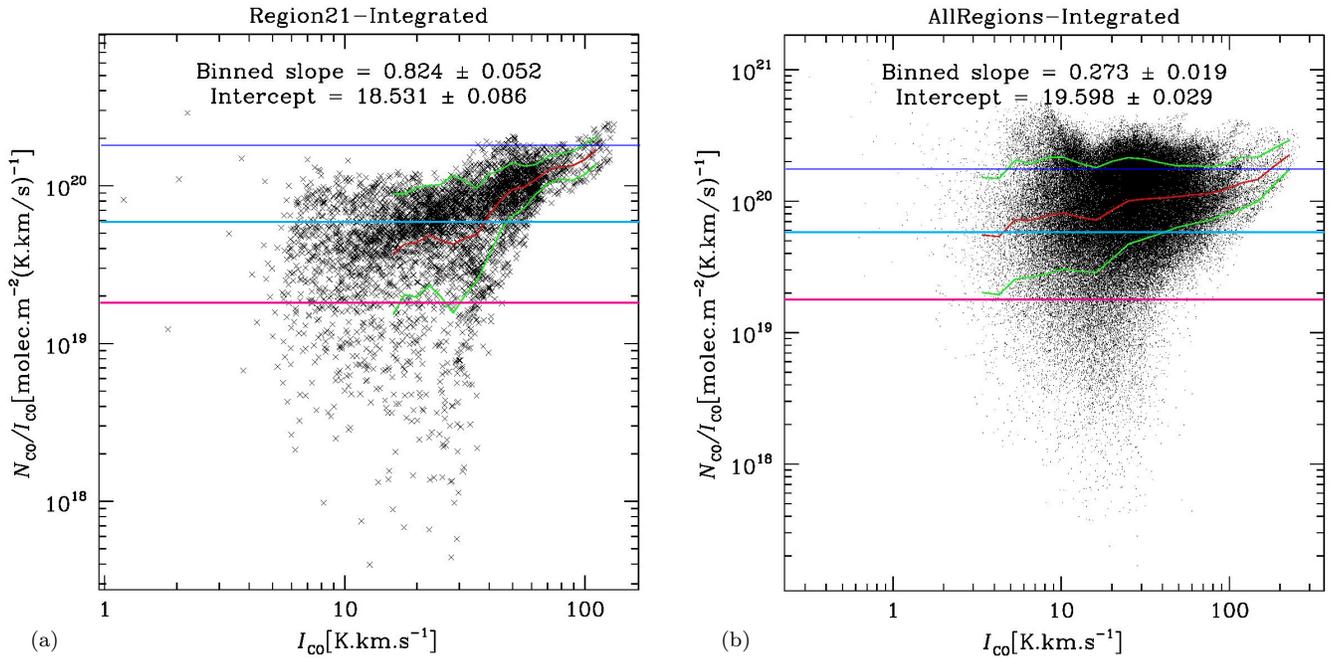

Even so, one could argue that the masked integration, while preserving the intrinsic physics of the radiative transfer, is not the same as what is used observationally, which is the total {\em un}masked \ico\ across all velocity channels with detectable \tco\ emission.  Dividing the computed \nco\ by this \ico\ will indeed lower the pseudo-$X$ quotient, especially for fainter pixels where the number of \ttco\ channels available to compute \nco\ will be (sometimes substantially) less than the number of \tco\ channels contributing to that integral.  Therefore, we plot unmasked versions of the same two panels in Figure \ref{fullint}, to illustrate practical conversion laws that could be used with \tco\ data alone.

As expected, the low-$I$ values of the $N$/$I$ ratio are much lower than in Figure \ref{maskint}, but more intriguing is that the net effect in Figure \ref{fullint}a is to give a single fitted slope for most of the \ico\ range in each Region.  Reassuringly, these fitted slopes are typically only slightly smaller (although with some scatter) than those derived from the full-resolution data in each Region, meaning that we do preserve the high-$I$ radiative transfer physics to some extent.  But averaged across all Regions, Figure \ref{fullint}b shows this power law again becomes more muted, although not completely flat as in Figure \ref{maskint}b.  At the same time, we see again in Figure \ref{fullint}b that the steeper power laws of individual Regions are still embedded in this diagram.  Much of the muting of individual Regions' power law behaviour, therefore, is due to different local normalisations of the inherently steeper power law relationship, governed by the radiative transfer physics, i.e., the dependence in this diagram on local values of both $\tau$ and \tex.  This is in addition to the flattening of the individual Regions' power laws due to velocity binning/integration.

Ultimately, this means that we should be careful to distinguish between local and average \tco\ $\rightarrow$ \htwo\ conversion laws.  For large-scale studies of the molecular ISM in disk galaxies, an average conversion is clearly appropriate.  Where local conversion laws can be established, they should be used in preference to an average one, since we see even in Figure \ref{fullint}b that local normalisations can vary by a factor of at least 2 around the mean.  But where local conversions are unknown or impractical to calibrate, an average law is also appropriate, while acknowledging the additional calibration uncertainty therein.

Our best estimate for an average conversion law is given by Figure \ref{fullint}b.  That fitted slope corresponds to a power law across almost 2 orders of magnitude in \ico\ (3--200\,K\kms), and with a dispersion of 0.3--0.1 dex (resp.) across this range,
\begin{eqnarray}   
	& &N_{\rm ^{12}CO} = N_0\left(\frac{I_{\rm ^{12}CO}}{\rm K\,km\,s^{-1}}\right)^p,~~~~~~{\rm where}~~~ \\
	& &N_0 = (4.0\pm0.3)\times10^{19}\,{\rm CO\,molecules\,m}^{-2},~~~{\rm and} \nonumber \\
	& &p = 1.27\pm0.02~~~. \nonumber   
\end{eqnarray}
The uncertainties quoted are the formal least-squares fitting errors to all 140,000 pixels over this range of \ico, not the intrinsic dispersion in the data.

\notetoeditor{}
\begin{figure*}[t]
\centerline{\includegraphics[angle=-90,scale=0.635]{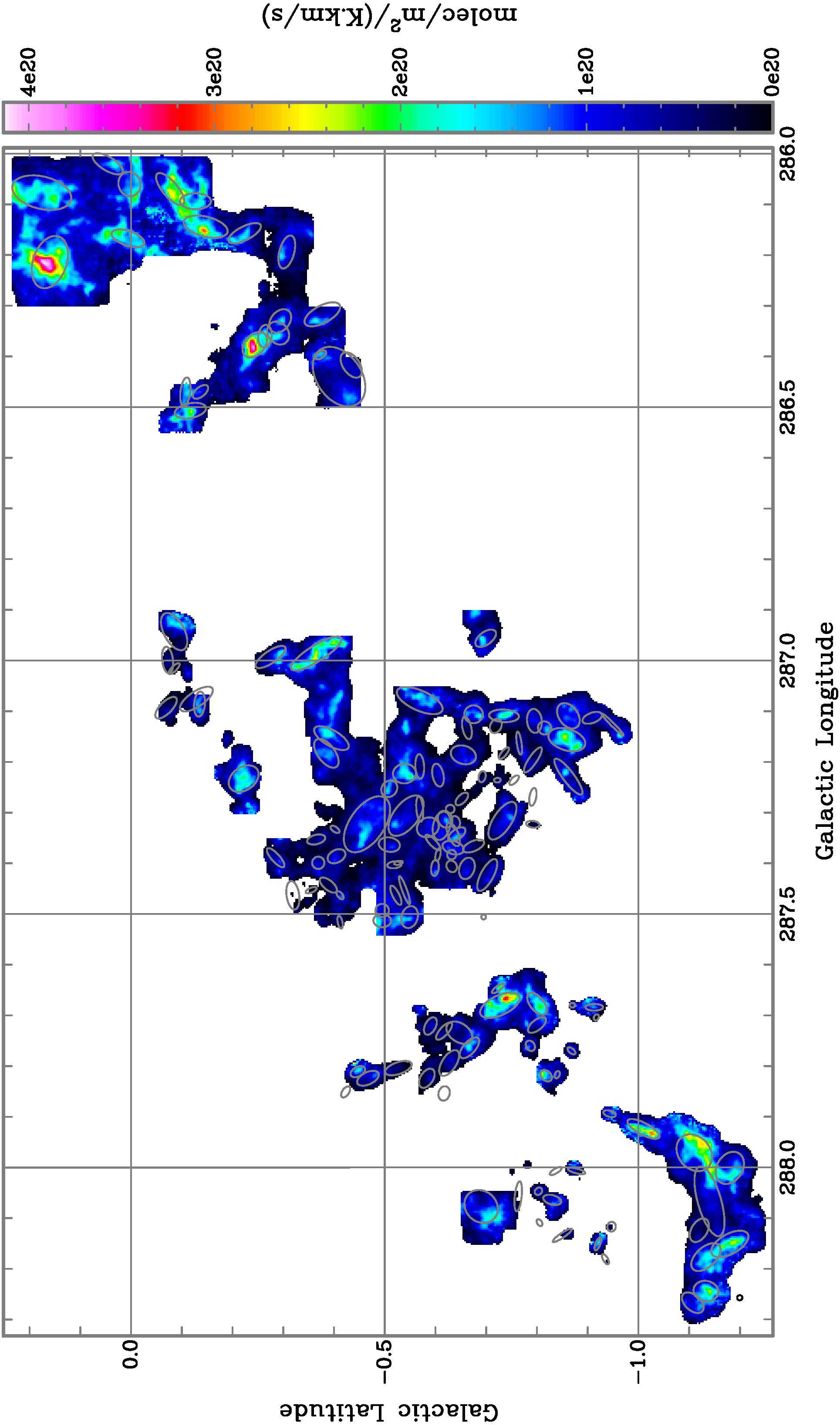}}
\caption{Sample map of the $X$ factor across the $\eta$ Car GMC (the same area as in Fig.\,\ref{etacar}), obtained from the ratio of integrals \nco/\ico\ as calculated in Fig.\,\ref{fullint}.}
\label{etacarXcl}
\vspace{-2mm}
\end{figure*}

The power law index in Eq.\,8 is now smaller than any in Figures \ref{convlaw}, \ref{convagg}, or from ThrUMMS.  Given that it is an integral relation, and that we have shown in Figures \ref{convlaw} and \ref{convagg} that the index becomes shallower with more velocity-binning/integration/spatial averaging, this should not be surprising, since the radiative transfer physics has been blurred, as previously discussed.  But this is the most sensitive, best-calibrated, and most practical parametrisation of the integral conversion law to date.  Also, Eq.\,8 still contains a super-linear power, which suggests that the CO abundance plays a role (see below).  Finally, the normalisation of this power law gives higher than expected column densities, as seen by the blue, cyan, \& magenta lines in these plots: the data generally lie towards the upper end of the range bracketed by these lines.  We show a sample mosaic of this conversion across the $\eta$ Carinae GMC in Figure \ref{etacarXcl}, the same area as displayed in Figure \ref{etacar}.  There one can see that the conversion factor qualitatively tracks the \ico\ distribution in many places, as required by the super-linear relation (Eq.\,8).

To see how $p$$>$1 connects to the astrochemical evidence of CO abundance variations \citep{gbs14}, note that relative to \htwo, the \tco\ abundance (= $R_{12}^{-1}$) varies from $\sim$10$^{-5}$ in dark clouds (where \ico\ is typically near the lower end of the range in Fig.\,\ref{fullint}b, perhaps 10\,K\kms), to $\sim$10$^{-4}$ near HII regions (where \ico\ tends to be bright, say 100\,K\kms).  Based on this, one could suppose an approximate relation like
\begin{equation}   
	R_{12}^{-1} = 10^{-6}I_{\rm ^{12}CO}/({\rm K\,km\,s}^{-1})~,~
\end{equation}
although the actual relation is probably weaker than this, since Eq.\,9 would imply that \ico\ directly traces only the \tco\ abundance.  Nevertheless, something like this relation would indeed be indicated if widespread CO-dark molecular gas layers exist around molecular clouds \citep{p13}, since the smaller self-shielding of CO reduces its abundance in the \htwo\ molecular layer near $A_V$ = 2$^m$, and changes the \tco\ $\rightarrow$ \htwo\ mass conversion.  Then, where an $R_{12}$-\ico\ relation holds, we might write
\begin{eqnarray}   
	N_{\rm H_{2}}	& = & R_{12}N_{\rm ^{12}CO}~,~{\rm or} \\
		& = & 10^{6}\,N_0\left(\frac{I_{\rm ^{12}CO}}{\rm K\,km\,s^{-1}}\right)^{p-1}~ \nonumber
\end{eqnarray}
in the case of the hypothetical Eq.\,9: this would produce a very large mass conversion indeed, and is probably unlikely as explained above.  Instead, we take a median $R_{12}$ = 3$\times$10$^4$ (as for the cyan lines in Figs.\,\ref{convlaw}, \ref{maskint}, \& \ref{fullint}).  In this case, Eq.\,10 then gives a median \tco\ $\rightarrow$ \htwo\ mass conversion for the CHaMP clouds of
\begin{equation}   
	N_{\rm H_{2}} = (1.2\pm0.1)\times10^{24}\,{\rm m}^{-2}\left(\frac{I_{\rm ^{12}CO}}{\rm K\,km\,s^{-1}}\right)^p~,
\end{equation}   
with $p$ as in Eq.\,8.  When allowing for the typical values of \ico, this conversion is 1.9$\pm$0.9 times greater than a flat $X$ factor with $R_{12}$ = 3$\times$10$^4$ (compare the distribution of points to the cyan line in Fig.\,\ref{fullint}b), and similar to what was found by \citet{bm15} for the ThrUMMS data.  This result supports the same global underestimation of the Milky Way's molecular mass budget as first described by \citet{bm15}, but is now based on a more sensitive and careful treatment.  The obvious next step is to actually evaluate $R_{12}$ among the CHaMP clouds, in order to pin down the numerical form of Eqs.\,9--11: see \cite{pbv18} for the first steps in such work.

On this basis, we recommend that studies converting integrated \tco\ emission to equivalent \tco\ column densities employ Eq.\,8, and if an appropriate $R_{12}$ is known, use that to convert to an \htwo\ column density with a relation like Eqs.\,10 or 11, rather than using a single $X$ factor for all \ico.  This recommendation extends to studies of molecular clouds in other disk galaxies that are analogues to the Milky Way (e.g., with similar metallicity).

\section{Differential Kinematics in Clumps}\label{kinem}
\subsection{Formalism}\label{formalkin}

We now introduce a new concept enabled by these sensitive multi-species maps.  On one hand, we have derived relatively unbiased maps of the total CO column density in each Region.  Due to the nesting of the iso-CO species in terms of the depths into which they probe molecular cloud conditions, these maps are a high dynamic range measure of this column density, from the lowest values that can be traced by \tco\ at this sensitivity, $\sim$3\,M\solar\,pc$^{-2}$ at 5$\sigma$, up to the highest column densities we trace, typical of dense clumps, $>$1000\,M\solar\,pc$^{-2}$.

\begin{figure}[t]
\vspace*{-2mm}\hspace{-4mm}
\includegraphics[angle=0,scale=0.33]{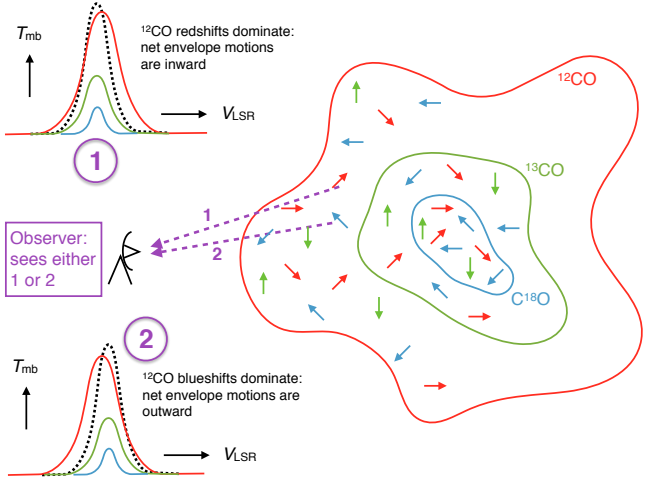}\vspace*{-2mm}
\caption{Schematic illustration (not to scale) of a typical molecular cloud, as traced by the three different isotopologues \tco\ (red), \ttco\ (green), and \ceto\ (blue), shown in outline.  The coloured arrows represent the line-of-sight velocities of parcels of gas (any species) which are in motion relative to the mean cloud \vlsr\ (red- or blue-shifted as appropriate).  Note that the motions probed by \tco\ are limited to parcels of gas on the side of the cloud envelope nearest to the observer, because of that species' high opacity.  The \ttco\ and \ceto\ are more likely to sample motions of most parcels of gas along each line of sight, due to their much lower opacities.  (However, neither \ttco\ nor \ceto\ necessarily trace the {\em total} column density by themselves.)  The sample spectra shown (coloured by species, except for the total column density spectrum, which is represented as a dotted black curve) are then expected to result when the envelope has a net inward (upper diagram) or outward (lower diagram) motion with respect to the cloud interior.}
\label{schematic}
\end{figure}

On the other hand, the \tco\ maps by themselves are of a species which is optically thick almost everywhere it is seen, and certainly where any \ttco\ is seen, with the exception of the small fraction of voxels with \ico\ $\sim$ 0.5\,K\kms\ and \ittco\ $<$ 0.06\,K\kms.  This represents a very different perspective than the conventional view, where it is thought that velocity differences in the gas might allow one to probe the whole cloud volume, even in an optically thick emission line.  Instead, it suggests that the \tco\ traces gas almost exclusively on the side of each molecular cloud that is {\em nearest} to us, with only a few percent of a cloud's volume or mass traced where $\tau_{12}$ \lapp\ 3 (see Fig.\,\ref{convlaw}b).  
An interesting comparison can then be made between this nearside gas, traced by a $\Sigma_{12}$ derived directly from the \tco\ cube via a fixed, local conversion law (following the discussion in \S\ref{genlaw}), and the gas from the bulk of the cloud as traced by the total $\Sigma_{\rm mol}$ (Eq.\,7), which is more accurate than the locally-derived conversion law fit, since the \nco\ computations are made on each voxel.\footnote
{In principle, the \ttco\ or \ceto\ cubes also approximately trace total $N$/$\Sigma$, due to their much lower opacities, but this is not exactly true, and will diverge most strongly at the highest column densities.  Therefore, comparing the $\Sigma_{12}$ foreground gas with the total $\Sigma_{\rm mol}$ is the most accurate way to explore the differential dynamics.}  
In this way, we measure accurate column densities separately for the cloud envelopes and interiors.  A schematic diagram showing these effects in a typical cloud is given in Figure \ref{schematic}.

We consider the velocity field of each, in particular, the difference in these two velocities.  At each pixel, if the mean velocity $V_{\Sigma_{12}}$ (measured by the 1st moment in the $\Sigma_{12}$ cube) is red-shifted compared to the mean velocity $V_{\Sigma_{\rm mol}}$ of the $\Sigma_{\rm mol}$ cube, then that would suggest that the nearside envelope material is moving {\em towards} the bulk of the cloud, at that location.  In this case ($\Delta$$V_{\rm env}$ $\equiv$ $V_{\Sigma_{12}}$--$V_{\Sigma_{\rm mol}}$ $>$ 0), the cloud would be {\em accreting} the nearside gas there.  Conversely, blue-shifted \tco\ ($\Delta$$V_{\rm env}$ $<$ 0) indicates nearside envelope gas that is {\em expanding} away from the bulk of the cloud, in the pixel under consideration.

In making this measurement, we want to actually find the differential {\em mass flux} in each pixel between the clump envelope and interior, rather than just the $\Delta$$V_{\rm env}$, in order to establish whether the clump is accreting or losing mass.  In other words, we want the mass-weighted version of $\Delta$$V_{\rm env}$.  We now carefully define the momentum surface density in \tco\ at each pixel, $p_{\Sigma{\rm ,env}}$, as $\Sigma_{12}$\,$\Delta$$V$, or more precisely, the integral (0$^{\rm th}$ moment) of a cube of such values,
\begin{eqnarray}   
	p_{\Sigma{\rm ,env}} & = & \int\Sigma_{12}(V_{\Sigma_{12}}-V_{\Sigma_{\rm mol}})\,{\rm d}V\\
			& = & \int\Sigma_{12}V_{\Sigma_{12}}\,{\rm d}V - \int\Sigma_{12}\,{\rm d}V \times V_{\Sigma_{\rm mol}} \nonumber \\
			& = & \Sigma_{12{\rm ,integr}}{<}{V}{>}_{\Sigma_{12}} -\ \Sigma_{12{\rm ,integr}}V_{\Sigma_{\rm mol}} \nonumber \\
			& = & \Sigma_{12{\rm ,integr}}\,\Delta{V}_{\rm env}~~~~, \nonumber
\end{eqnarray}
where each line in Eq.\,12 illustrates how the various quantities can be separated.  $\Sigma_{12}$ refers to quantities from the \tco-derived $\Sigma$ cube (e.g., $<$$V$$>_{\Sigma_{12}}$ is the 1st moment of this cube); and $V_{\Sigma_{\rm mol}}$ is the 1st moment of the total $\Sigma_{\rm mol}$ cube (i.e., derived from the iso-CO data), which is already single-valued at each pixel.  The velocity difference inside the integral sign in the first line of Eq.\,12, is that between each channel of $\Sigma_{12}$, and the single number from the iso-CO $\Sigma_{\rm mol}$ velocity field $V_{\Sigma_{\rm mol}}$.  For reference, we can also define the $\Sigma$-weighted mean velocity-difference as $\Delta$$V_{\rm env}$ = $p_{\Sigma{\rm ,env}}$/$\Sigma_{12{\rm ,integr}}$.

Calculated in this way, $p_{\Sigma{\rm ,env}}$ gives a net algebraic sum of the momentum across all velocity channels in a given pixel, whether positive or negative.  Suitably normalised, this has units of M\solar\,pc$^{-1}$\,Myr$^{-1}$.  To convert this into a true net mass flux (whether accreting or dispersing), we need to divide $p_{\Sigma{\rm ,env}}$ by an estimate for the length scale along the line of sight over which this mass is moving. 

\notetoeditor{}

\notetoeditor{}
\begin{figure*}[ht]
\centerline{\hspace{5mm}{\large $\Delta{V}_{\rm env}$}
		\hspace{50mm}{\large $\dot{\Sigma}$}
		\hspace{45mm}{\large $t_{\rm accr}$ \& $t_{\rm disp}$}}
\centerline{\includegraphics[angle=0,scale=0.31]{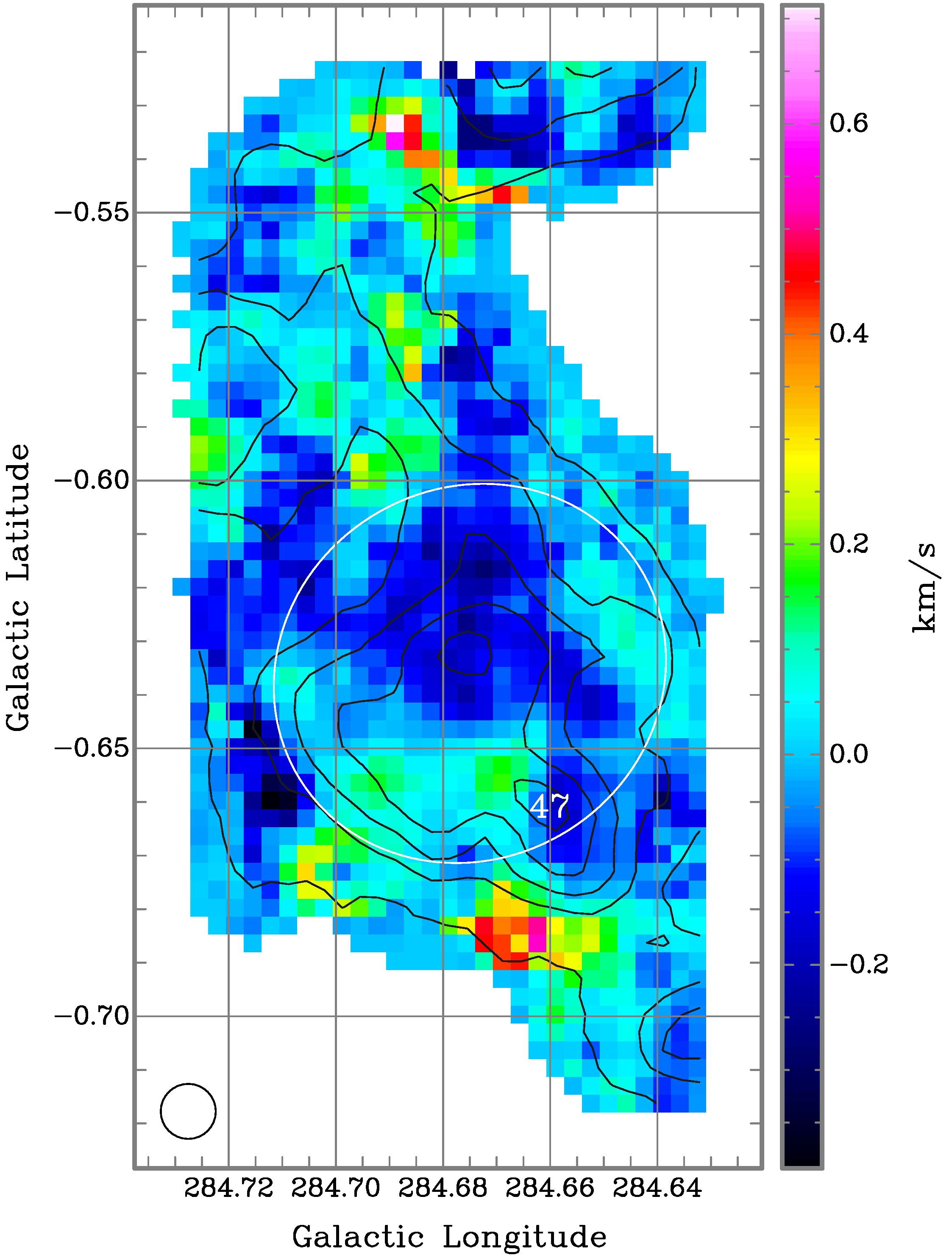}
		\includegraphics[angle=0,scale=0.31]{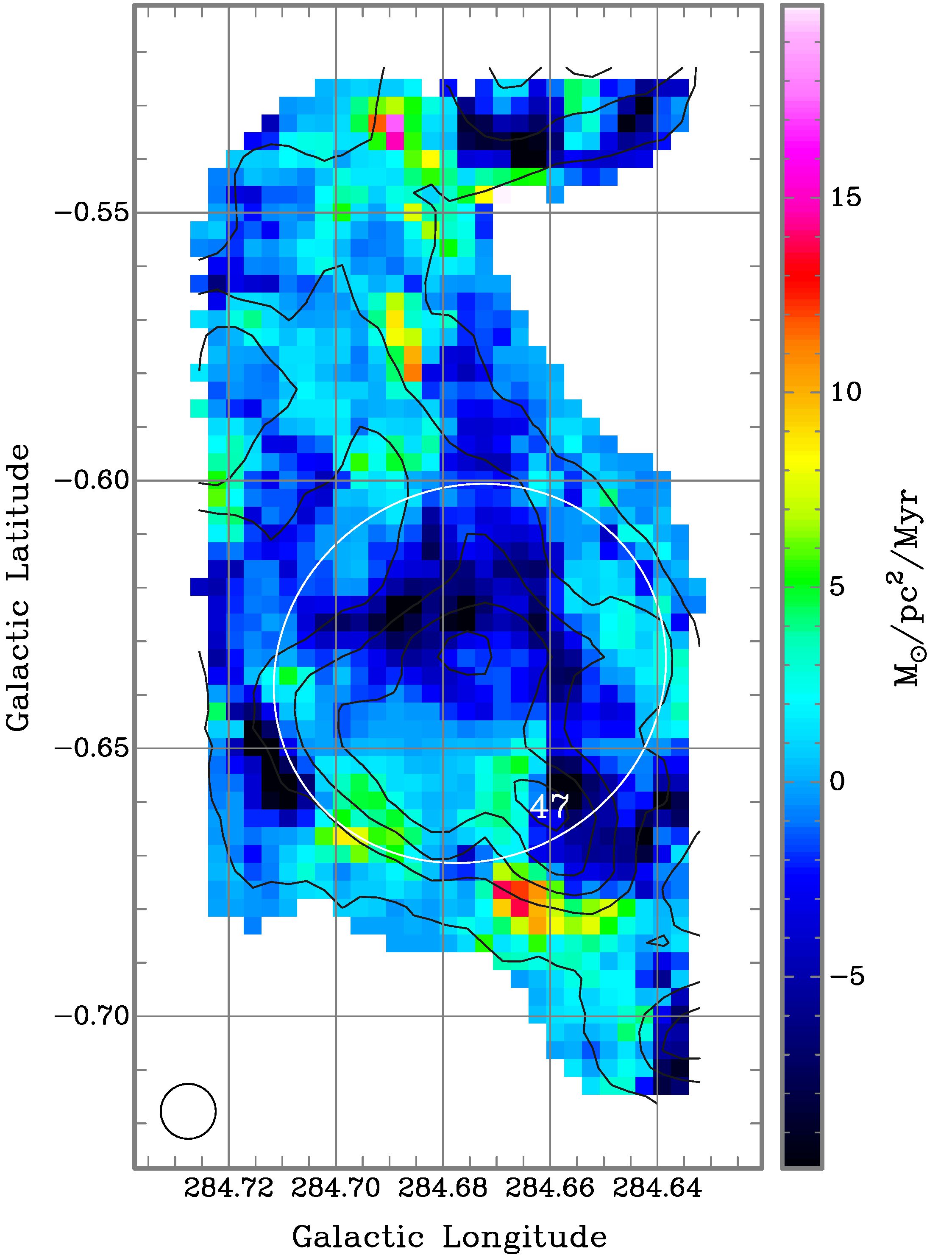}
		\includegraphics[angle=0,scale=0.31]{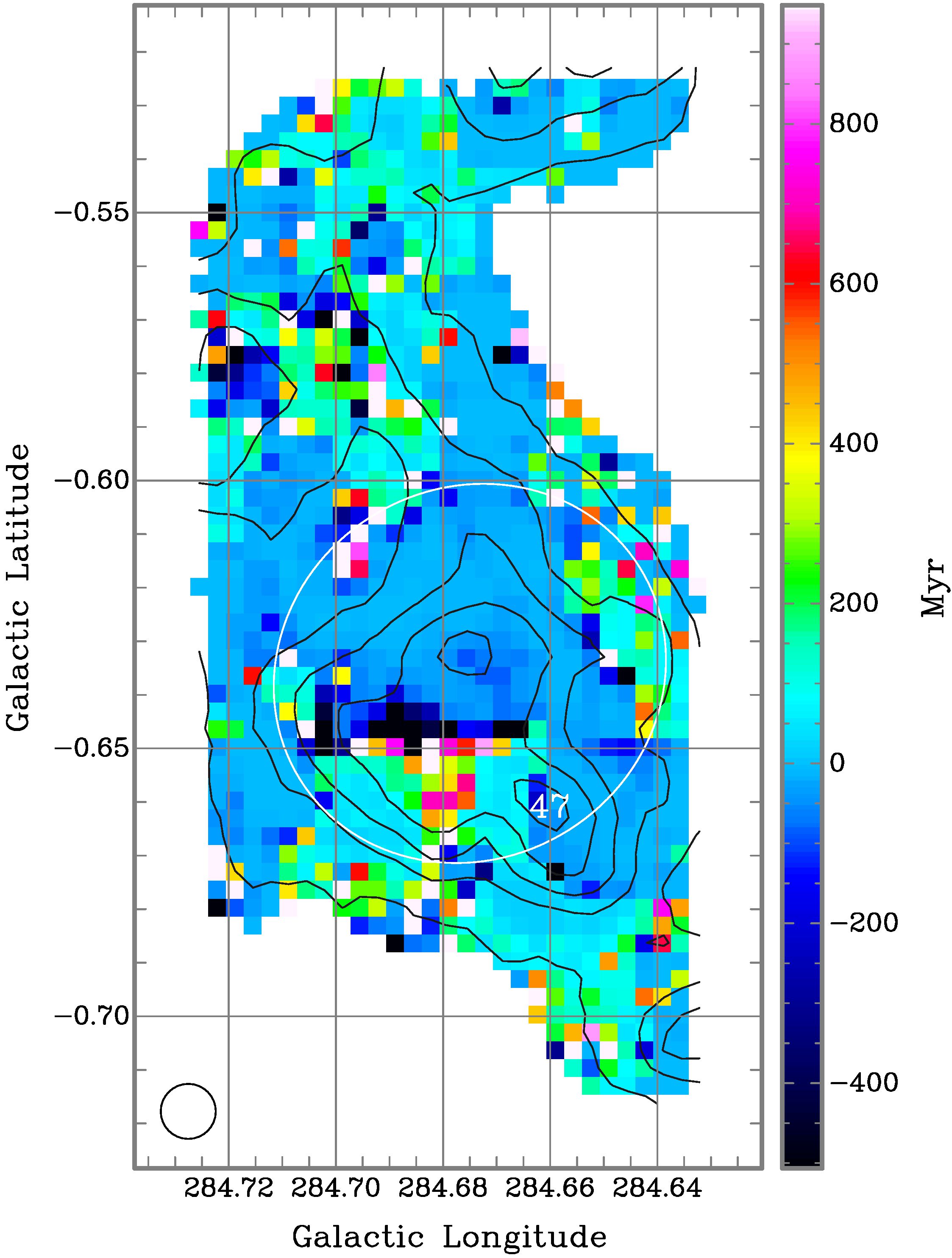}}
\caption{Sample dynamics maps for Region 7 as labelled, and computed according to Eqs.\,12--15.  The colour scale is adjusted in each panel so that zero values lie between cyan and dark blue, which provides an intuitive scheme to visually distinguish accretive motions (cyan and warmer colours) from dispersive ones (cooler blues).}
\label{dynsample}
\end{figure*}

Under the assumption that the depth of any given structure (not just a cloud) can be approximated by its geometric mean size $\sqrt{b_{\rm major}b_{\rm minor}}$ projected on the sky, and more importantly, that this size scale can be further approximated by the inverse relative gradient $R_{\Sigma_{\rm mol}}$ = $\Sigma_{\rm mol}$/$\nabla\Sigma_{\rm mol}$ of the projected mass distribution in that structure, then we can calculate the mass flux in a spatially resolved way,
\begin{equation}   
	\dot{\Sigma} = \frac{{\rm d}\Sigma_{12}}{{\rm d}t } = \frac{p_{\Sigma{\rm ,env}}}{R_{\Sigma_{\rm mol}}}
\end{equation}
in each pixel.  Optionally, this can be summed over any area of interest (such as a clump) to obtain a net mass flux rate (in units of M\solar\,Myr$^{-1}$) for that structure, but Eq.\,13 holds for each pixel.  Finally, we can define a mass flux timescale in each pixel as
\begin{equation}   
	t_{\dot{\Sigma}} = \frac{\Sigma_{\rm mol}}{\dot{\Sigma}} = \frac{R_{\Sigma_{\rm mol}}}{\Delta V_{\rm env}}~,
\end{equation}
where the second expression recognises the appropriate timescale for the envelope motions observed.  If the envelope mass flux ${\dot{\Sigma}}$ $>$ 0, this timescale is for mass accretion through each pixel or, when integrated over the appropriate area, for each clump or structure.  If ${\dot{\Sigma}}$ $<$ 0, this corresponds instead to a mass dispersal timescale.

\begin{figure}[t]
\centerline{\includegraphics[angle=0,scale=0.21]{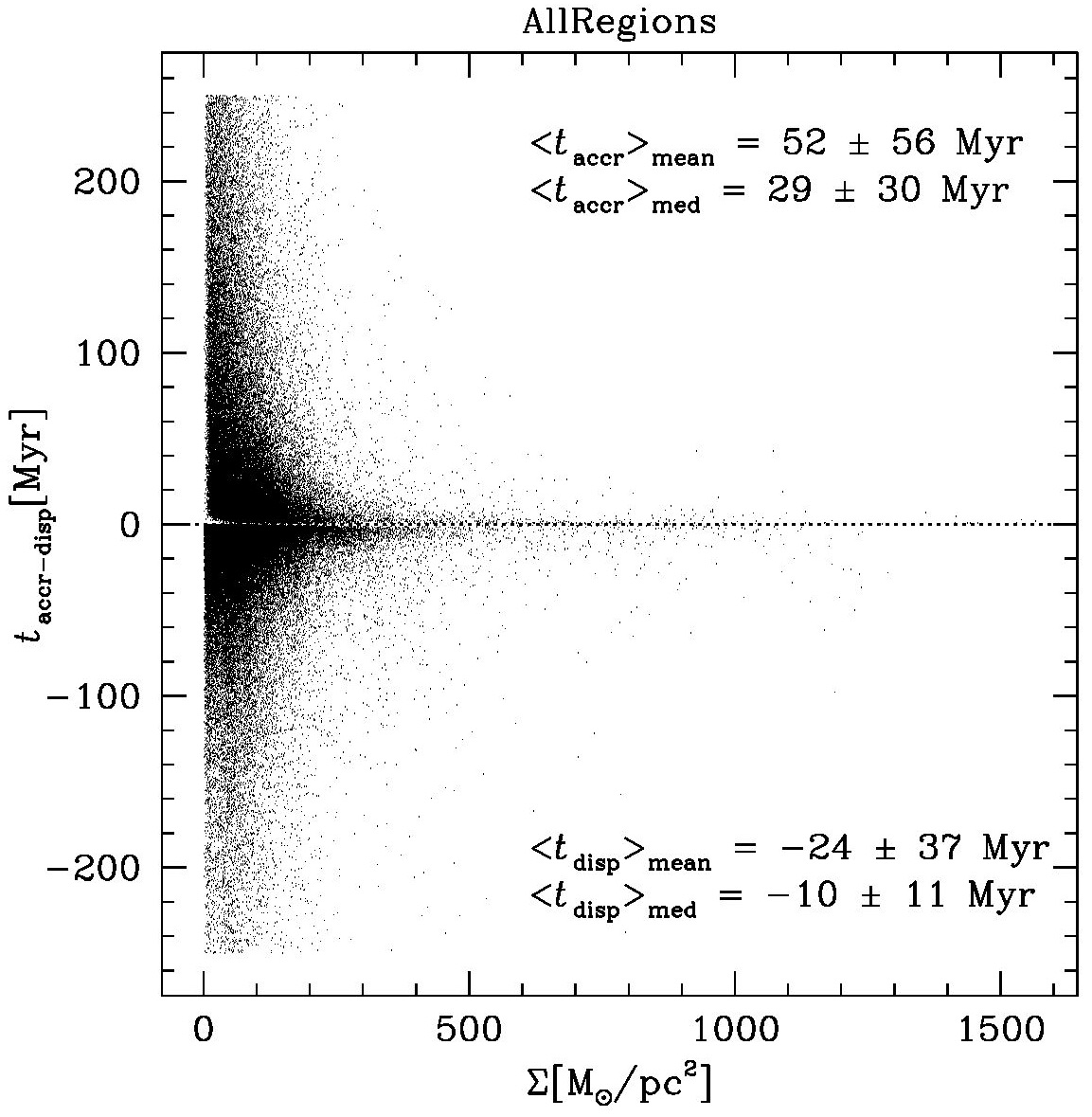}}
\caption{Plot of accretion/dispersal timescales vs the mass surface density, in each pixel and aggregated across all Region maps, on linear scales for both axes.  As defined in Eq.\,15, accretion and dispersal timescales are shown as positive- and negative-valued, respectively, and the separation between them is indicated by a dotted line at $t$=0.  The labelled mean and median values are computed separately for each timescale.}
\label{tcomb}
\end{figure}

However, for purposes of measuring the influence of this flux on star formation, a possibly more useful timescale is that required to accrete some additional standard mass surface density, above which clear evidence of ongoing star formation (e.g., presence of an embedded cluster, etc.) is commonly seen.  Here we conservatively take this threshold for addition to be $\Sigma_{\rm thresh}$ = 200\,M\solar\,pc$^{-2}$, although it may actually be somewhat larger, $\sim$300--1000\,M\solar\,pc$^{-2}$ according to the results in Paper III, with a corresponding increase in the resulting timescale.  Using the slightly smaller threshold means we are computing firm lower limits to the timescales; if they are even longer than what is discussed below, the results and implications are only strengthened. We now combine these
\begin{eqnarray}   
{\rm evolution~timescales}~~&\equiv&~t_{\rm accr} = \frac{\Sigma_{\rm thresh}}{\dot{\Sigma}}~,~~{\dot{\Sigma}}>0~~~~~~ \nonumber \\
				&\equiv&~t_{\rm disp} = {\Sigma_{\rm mol}}/{\dot{\Sigma}}~,~~{\dot{\Sigma}}<0
\end{eqnarray}
so we can track both an accretion and a dispersal timescale in one function.  In 
Figure \ref{dynsample} we provide sample images of $\Delta$$V_{\rm env}$, 
$\dot{\Sigma}$, and $t_{\rm accr/disp}$ for Region 7, as defined above.  We also compare $t_{\rm accr/disp}$ with $\Sigma_{\rm mol}$ in each Region, as shown in Appendix \ref{dynmaps}; an aggregate $t$--$\Sigma$ plot for all Regions is shown in Figure \ref{tcomb}.

\notetoeditor{}

\notetoeditor{}
\begin{figure*}[ht]
\centerline{\includegraphics[angle=0,scale=0.21]{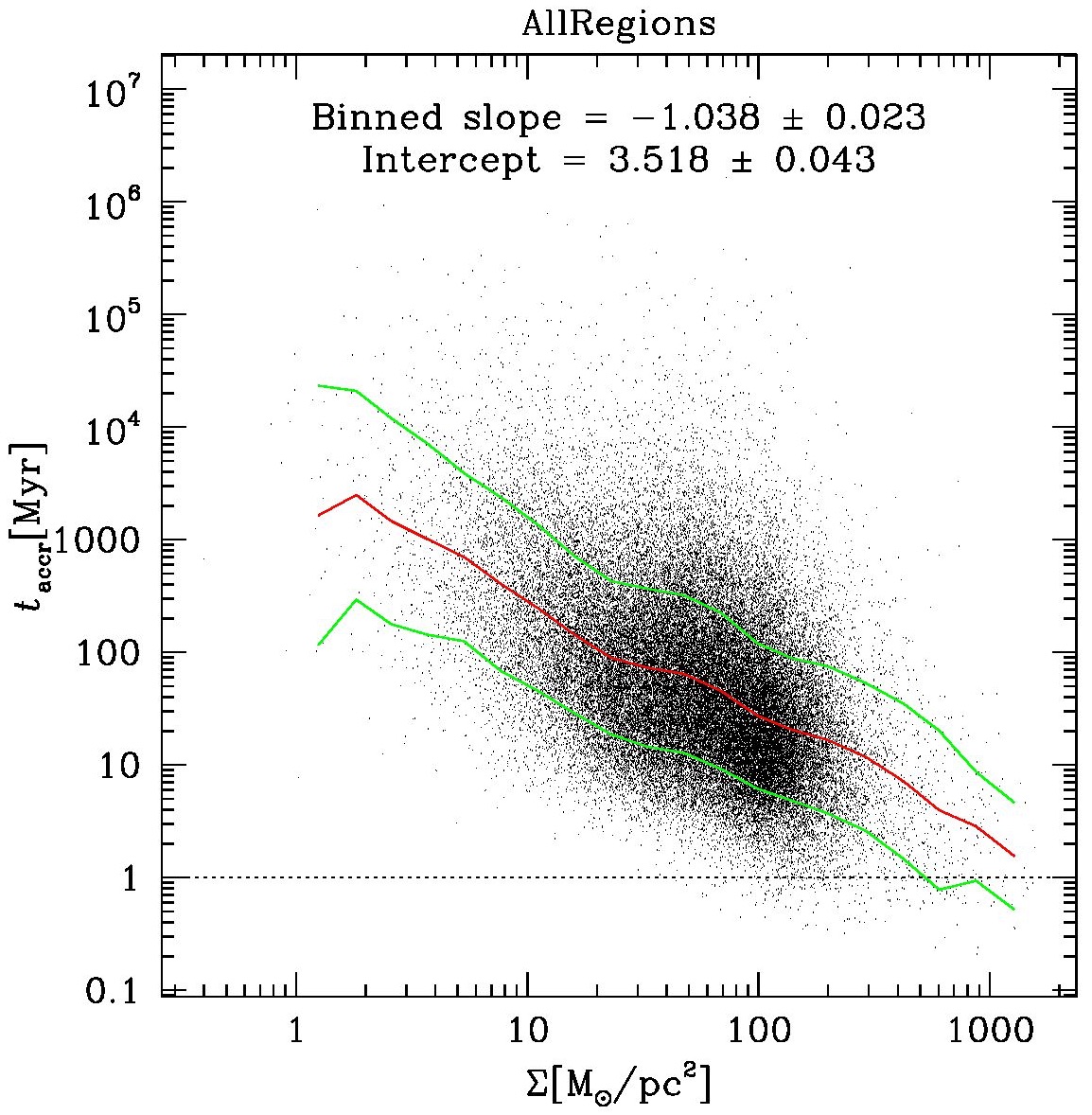}\hspace{4mm}
		\includegraphics[angle=0,scale=0.21]{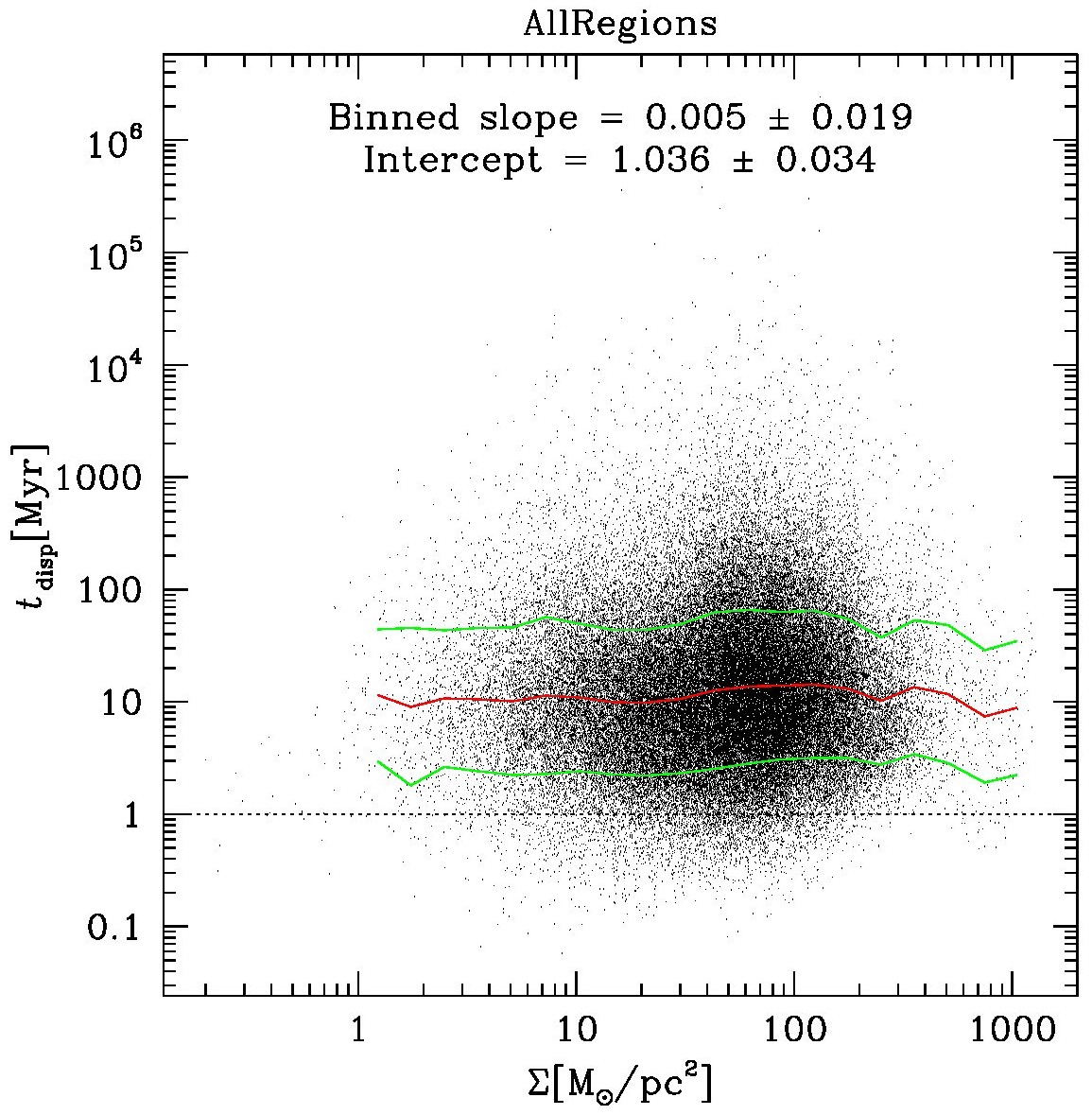}}
\vspace{-4mm}\hspace{5mm}(a)\hspace{88mm}(b)
\caption{(a) Accretion timescales vs mass surface density as in Fig.\,\ref{tcomb}, but on logarithmic scales for both axes.  Overlaid as red and green lines are trends in the mean $\pm$1$\sigma$ in log($t_{\rm accr}$) for small bins of log($\Sigma$), while a least-squares fit to the binned values is as labelled.  This fit has a 
reduced $\chi^2$ = 0.16 and Pearson's correlation coefficient $r$ = --0.995.  The dotted horizontal line is set at $t$ = 1 Myr.  (b) Similar to panel a, but for dispersal timescales (now shown as positive-valued).  The labelled least-squares fit has 
reduced $\chi^2$ = 0.10 and $r$ = 0.06.
}
\label{tad}
\end{figure*}

\subsection{Mass Accretion}

The main result of this analysis is that, throughout most of the maps, there are large contiguous areas which are either accreting material from the envelopes onto the clumps' interiors (the net mass flux is {\em inwards} to the bulk of the clump), or where the clumps' envelopes are undergoing dispersal (the net mass flux is {\em outwards} from the bulk of the clump).  At first glance, there seems to be no systematic correlation between the clump locations (i.e., at the higher surface densities) and whether material is preferentially being accreted or dispersed.  In other words, the clump population seems roughly evenly divided between those accreting (58,000 pixels) and those dispersing (81,400 pixels).  What does seem systematic is that the dispersal times are typically in the 10 Myr range, while the accretion times can be quite long, especially at low surface densities, up to several 10s, or even 100s, of Myr (although the largest values in each timescale do tend to occur where the uncertainties are highest, e.g., where $\Delta$$V_{\rm env}$ $\approx$ 0).  

To some extent, this difference between longer accretion times and shorter dispersal times is by construction, since $t_{\rm accr}$ is defined against a fixed mass surface density, and one which is somewhat larger than in most pixels.  In contrast, $t_{\rm disp}$ is self-scaled, i.e., it is the time required to disperse the mass that exists at that position, whatever its value, which for most pixels is smaller than the accretion threshold.  But this difference becomes much more interesting when we examine the accretion and dispersal timescales separately on logarithmic scales.

In Figure \ref{tad}a, we see a remarkable result:\footnote
{Note that Figure \ref{tad}a does not imply $t$ $\propto$ $\Sigma^{-z}$, or $\Sigma$ $\propto$ $t^{-1/z}$, which would imply mass loss, not accretion.  The ordinate is the {\em remaining} time required to accumulate an additional $\Sigma_{\rm thresh}$ at the observed $\dot{\Sigma}$, and necessarily decreases with chronological time.  It is not an independent variable for age or time.} 
the time to accrete enough molecular gas to pass the assumed star formation threshold is strongly correlated with $\Sigma_{\rm mol}$.  We can examine this algebraically if we assume that each pixel fairly samples an underlying mass accretion law, with variables for the accretion rate and column density as interdependent functions of the time needed to accrete an additional threshold mass:
\begin{eqnarray}   
	t_{\rm accr} & = & \frac{\Sigma_{\rm thresh}}{\dot{\Sigma}(t)} = t_0~\left(\frac{\Sigma_{\rm mol}}{{\rm M_{\odot}\,pc}^{-2}}\right)^{-z},~{\rm where}~~~ \\
	\Sigma_{\rm thresh} & \equiv & 200\,{\rm M_{\odot}\,pc}^{-2}~, \nonumber \\
	t_0 & = & 3290\pm330~{\rm Myr}~,~~{\rm and} \nonumber \\ 
	z & = & 1.038\pm0.023~. \nonumber 
\end{eqnarray} 
Here $t_0$ and $z$ are obtained from the least-squares fit, although there is a large scatter in this relationship, $\sigma$ $\sim$ 0.5\,dex as seen in Figure \ref{tad}a.  This large scatter is partly due to uncertainties in the way the timescales are calculated, e.g., in the exact mass conversion law used to evaluate $\Sigma_{12}$, and in the values computed for the spatially resolved size scale $R_{\Sigma}$.  Despite this scatter, we have confidence that the method is sound, since there is good spatial consistency in the mass flows derived across each map.  Therefore, we believe the scatter is mostly inherent, and not due to systematics.  Eq.\,16 means that
\begin{equation}   
	\dot{\Sigma}(t) = \frac{\Sigma_{\rm thresh}}{t_0}\left(\frac{\Sigma_{\rm mol}}{{\rm M_{\odot}\,pc}^{-2}}\right)^z~.
\end{equation}
Because $z$ is so close to 1, this means that the mass accretion rate is approximately proportional to the accreted mass.  If we further assume that this mass accretion law remains valid over time for a given pixel or cloud, or is generally valid for all pixels, we infer that mass accumulates almost exponentially:
\begin{eqnarray}   
	\Sigma(t) & = & \Sigma_{\rm mol}\,e^{t/t_e}~,~~{\rm where} \\
			t_e & = & t_0/\Sigma_{\rm thresh} \nonumber \\
				& = & 16.5\pm1.6~{\rm Myr}~. \nonumber 
\end{eqnarray}
With Eq.\,18, the value of the threshold is not significant: changing that assumption merely shifts this relationship by a constant factor in $t_{\rm accr}$ and $t_0$, or by the inverse factor in $\dot{\Sigma}$.  That is, the value of $t_e$ doesn't depend on the details of this analysis, nor on the value of $\Sigma_{\rm mol}$.

The suggestion of almost exponential cloud growth is surprising, however, because the star formation that results from these clouds is so inefficient, primarily due to feedback effects in the ISM.  But our result is not inconsistent with slow star formation once we note the long time constant $t_0$ $>$ 1\,Gyr in the above formalism, which is the timescale that {\em would} be needed to build up a molecular cloud from the smallest discernible masses, $\sim$2\,M\solar/pc$^2$, if this process operated continuously over that mass range.  Given the turbulent nature of the ISM, especially at low densities, this is unlikely, to say the least.  The relevant time constant for more typical clouds at the average mapped $\Sigma_{\rm mol}$ $\sim$ 100--200\,M\solar\,pc$^{-2}$ is given by evaluating the constants in Eq.\,18.  Then, building up such clumps from roughly ambient levels (5--10\,M\solar) would take 3 $e$-folding times, or $\sim$50\,Myr.

\subsection{Mass Dispersal}

We also examine the result in Figure \ref{tad}b, which suggests that the time to disperse a given cloud is also approximately constant with $\Sigma_{\rm mol}$. 
This actually implies the inverse process as for accretion, namely a rapidly slowing dispersal akin to a pseudo-exponential decay.  Indeed, the slope in this panel is indistinguishable from 0; then
\begin{eqnarray}   
	\Sigma(t) & = & \Sigma_{\rm mol}\,e^{-t/t_1}~,~~{\rm where} \\
	t_1 & = & 10.9\pm0.9~{\rm Myr}~. \nonumber 
\end{eqnarray}
This means the dispersal timescale is approximately the same at all $\Sigma_{\rm mol}$, so in reality this dispersal must slow down over time.  For example, reducing a clump from 100--200\,M\solar\ to roughly ambient levels (5--10\,M\solar) would take 3 $e$-folding times, or $\sim$33\,Myr.  This dispersal relation also has a similar overall scatter to Eq.\,18, $\sim$0.6\,dex.

Although the mathematical forms seem similar, the physical situation between the accreting pixels and the dispersing ones is very different.  This is because the accretion of mass to the clumps {\em accelerates inwards} over time as the surface density increases, while the dispersal {\em decelerates outwards} as the surface density drops.  So for a cloud over which accretion and dispersal are occurring in different locations, over a long enough time interval the accretion should usually dominate over dispersal.  It is also interesting to note in Figure \ref{tad}a the $\Sigma$-distribution of all accreting pixels: there seems to be a distinct peak near $\Sigma_{\rm mol}$ = 100\,M\solar/pc$^2$.  Below this modal value down to ambient levels, the number of pixels per unit log($\Sigma_{\rm mol}$) tails off gently; above this value, however, the incidence of high-$\Sigma_{\rm mol}$ pixels drops off dramatically.  This may be closely related to where gas accretion seems to dominate dispersal: it is probably not a coincidence that the average accretion timescale matches the average dispersal timescale at this $\Sigma_{\rm mol}$.  Thus, it is tempting to suppose this is all connected: once accretion can overwhelm dispersal, the accumulation of gas (and consequent star formation) can proceed rapidly, potentially accounting for the star formation threshold $\Sigma_{\rm thresh}$ just above this level.

In short, the iso-CO data seem to reveal a detailed, semi-analytical process for slow, somewhat chaotic mass accretion onto molecular clouds, a process which one might call ``sedimentation,'' as denser parcels of gas gradually accrete onto each other despite jostling from nearby, dispersing parcels.  The mathematical form of the accretion/dispersal relations suggest strongly that gravity must play an important role in both processes, since it is the only force capable of producing an inwards acceleration in both situations.

\section{Discussion}\label{disc}
\subsection{Implications for Timescales}

Encouragingly, numerical simulations of star-forming molecular clumps, but without radiative transfer \citep[][and earlier work referenced therein]{pp16,cv16}, show distinct signs of contraction and dispersal in the 3D velocity field, and in a similar ratio of about half the clumps contracting and half dispersing, as in our derived mass flux maps (Fig.\,\ref{tad}).  Also, in some models \citep[e.g.,][]{zv14}, the star formation efficiency for clumps in our mass range rises rapidly towards later times, in agreement with our delayed star formation picture, and qualitatively with other observational \citep{hbh12} or numerical \citep{pp16} studies.  However, even the longer timescales of gravity- or turbulence-driven models, $\sim$10\,Myr \citep{hbh12,pp16} to 20--30\,Myr \citep[e.g.,][]{zv14,cv16}, fall short of our preferred range.  Some of these simulations do not include feedback, which may be significant: short of disrupting clouds, moderate feedback injects energy into the gas, delays collapse, lowers the SFE, and lengthens the effective SF timescale.  Feedback-oriented models \citep{i15,ki17,kk18} therefore align well with our $\sim$50\,Myr timescales.  We anticipate that once more feedback is included in other models, their timescales might also extend to 30--50\,Myr or longer, possibly matching our results better.

This potential convergence means our results could be remarkable.  For the first time, they may directly measure the assembly of massive, dense molecular clumps as first debated by \citet[][smooth, large velocity gradients]{gk74} and \citet[][turbulent ``swiss cheese'' structures and velocities]{ze75} (hereafter GK and ZE, resp.).  The pattern of assembly is intermediate between those debated scenarios, but concordant with modern simulations.  More significantly, {\em we have clearly defined the average timescales involved}, albeit with some local uncertainties.  In particular, the mass-accretion times needed to reach reasonable star-formation thresholds are very near the range of clump lifetimes first mooted by the purely demographic analysis that appeared in Paper I.  Also, the accelerating mass accretion onto clumps fits perfectly with the demographics of embedded clusters, which seem to arise only in the last $\sim$10\% of a cloud's lifetime, when massive star formation especially is enhanced, as described in Paper I and \citet{b13}, and supported by \citet{zv14}.

Even though the trends in Figure \ref{tad} seem very clean, the scatter from pixel to pixel and from Region to Region is still rather large, which means that local conditions can easily mask out this statistical sedimentation.  This is also manifest in the $t_{\rm accr/disp}$ maps (Fig.\,\ref{dynsample}), where the apparent pattern of accretion or dispersal seems to have little correlation with the column density or emission line structure of individual clouds (Figs.\,\ref{etacar},\ref{phsample},\ref{zgsample}).  It is then rather impressive that the sedimentation signal in Figure \ref{tad} is so clear.



To test these results, we should examine the assumptions built in to our plane-parallel LTE radiative transfer and cloud kinematics analysis.  For example, a numerical model with radiative transfer post-processing to simulate observed spectral lines \citep[e.g., like those of ][]{pc17,pc18,w18} could be analysed in our pipeline.  Then we can ask what conversion laws are derived under what conditions (cloud density/temperature profiles; non-LTE; radiation field; astrochemistry), and how any such variations change the implied mass flux rates or isotopologue ratios, as opposed to the true values in the physical models themselves.  (Is the pipeline itself valid?  Are the separate envelope/interior motions traceable as assumed?)  Such questions are not easy to answer, and are more properly the domain of separate studies, such as the recent work already cited.  While we construct tests of this scenario ourselves (Loughnane et al., in prep; Salome et al., in prep), we also encourage other workers to look into the same questions.

Fundamentally, however, our sedimentation picture of the assembly of molecular clouds is on a longer time scale (i.e., $>$50\,Myr, up to 100\,Myr) than in some prior observational studies of GMCs \citep[e.g.,][]{kmm09,mkt12}, or models which tend to focus on the molecular core scale near the end of the process described above \citep[e.g.,][]{mo07,LK14}.  Here it is important to distinguish between {\em clump} and {\em core} evolution timescales: we focus on the former.  Besides arguments based on demographics (Paper I), star formation activity \citep{b13}, or cloud stability (Paper III), our long timescale picture also provides dynamical evidence for gravity as the irresistible force that, on longer times and larger scales \& cloud samples, eventually overcomes the sometimes considerable local forces (turbulence, ionisation, winds, collisions) which act to disperse the cloud material.

The timescale of molecular cloud evolution has enjoyed a vigorous debate for many years after GK and ZE.  Several observational and theoretical studies of molecular clouds as a population \citep{bgp77,bs80} argue for cloud lifetimes $\sim$20\,Myr, or slightly longer or shorter, depending on the circumstances.  Upon re-examining these arguments in light of our more sensitive data and complete clump demographics \citep[Papers I, III,][]{b13}, we estimate that such lifetime estimates are typically raised by factors \gapp2, largely reconciling these studies with ours.  This increase arises mainly in the fact that we see more molecular material than before (much of it starless), without changing the overall star formation rate, thereby effectively lengthening the gas depletion timescale.

Other studies looking in detail at the embedded stellar population emerging from molecular clouds \citep{bhv99,e00,hbb01,bh07} provide a stronger contrast to our results.  They generally conclude that cloud evolution must be quick (a few Myr) because of the ``post-T Tauri problem,'' where typically, few stars of age \gapp5\,Myr are found in molecular cloud environments.  Some of these arguments focus on rapid {\em core} evolution timescales, upon which we cannot comment.  For {\em clumps}, however, we can now strongly counter such arguments thanks to a combination of new evidence: (1) the telling demographics of starless clumps \citep[Paper I,][]{b13}, a vast population only recently recognised; (2) their pressure-stabilisation from massive envelopes (Paper III), reducing the probability of starless clumps being ephemeral; (3) the much slower {\em differential} envelope/interior kinematics (sedimentation) seen here, compared to (e.g.) turbulent or bulk motions of clouds; and (4) the idea of slowly accelerating SF during clump evolution, supported both observationally via our mass accretion rates, and theoretically via the \citet{zv14} simulations.

\subsection{Comparison with Models}

Many approaches to modelling the properties and star formation activity of molecular clouds have appeared in the literature \citep[e.g., see reviews of][]{mo07,pf14}.  For example, large-scale ISM turbulence that cascades down to cluster/clump scales can reproduce many observed properties of molecular clouds, but requires most pc-scale clumps to be relatively ephemeral, and tends to generate prompt star formation in 5--10\,Myr \citep{pp16}.  Turbulent cloud simulations also tend to over-produce stars, with SFRs 2--4$\times$ higher than observed \citep{f15}, and have other inconsistencies with observations \citep{kf17}.  Ambipolar diffusion models support longer timescales and lower SFRs \citep{tm04,mt06}, but do not typically accommodate the chaotic environments of real clouds, and predictions of the $B$-$n$ relation are moderately at odds with observations \citep{c12}.  While understanding the role of turbulence and magnetic fields is important, we defer further comparisons between CHaMP data and MHD/turbulence theory for pending studies (Hernandez et al, in prep; Schap et al, in prep).

Still, \citet{bh04} and several subsequent studies found that gravitationally-driven hierarchical cloud collapse can explain many aspects of star formation.  These studies explored a wide range of physics, including colliding flows, magnetic fields, etc.  Starting with \citet{hh08}, such work collectively showed that timescales for cloud assembly and evolution were somewhat extended (\gapp10\,Myr) on GMC/cloud (10--100\,pc) and clump (1--10\,pc) scales, compared to the shorter core timescales ({\lapp}a few Myr).  Additionally, \citet{bhv11} argued that free-fall collapse of clouds can mimic virial equilibrium, such that the mean virial parameter $<$$\alpha$$>$ $\approx$ 2.  But they also argued that in this picture, external pressure confinement was not necessary to explain clouds' turbulent state (viewed instead as a {\em consequence} of gravitational collapse), contrary to our \tco\ results (Paper III).  \citet{zv14} showed that star formation accelerates during GMC evolution, while \citet{hvb15} showed that HI envelopes of molecular clouds participate in the large-scale gravitational accumulation of matter.

Another class of models focuses on cloud collisions and gas ``resurrection'' \citep{i15,ki17,kk18}, where star formation occurs in the presence of large-scale bubbles from SNe, OB associations, and HII regions that collect large amounts of gas in a self-sustaining process, over a long timescale of several 10s of Myr.  The inflow of material is also seen in filamentary models from the pc scale \citep[][and references therein]{a14} to Galactic shear models on the kpc scale \citep{s14,dd17}.

Among this limited sample, models partially match our CHaMP physics and similar results on long timescales derived in other contexts \citep[e.g.,][]{kss09}.  Given the now extensive suite of CHaMP results favouring long times for molecular cloud evolution, we prefer models that can incorporate such lifetimes into a consistent framework.  From our point of view, these 3 classes of models --- the hierarchical collapse \citep{zv14,cv16}, gas resurrection \citep{ki17}, and large-scale \citep{s14,dd17} models --- merit further attention.

In the latter case, the masses and flows involved are for spiral arm segments and GMFs (giant molecular filaments), so on a much larger scale than considered here.  But these provide the Galactic context in which pressure-confinement can be seen to operate over the long timescales available between passages of material through the spiral arm potential, $\sim$100\,Myr.  These periods permit the initially mostly atomic clouds to progressively increase their molecular fraction as they react to their changing environment, and we propose that this timing provides the setting for the clump evolution we examine here.

At intermediate scales, demographic evidence shows that GMCs in the Large Magellanic Cloud \citep{kmm09} and M33 \citep{mkt12}, evolve through 3 or 4 stages of star formation activity, lasting a total of 20--40\,Myr from early to late phases.  However, these studies were not sensitive to individual clumps, especially in M33 \citep{mkt12}.  We also consider the LMC clouds' ``early'' phase \citep{kmm09} more evolved than most of our clumps, which are more quiescent overall.  Thus, we consider the results of these studies to be consistent with the more general timescales we derive in \S\ref{kinem}.

We believe we can also link our long timescales to the much shorter phase of active cluster/massive star formation \citep[\lapp\,10\,Myr;][]{hbh12}, through a clump mass accretion rate which rises over the correct, longer timescale to produce the star formation we see at late times.  In this, the hierarchical collapse models do well \citep{zv14,hvb15,cv16}, since they incorporate both the temporal acceleration and the mechanism (gravity) to operate the process and mimic our results.  However, the timescales for those models are still somewhat shorter (around 20\,Myr) than in \S\ref{kinem}.  A reconciliation is achieved (1) at the higher $\Sigma_{\rm mol}$ \gapp\ 200\,M\solar\,pc$^{-2}$, where the accretion timescale does indeed drop below this range of model timescales, or (2) if all timescales are extended once the models include feedback.  For example, the cloud collision/gas resurrection \citep{i15,ki17,kk18} models match better our longer timing and the low overall SFE of molecular gas, and although their stochastic nature does not predict any systematic flows, their stochasticity at least matches the wide scatter in flow rates that we do see in our data.


The original observations of molecular cloud linewidths and cloud-to-cloud velocity dispersions of a few \kms\ suggest some sort of kinematic evolution over a few parsec scale in only 1\,Myr ({\em cf.}\ the original debate of GK and ZE).  The kinematic and post-T Tauri arguments have largely driven the debate over cloud (core) lifetimes for a considerable time.  In contrast, we now see that the {\em differential} motions between clump envelopes and their denser interiors may be dynamically significant, while the weight of the overlying envelopes seems sufficient to stabilise clump interiors against dispersal on any shorter timescale, as explained in Paper III.

As well, there are both observational and theoretical grounds to support a longer, but accelerating, star-formation timescale.  This includes the observed age spreads in young clusters, where the median age among PMS stars in an embedded cluster drops as the stellar mass rises, suggesting that the more massive stars form later \citep[][in NGC\,2024 and Wd2, resp.]{LL06,zg16}, and simulations showing the appropriate mass accretion can be quite extended in inter-arm clouds \citep{dpn14}, or star formation itself rising dramatically at late evolutionary times \citep{zv14}.  Furthermore, we can see directly in NIR images of our clumps \citep[][Jeram et al.\ in prep]{b13} that most do not reveal any discernible embedded star formation, suggesting again that the apparent post-T Tauri problem may in part be a selection effect.

Therefore, we believe our view of long latency periods for star formation during clump evolution may largely resolve some of the most vigorous debates about this process that have persisted for the previous 40 years, such as the role of turbulence and its naive implications for shorter cloud lifetimes, vs.\ the low star formation efficiency per free-fall time in clouds and the consequent requirement for longer evolutionary timescales.

This new sedimentation scenario also makes some valuable and timely predictions.  While we argue that \tco\ clump envelopes stabilise dense clump interiors, and the envelopes are also dynamically important for long-term evolution, there is already evidence that further layer(s) must contribute to and participate in this evolution.  These layers are the atomic material out of which the \htwo\ forms, and the dark molecular gas (DMG) layer, between the HI and fully molecular gas where \tco\ self-shields and is easily mapped.  Models that show \htwo\ forming from HI in gas participating in the gravitational inflow \citep{gm07a,gm07b}, and evidence that the DMG is widespread \citep{p13}, have already appeared, but for both the DMG and HI, we have not yet had the sensitivity to directly map their contribution to clump formation and evolution at the level described herein, and so to test further the above picture.

However, new instrumentation promises to enable such tests in the very near future.  Far-IR spectroscopy from {\em Herschel} and SOFIA \citep{LV14} is starting to reveal the detailed distribution of the DMG, while GASKAP \citep{d13} should soon be producing HI maps of sufficient sensitivity and resolution to be directly comparable to our Mopra data for the first time.  These additional comparisons should afford us the opportunity to start converging on an overarching paradigm for molecular cloud evolution and star formation, which in our view should include the large-scale (\gapp\,100\,pc), long-term (50--100\,Myr) evolution of the multi-phase ISM.

\section{conclusions}\label{conclude}

In this paper we have presented new results from complete iso-CO molecular line mapping (i.e., \tco, \ttco, and \ceto) of the CHaMP clouds, using the Mopra radio telescope.  The data were calibrated and reduced using the enhanced pipeline we developed and described in Paper III, including the SAM (smooth-and-mask) technique for computing moments from the data cubes.  This data release constitutes the most thoroughly-examined such sample of which we are aware: a comprehensive (all clouds containing dense gas over 120\,deg$^2$ towards the Carina arm) and sensitive (\tmb(rms) = 0.4--0.7\,K per 0.09\kms\ channel) survey of line emission in all three \joz\ lines from a large sample ($\sim$300 clouds in 36 Regions) of Galactic cluster-forming molecular clumps.  The main results of this work are:

\hspace{3mm} \put(5,3){\circle*{3}} \hspace{2mm} We have applied the radiative transfer analysis of \citet{bm15} to the iso-CO data in order to derive spatially-resolved maps of line ratios, optical depths, excitation temperatures, total CO column densities, relative abundances, mass surface densities, and kinematics (unbiased by the above effects) for these molecular clumps.

\hspace{3mm} \put(5,3){\circle*{3}} \hspace{2mm} We have re-examined the CO $\rightarrow$ \htwo\ mass conversion law first derived from the ThrUMMS analysis \citep{bm15}, and found that our more sensitive data allow us to explore and understand the radiative transfer physics behind this law, in a deeper way than has been possible previously.  The true, velocity-resolved mass conversion law, $N$ $\propto$ $I^p$, is a much steeper function of the \tco\ integrated intensity ($p$ almost 2) than the single standard $X$ factor ($p$ = 1), or even the ThrUMMS law ($p$ $\approx$ 1.4).  The radiative transfer physics is embedded in the above law, since we also find \nco\ $\propto$ $\tau$\tex.

\hspace{3mm} \put(5,3){\circle*{3}} \hspace{2mm} When integrated across the \tco\ line, the effective conversion law (i.e., one that can be used in other situations without our high velocity resolution) was much less steep ($p$ $\approx$ 1.3) than the velocity-resolved law.  This flatter law arises partially in the velocity binning or integration, and partially in the averaging across clouds with different physical conditions.  It is true in the mean, but misses the essential radiative transfer physics.

\hspace{3mm} \put(5,3){\circle*{3}} \hspace{2mm} In all cases, however, the conversion laws require a higher normalisation than implied by the standard $X$ factor, indicating (as found with ThrUMMS) that the Milky Way's molecular mass budget is underestimated by a factor of $\sim$2 compared to the $X$ factor approach.  This result includes suggestions that the average \tco\ abundance relative to \htwo\ in molecular clouds is $\sim$3$\times$ lower than the widely-used 10$^{-4}$.

\hspace{3mm} \put(5,3){\circle*{3}} \hspace{2mm} We use our new conversion law and sensitive data to study the differential mass distribution and kinematics in clouds, between the bulk of the mass in the clumps' interiors, and their envelopes as primarily traced by \tco.  Pending validation of this method by numerical simulations, we provisionally find that there are slow ($\sim$0.2\kms) but systematic line-of-sight velocity shifts between these layers, suggesting mass accretion and dispersal around the clumps (``sedimentation'') on long timescales, 10--100\,Myr.

\hspace{3mm} \put(5,3){\circle*{3}} \hspace{2mm} The spatial distribution of these mass flows appears uncorrelated with the clumps' emission or column density structure, when viewed clump by clump or Region by Region.  Aggregated over all the CHaMP clouds, however, the inferred mass accretion and dispersal timescales seem almost independent of the mass surface density; both relations have a large scatter.  Over long enough timescales, these results suggest accretion will eventually dominate over dispersal, since the accretion accelerates over time, while the dispersal decays over time.

In summary, we may have found the first direct evidence that star formation is enabled in molecular clouds by the slow, yet inexorable and global accumulation of mass into cluster-scale clumps, even though this accumulation can be masked by semi-random local conditions due to Galactic shear, cloud turbulence, and feedback from the star formation process itself.  If confirmed, this picture of a grand assembly through ``sedimentation'' ends when the accelerating inflow of material produces a crescendo of star formation activity, disrupting the inflow and revealing the newly-formed cluster.


\acknowledgments

We thank the anonymous referee for a thorough and constructive report, including several helpful suggestions and critiques, which broadened and clarified the discussion, and improved the paper.  We also thank the ATNF staff for their support of the Mopra telescope.  We acknowledge support from grants NSF-AST1312597 (PJB), NASA-ADAP NNX15AF64G (PJB and RLP), and NSF-AST1517573 (AKH).  PJB also thanks Prof Enrique V\'azquez-Semadeni and the star formation research group at IRyA-UNAM for many stimulating discussions.

Facilities: \facility{Mopra(MOPS)}.


\clearpage

\appendix

\section{CHaMP iso-CO Composite Images, Line Ratios, and X-factors}\label{rgbimages}

For each Region, we show here two 3-colour overlays of 2D projected integrals from the respective 3D line data cubes, plus another map and three plots from the analysis of the line cube data (sometimes spread over 2 {\em Journal} pages).  The top or top-left panel is an overlay of the velocity-integrated intensity (0$^{\rm th}$ moment) for \tco\ (red), \ttco\ (green), and \ceto\ (blue).  The second or top-right panel is a similar overlay but integrated across one spatial coordinate, so a 3-colour position-velocity diagram (either $lV$ or $Vb$, depending on the Region).  For both of these panels, grey contours are in \nco\ at intervals of 10\%--20\% of the peak (see Appendix \ref{physmaps} or \ref{zmaps} for numerical scales), and labelled white ellipses in the first panel show the clumps' half-power sizes and orientations as measured in \tco\ from Paper III.

The third panel plots, in the iso-CO ratio-ratio diagram (RRD), the distribution of voxels with high S/N data in all three cubes, similarly to Figure \ref{rrd}.  The curved grid lines in $R_{18}$ and $\tau_{18}$ indicate loci of the radiative transfer solutions, as labelled. 
The fourth panel plots the ratio \nco/\ico\ vs \ico\ for all voxels which allow an \nco\ solution, similarly to Figure \ref{convlaw}.  This is mostly limited by a S/N threshold in the \ttco\ data; therefore, this panel shows more points than the third panel.  The curved black grid lines again indicate loci of the radiative transfer solutions, this time in $\tau_{12}$ and \tex.  The parameters labelled in red are for a power-law fit to the binned data, typically above a 10$\sigma$ \ico\ noise limit (solid red and green curves); below this, the distribution of points becomes less complete and no fit is made (dotted curves).  These red and green curves respectively connect the $I$-bins' mean and $\pm\sigma$ \nco/\ico\ values.

The fifth panel gives a map of the equivalent $X$ factor formed from the ratio of integrals $\int$\nco\,d$V$/$\int$\ico\,d$V$ across the line emission in each pixel, and overlaid by the same \nco\ contours and ellipses as in the first panel.  Finally, the sixth panel plots this integral $X$ factor vs the velocity-integrated (moment-0) \ico\ for each pixel, analogously to panel four except for the integrations.  In this panel, no radiative transfer grid is shown since the integration range varies from pixel to pixel, effectively smearing the gridlines along the $x$ axis.

\vspace{3mm}The full Appendix A (87 Mb) is available at http://www.astro.ufl.edu/$\sim$pjb/research/champ/papers/champIV-appxA.pdf


\section{CHaMP iso-CO Maps of Physical Quantities}\label{physmaps}

Here we present moment maps for each Region of four derived parameters from the radiative transfer calculations: from top to bottom, the peak optical depth in the \tco\ line, peak excitation temperature, integrated \tco\ column density, and intensity-weighted mean abundance ratio $R_{18}$; these are continued to a second page in some cases.  The physical maps are shown in the left column, with an estimated error map for each in the right column.  Contours (10\%--20\% of the peak \nco, third row) and ellipses are the same as in Appendices \ref{rgbimages}, \ref{zmaps}, and \ref{dynmaps}.

\vspace{3mm}The full Appendix B (160 Mb) is available at http://www.astro.ufl.edu/$\sim$pjb/research/champ/papers/champIV-appxB.pdf


\section{CHaMP iso-CO Column Density Moments}\label{zmaps}

We provide here a standard set of moment maps for the mass surface density $\Sigma$ ($\propto$ \nco\ by Eq.\,7) in each Region.  Each map in a given Region is overlaid by the same set of contours in $\Sigma$, spaced by 10\%--20\% of the peak for that Region (see scale bar for values).  From top to bottom on the left hand side are the integrated $\Sigma$, the $\Sigma$-weighted mean \vlsr, and velocity dispersion $\sigma_V$.  On the right hand side we show the $\Sigma$ and velocity error maps.  The ellipses are the same as in Appendix \ref{rgbimages}.

\vspace{3mm}The full Appendix C (97 Mb) is available at http://www.astro.ufl.edu/$\sim$pjb/research/champ/papers/champIV-appxC.pdf


\section{CHaMP Clumps' Differential Dynamics}\label{dynmaps}

In these maps we illustrate the results of the differential kinematics and dynamics calculations for the clumps, as described in the text (\S\ref{kinem}), with the same contours and ellipses as in Appendices \ref{rgbimages}--\ref{zmaps}.  For the momentum, differential velocity, flux, and timescale panels, the zero-point colour is chosen (where the scale allows) to lie between between light and dark blue, in order to give a visually intuitive impression of the sign of the mass flow.  Positive values (light blue and ``warmer'' colours) correspond to mass accretion, while negative values (darker blue) correspond to mass loss.  The sixth panel is a pixel-by-pixel plot of the timescale (Eq.\,15) vs mass surface density (Appendix \ref{zmaps}).

\vspace{3mm}The full Appendix D (99 Mb) is available at http://www.astro.ufl.edu/$\sim$pjb/research/champ/papers/champIV-appxD.pdf

\clearpage

\end{document}